\documentclass[10pt,journal,final,oneside,twocolumn,nofonttune]{IEEEtran}
\usepackage{mathrsfs}

\usepackage{graphicx}
\usepackage{cite}
\usepackage{color}
\usepackage{colortbl}
\usepackage{psfrag}
\usepackage{subfigure}
\usepackage{amssymb}
\usepackage{epsfig}
\usepackage{pifont}
\usepackage{amsmath}
\usepackage{array}
\usepackage{multicol}
\usepackage{pifont}
\usepackage{indentfirst} 
\usepackage{amsfonts}
\usepackage{fancyhdr}
\usepackage{amscd}
\usepackage{bm}
\usepackage{extarrows}
\usepackage{multirow}
\usepackage{algorithm}
\usepackage{algorithmic}

\newtheorem{theorem}{Theorem}
\newtheorem{proposition}{Proposition}
\newtheorem{definition}{Definition}
\newtheorem{corollary}{Corollary}
\newtheorem{lemma}{Lemma}
\newtheorem{remark}{Remark}

\newtheorem{property}[proposition]{Property}

\makeatletter
\def\ifundefined{\@ifundefined}
\makeatother 

\makeatletter
\renewcommand{\citepunct}{,\penalty\@m\hskip.13emplus.1emminus.1em}
\renewcommand{\citedash}{\hbox{--}\penalty\@m}
\makeatother
\usepackage[bookmarks=false]{hyperref}

\begin{document}


\title{Genie Tree and Degrees of Freedom of the Symmetric MIMO Interfering Broadcast Channel}
\author{{ Tingting Liu}, {\em Member, IEEE} and { Chenyang Yang}, {\em Senior Member, IEEE}
\thanks{T. Liu and C. Yang are with the School of Electronics and Information Engineering, Beihang University, Beijing 100191, China. (E-mail: \{ttliu,cyyang\}@buaa.edu.cn)}}

\maketitle
\begin{abstract}
In this paper, we study the information theoretic maximal degrees of freedom (DoF) for the symmetric multi-input-multi-output (MIMO) interfering broadcast channel (IBC) with arbitrary antenna configurations. For the $G$-cell $K$-user $M\times N$ MIMO-IBC network, we find that the information theoretic maximal DoF per user are related to three DoF bounds: 1) the decomposition DoF bound $d^{\mathrm{Decom}}=MN/(M+KN)$, a lower-bound of linear interference alignment (IA) with infinite time/frequency extensions (called asymptotic IA); 2) the proper DoF bound $d^{\mathrm{Proper}}=(M+N)/(GK+1)$, an upper-bound of linear IA without time/frequency extensions (called linear IA); and 3) the quantity DoF bound $d^{\mathrm{Quan}}$, a zigzag piecewise linear function of $M$ and $N$. The whole region of $M/N$ can be divided into two regions, Region I: $\mathcal{R}^{\mathrm{I}}=\{M/N | d^{\mathrm{Proper}} < d^{\mathrm{Decom}} \}$ and Region II: $\mathcal{R}^{\mathrm{II}}=\{M/N | d^{\mathrm{Proper}} \geq d^{\mathrm{Decom}}\}$. Specifically, for most configurations in Region I, 
the information theoretic maximal DoF are the decomposition DoF bound and can be achieved by the asymptotic IA. For all configurations in Region II, the information theoretic maximal DoF are the quantity DoF bound and can be achieved by the linear IA.

To obtain the tight upper-bound, we propose a unified way to construct genies, where the genies help each base station or user resolve the maximal number of interference. According to the feature that the designed genies with the same dimension can derive identical DoF upper-bound, we convert the information theoretic DoF upper-bound problem into a linear algebra problem and obtain the closed-form DoF upper-bound expression. Moreover, we develop a non-iterative linear IA transceiver to achieve the DoF upper-bound for the networks with antenna configurations in Region II, which means that the derived DoF upper-bound is tightest. The basic principles to derive the DoF upper-bound and design the linear IA transceiver to achieve the DoF upper-bound can be extended into general asymmetric networks.
\end{abstract}


\setlength \arraycolsep{1pt}
\section{Introduction}
Inter-cell interference (ICI) is a bottleneck of improving the sum capacity of multi-cell networks, especially when multi-input-multi-output (MIMO) techniques are used. The degrees of freedom (DoF) can reflect the potential of interference networks, which are the first-order approximation of sum capacity at high signal-to-noise ratio (SNR) regime~\cite{Jafar2007b,Jafar_3cell}. Recently, significant research efforts have been devoted to find the DoF for MIMO interference channel (MIMO-IC)~\cite{Jafar2007b,Jafar_3cell,Cadambe2008b,Tse_3cell_2011,Jafar2010DoF,Ghasemi2010,Jafar_IAchainKcell_all,Gonzalez2014,Ruan2013}, MIMO interfering broadcast channel (MIMO-IBC)~\cite{YuWei2013,Zhuguangxi2010,Park2011}, and MIMO interfering multiple access (MIMO-IMAC)~\cite{IBC_L_cell_Lovel} networks.

For the symmetric $G$-cell $M \times N$ MIMO-IC network where each base station (BS) and each mobile station (MS) have $M$ and $N$ antennas, respectively, when the channels are time/frequency varying and $M = N$, the information theoretic maximal DoF per user are $M/2$. This implies that the sum DoF can linearly increase with $G$, and the performance of the interference network is not interference-limited~\cite{Cadambe2008b}.
Encouraged by such a promising result, many recent works strive to analyze the information theoretic maximal DoF for the MIMO-IC~\cite{Jafar_3cell,Tse_3cell_2011,Cadambe2008b,Jafar2007b,Jafar2010DoF,Ghasemi2010,Jafar_IAchainKcell_all} and MIMO-IBC~\cite{YuWei2013,Zhuguangxi2010,Park2011} networks with various antenna configurations and devise interference management techniques to achieve the information theoretic maximal DoF.

For the two-cell MIMO-IC network, the information theoretic maximal DoF were obtained with all antenna configurations in \cite{Jafar2007b} and can be achieved by the linear interference alignment (IA). For the symmetric three-cell $M\times N$ MIMO-IC network, the information theoretic maximal DoF were obtained as a piecewise linear function of $M$ and $N$ in \cite{Jafar_3cell} and achievable by the linear IA. For the symmetric two-cell $K$-user $M \times N$ MIMO-IBC network where $K=2,3$, the information theoretic maximal DoF were obtained with all antenna configurations in \cite{YuWei2013} and achievable by the linear IA. These results show that when $G=2$, $K=1,2,3$ and $G=3$, $K=1$, the linear IA can achieve the information theoretic maximal DoF.

However, for the networks where $G=2,~K\geq 4$, $G=3,~K\geq 2$, and $G\geq 4,~K\geq 1$, the information theoretic maximal DoF were only obtained with special configurations.\footnote{The configurations are summarized in Table \ref{Table:Existing_DoF}.} For many antenna configurations in these networks, the information theoretic maximal DoF cannot be achieved by the linear IA, but can be achieved by the asymptotic IA in \cite{Cadambe2008b}.

Considering that infinite time/frequency extensions of asymptotic IA are not viable for practical networks, many studies investigated the maximal DoF achieved by the linear IA. For the MIMO-IC network with constant coefficients, the feasible conditions of linear IA, called \emph{IA conditions} for short, were derived in~\cite{Gomadam2008}. The IA conditions contain a class of rank constraints to receive the desired signals and a class of zero-forcing constraints to eliminate all ICIs. A \emph{proper condition} was first proposed in~\cite{Yetis2010}, which requires that the number of variables to cancel the ICIs is greater than or equal to the number of ICIs.
When the channels in the MIMO-IC network are \emph{generic} (i.e., drawn from a continuous probability distribution), the authors in~\cite{Tse2011,Luo2012} proved that the proper condition is a necessary condition for the feasibility of linear IA. The proper condition was proved to be sufficient for two cases: 1) $M=N$ in \cite{Tse2011} and 2) either $M$ or $N$ is divisible by the number of data streams per user $d$~\cite{Luo2012}. The study in \cite{Ruan2013} investigated the sufficient conditions of linear IA feasibility for the asymmetric MIMO-IC. In \cite{Gonzalez2014}, the authors developed a polynomial complexity test that allows a complete characterization of the achievable DoF region by the linear IA for the symmetric MIMO-IC with
arbitrary antenna configuration. For the MIMO-IBC network with generic channels, the proper condition was proved to be necessary in~\cite{IBC_Honig,TTTSP2013} and to be sufficient for two cases: 1) $M=N+(K-1)d$ by extending the result in \cite{Tse2011} and 2) either $M$ or $N$ is divisible by $d$ in~\cite{TTTSP2013}. For general cases, these results did not answer when the proper condition is sufficient and when the linear IA can achieve the information theoretic maximal DoF.

So far, the following questions are still open for general MIMO-IC and MIMO-IBC networks: \begin{enumerate}
 \item What are the information theoretic maximal DoF?
 \item How to derive the information theoretic maximal DoF? How to design IA transceiver to achieve the information theoretic maximal DoF?
 \item Why some linear IA transceivers that satisfy the proper condition are infeasible?
\end{enumerate}

This paper strives to answer these questions. The main contributions are summarized as follows,
\begin{enumerate}
\item We obtain the information theoretic maximal DoF for the symmetric $G$-cell $K$-user $M \times N$ MIMO-IBC network. The whole region of $M/N$ is divided into two parts, Region I and Region II. For most configurations in Region I, the information theoretic maximal DoF are $MN/(M+KN)$, which are achievable by the asymptotic IA.\footnote{This study considers the \emph{asymptotic IA} proposed in \cite{Cadambe2008b}, which is a linear IA with infinite time/frequency extensions.} For all configurations in Region II, the information theoretic maximal DoF are the piecewise linearly function of $M$ and $N$, which are achievable by the linear IA.\footnote{This study considers the \emph{linear IA} discussed in \cite{Yetis2010}, which is a linear IA with no time/frequency extensions. The finite spatial extensions~\cite{Jafar_3cell} are allowed to avoid the loss of DoF caused by restricting the value of DoF to an integer.}
\item We provide a unified method to construct genies to derive the tight DoF upper-bound for the networks with general configurations, which is different from the methods in \cite{Jafar_IAchainKcell_all,Jafar_3cell} where different ways were developed for different configurations. By dividing the ICIs into resolvable ICIs and irresolvable ICIs, we construct the genie to assist each BS or user to resolve the maximal number of irresolvable ICIs in order to derive the tightest upper-bound. By using the feature that identical DoF upper-bound can be obtained from the genies with the same dimension, we convert the information theoretic DoF upper-bound problem into a linear algebra problem to obtain the closed-form DoF upper-bound expression. According to the structure of genie tree, we obtain the tightest DoF upper-bound and develop a non-iterative linear IA transceiver to achieve the DoF upper-bound with all antenna configurations in Region II.
\item We find that there exists a kind of ICIs\footnote{The ICIs are irreducible ICIs defined in \cite{TTTSP2013}.} that must be solely eliminated by the BSs or the users but cannot be jointly eliminated by the BSs and users. The proper condition does not ensure the system to provide enough antennas to eliminate this kind of ICIs, which explains why some linear IA transceivers are proper but infeasible.
\end{enumerate}

The basic principles of constructing the genie tree to derive the DoF upper-bound and design the feasible linear IA transceiver can be extended into asymmetric MIMO-IC, MIMO-IBC and MIMO-IMAC networks with general antenna configurations.

The rest of the paper is organized as follows. We describe the system model in Section~\ref{Sec:System model}. We present the information theoretic maximal DoF in Section~\ref{Sec:DoF} and prove the results in Sections~\ref{Sec:Proof1} and~\ref{Sec:Proof2}, respectively. Conclusions are given in the last section.

\emph{Notations:} Transpose, Hermitian transpose and expectation are represented by $(\cdot)^{T}$, $(\cdot)^{H}$, and $\mathbb{E}\{\cdot\}$, respectively. $\mathrm{diag}\{\cdot\}$ is a block diagonal matrix.
For the vector $\pmb{x}$, $\pmb{x}(i)$ denotes its $i$th variable, and $\pmb{x}(i:j)$ denotes a row vector obtained from its $i$th to $j$th variables. Similarly, for the matrix $\pmb{X}$, $\pmb{X}(i)$ denotes its $i$th column vector, and $\pmb{X}(i:j)$ denotes a matrix obtained from its $i$th to $j$th columns.

\section{System Model and Definition}\label{Sec:System model}
\subsection{System Model}
Consider a $G$-cell multiuser MIMO network where $G$ BSs each with $M$ antennas serves $K$ users each with $N$ antennas. All channel state information is assumed available at all nodes. In information theory terminology, this is a symmetric MIMO-IBC network, denoted as the $G$-cell $K$-user $M \times N$ MIMO-IBC network.

The information theoretic maximal DoF for the MIMO-IBC network are equal to those for the MIMO-IMAC network. Nevertheless, since the signal models of both MIMO-IBC and MIMO-IMAC networks are necessary to derive the DoF, we provide the signal models of both networks.

In the MIMO-IMAC network, the received signals of the $j$th BS (denoted by BS$_j$) can be expressed as
\begin{align}\label{Eq:Received_signal_IMAC}
\pmb{y}_j
=&\sum_{i=1}^{G}\sum_{k=1}^{K}\pmb{H}_{j,i_k}\pmb{x}_{i_k}+\pmb{n}_j
\end{align}
where $\pmb{x}_{i_k} \in \mathbb{C} ^{N \times 1}$ is a vector of transmit signals of the $k$th MS in the $i$th cell (denoted by MS$_{i_k}$), $\pmb{H}_{j,i_k} \in \mathbb{C}^{M\times N}$ is the channel matrix from MS$_{i_k}$ to BS$_j$, whose elements are independent and identically distributed (i.i.d.) random variables with a continuous distribution, and $\pmb{n}_{j}\in \mathbb{C}^{M\times 1}$ is the zero-mean additive white Gaussian noise of BS$_j$ with covariance matrix $\mathbb{E}\left\{\pmb{n}_{j}\pmb{n}_{j}^{H}\right\}=\sigma_{n}^{2}\pmb{I}_M$.

In the MIMO-IBC network, the received signals of MS$_{i_k}$ can be expressed as
\begin{align}\label{Eq:Received_signal_IBC}
\pmb{y}_{i_k}
=&
\sum_{j=1}^{G}\pmb{H}_{j,i_k}^{T}\pmb{x}_j+\pmb{n}_{i_k}
\end{align}
where $\pmb{x}_j \in \mathbb{C} ^{M \times 1}$ is a vector of transmit signals of BS$_j$, and $\pmb{n}_{i_k}\in \mathbb{C}^{N\times 1}$ is the zero-mean additive white Gaussian noise of MS$_{i_k}$ with covariance matrix $\mathbb{E}\left\{\pmb{n}_{i_k}\pmb{n}_{i_k}^{H}\right\}=\sigma_{n}^{2}\pmb{I}_N$.

%
%

%

\begin{definition}\label{DoF with finite spatial extension}
The \emph{DoF per user with finite spatial extensions} are defined as \cite{Jafar_3cell}
\begin{align}\label{Eq:DoF_Define_Spatial_Extension}
d(M,N)\triangleq\max_{m\in \mathbb{Z}^{+}}\left\{\frac{ \bar{d}(m M,m N)}{m}\right\}
\end{align}
where $\bar{d}(M,N)\triangleq \lim_{\gamma \rightarrow \infty} {R(\gamma)}/{ \log \gamma}$ are the DoF per user without spatial extensions, $R(\gamma)$ is the achievable rate per user for the given SNR $\gamma$, and $m$ is a finite integer and $\mathbb{Z}^{+}$ is the set of positive integers. It means that the DoF of $m{d}(M,N)$ are achievable for each user when each BS and each user are equipped with $m M$ and $m N$ antennas. With finite spatial extensions, the value of DoF per user is not necessary to be an integer, which avoids the loss of DoF due to rounding.
\end{definition}

For notational simplicity, in the sequel we omit $M,N$ and use $d$ to denote the DoF per user as well as the number of data streams able to be supported for each user.

\subsection{Related Notions}
In the forthcoming analysis, we will show that the expression and derivation of information theoretic maximal DoF are related to the notions of \emph{generalized continued fraction sequence}~\cite{Continued_Fractions1980}
and \emph{tree} in graph theory~\cite{Graph_Theory}. For readers' convenience, we first briefly introduce these notions. 

\subsubsection{Continued Fraction}
\begin{definition}\label{Def:Cont_frac}
A generalized continued fraction sequence is a sequence in the form~\cite{Continued_Fractions1980}
\begin{align}\label{Eq:Def_Cont_frac}
x_{n}=\beta_{0}+\cfrac{\alpha_1}{\beta_{1}+\cfrac{\alpha_2}{\beta_{2}+\cfrac{\alpha_3}{ \cdots + \cfrac{\alpha_{n}}{\beta_{n}}}}},\forall n\in \mathbb{Z}^{+}
\end{align}
%
\end{definition}

The generalized continue fraction sequence has the following properties~\cite{Continued_Fractions1980}.
\begin{property}\label{Property:Fibonacci_sequence}
The generalized continue fraction sequence in \eqref{Eq:Def_Cont_frac} can be expressed as a ratio of a generalized Fibonacci sequence-pair, i.e.,
\begin{align}\label{Eq:Def_Fibonacci}
 x_{n}={q_{n}}/{p_{n}}
\end{align}
where $\{q_{n},p_{n}\}$ is the generalized Fibonacci sequence-pair satisfying $\{q_{-1},p_{-1}\}=\{1,0\}$, $\{q_{0},p_{0}\}=\{\beta_{0},1\}$, and
\begin{align*}
\{q_{n},p_{n}\} =(\beta_{n}q_{n-1}+\alpha_{n}q_{n-2},\beta_{n}p_{n-1}+\alpha_{n}p_{n-2}),\forall n\in \mathbb{Z}^{+}
\end{align*}
\end{property}

\begin{property}\label{Property:Series}
The generalized continue fraction sequence in \eqref{Eq:Def_Cont_frac} can be expressed in the form of series as
\begin{align}\label{Eq:Def_Series}
 x_{n}=\beta_{0}+\sum_{m=1}^{n-1}\frac{(-1)^{m}\prod_{i=1}^{m-1}\alpha_{i}}{p_{m}p_{m+1}}, \forall n\in \mathbb{Z}^{+}
\end{align}
\end{property}

\subsubsection{Tree}
\begin{figure}[htb!]
\centering
\includegraphics[width=0.5\linewidth]{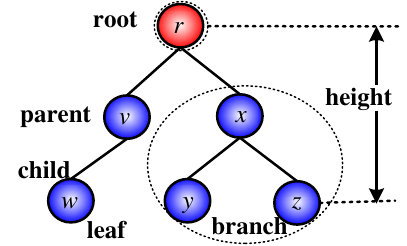}
\caption{Basic notions of rooted tree in graph theory.} \label{fig:tree}
\end{figure}

In graph theory~\cite{Graph_Theory}, a \emph{tree} is a connected graph without cycles, as shown in Fig. \ref{fig:tree}. A \emph{rooted tree} is a tree with a designated vertex called the \emph{root}, e.g., vertex $r$. In a rooted tree, the \emph{depth} of vertex $v$ is the distance from the root to the vertex. The \emph{height} is the greatest depth. The height of the tree in Fig. \ref{fig:tree} is two. If vertex $v$ immediately precedes vertex $w$ on the path from the root to $w$, then $v$ is the \emph{parent} of $w$ and $w$ is the \emph{child} of $v$. A \emph{leaf} is any vertex having no children, e.g., vertexes $w$, $y$, and $z$. A \emph{branch} is a subtree without the root. A branch with the greatest height is called the \emph{maximal branch}. In Fig. \ref{fig:tree}, vertexes $x$, $y$, and $z$ constitute a maximal branch.

\section{Information Theoretic Maximal DoF}\label{Sec:DoF}
In this section, we first present and compare several DoF bounds for symmetric MIMO-IBC networks. Then, we present the information theoretic maximal DoF to be proved later and show the connection with existing results in the literature.

\subsection{DoF Bounds}
\subsubsection{Decomposition DoF Bound}
For the $G$-cell $M \times N$ MIMO-IC, the \emph{decomposition DoF bound} was derived as $d^{\mathrm{Decom}}=MN/(M+N)$. It was a lower-bound of DoF per user introduced in~\cite{Jafar2010DoF} and achievable by the asymptotic IA through decomposing the antennas of both BSs and users.

Using the similar way of decomposing the antennas of the BSs, a $G$-cell $K$-user $M \times N$ MIMO-IBC network can be decomposed to a $GK$-cell $M/K\times N$ MIMO-IC network.\footnote{Since we consider finite spatial extensions, the number of antennas is no necessary to be an integer.} Then, the \emph{decomposition DoF bound} for the symmetric MIMO-IBC network can be obtain as
\begin{align}\label{Eq:Decomposition DoF}
 d^{\mathrm{Decom}} = \frac{M/K\cdot N}{M/K+N}= \frac{MN}{M+KN}
\end{align}
which is achievable by the asymptotic IA designed for the MIMO-IC network in \cite{Cadambe2008b}.

From \eqref{Eq:Decomposition DoF}, we know that the achievable DoF per user are independent of the number of cells $G$. It implies that the achieved sum DoF of the network can linearly increase with $G$.

\subsubsection{Proper DoF Bound}
In~\cite{Yetis2010}, a \emph{proper condition} was proposed, which has been proved to be necessary for the feasibility of linear IA in the MIMO-IC~\cite{Tse2011,Luo2012} and MIMO-IBC networks~\cite{TTTSP2013}, when the channels are generic. For the symmetric MIMO-IBC network, the proper condition was~\cite{TTTSP2013}
\begin{align}\label{Eq:Proper_condition}
 M+N\geq (GK+1)d
\end{align}

We call the upper-bound of DoF per user obtained from the proper condition the \emph{proper DoF bound}. From \eqref{Eq:Proper_condition}, the \emph{proper DoF bound} is obtained as
\begin{align}\label{Eq:Proper DoF}
 d^{\mathrm{Proper}}= \frac{M+N}{GK+1}
\end{align}
Then, the upper-bound of the sum DoF is $GK(M+N)/(GK+1)$. As $G$ or $K$ increases, the sum DoF are bounded by $M+N$.

%
\begin{figure}[htb!]
\centering
\includegraphics[width=0.9\linewidth]{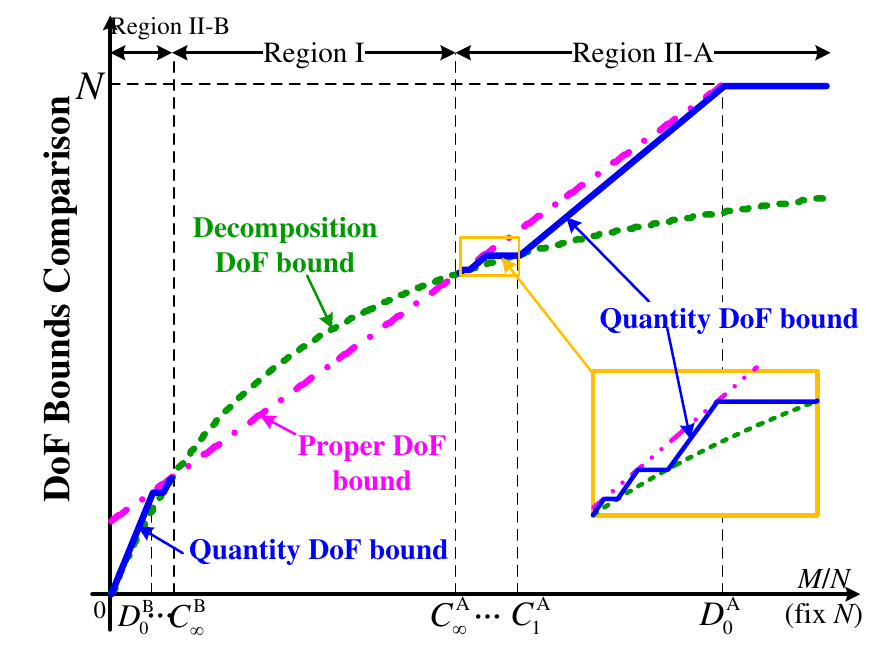}
\caption{Comparison of three DoF bounds for the symmetric MIMO-IBC.} \label{fig:DoF_Achievable}
\end{figure}

In Fig.~\ref{fig:DoF_Achievable}, we compare the decomposition DoF bound and the proper DoF bound with different values of $M/N$. For some values of $M/N$, the decomposition DoF bound is greater than the proper DoF bound, while for others, the decomposition DoF bound is less than the proper DoF bound. For easy exposition, we divide the whole region of $M/N$ into two parts, called Region I and Region II and denoted by $\mathcal{R}^{\mathrm{I}}$ and $\mathcal{R}^{\mathrm{II}}$, respectively.
Specifically, we define the two regions as
\begin{subequations}\label{Eq:Def_Region}
\begin{align}\label{Eq:Def_RegionI}
\mathcal{R}^{\mathrm{I}}&\triangleq\left\{M/N | d^{\mathrm{Proper}} < d^{\mathrm{Decom}} \right\}\\
\label{Eq:Def_RegionII}
\mathcal{R}^{\mathrm{II}}&\triangleq\left\{M/N | d^{\mathrm{Proper}} \geq d^{\mathrm{Decom}}\right\}
\end{align}
\end{subequations}

By substituting \eqref{Eq:Decomposition DoF} and \eqref{Eq:Proper DoF} into \eqref{Eq:Def_Region}, it is not difficult to express Regions I and II as
\begin{subequations}\label{Eq:Region}
\begin{align}\label{Eq:RegionI}
\mathcal{R}^{\mathrm{I}}&= \left(\xi^{\mathrm{B}},\xi^{\mathrm{A}}\right),~\forall
\left\{
\begin{array}{l}
 K\geq 4, G=2 \\
 K\geq 1, G\geq 3
\end{array}
\right.\\
\label{Eq:RegionII}
\mathcal{R}^{\mathrm{II}}&=
\left\{ \begin{array}{ll}
\left(0,\infty\right),&~\forall K\leq 3, G=2\\
\left(0,\xi^{\mathrm{B}}\right]\cup \left[\xi^{\mathrm{A}},\infty\right),& ~\mathrm{otherwise.} \\
\end{array}
 \right.
\end{align}
\end{subequations}
where $\xi^{\mathrm{A}}=((G-1)K+\sqrt{(G-1)^2K^2-4K})/2$ and $\xi^{\mathrm{B}}=((G-1)K-\sqrt{(G-1)^2K^2-4K})/2$.

\subsubsection{Quantity DoF Bound}
In the forthcoming analysis, we will provide a DoF bound that is a piecewise linear function of $M$ and $N$ and prove that the DoF bound is the information theoretical maximal DoF. When $G=3$ and $K=1$, the DoF bound reduces to the \emph{quantity DoF} introduced in~\cite{Jafar_3cell}. Therefore, we call it the \emph{quantity DoF bound}.

For the symmetric MIMO-IBC network, the \emph{quantity DoF bound} to be proved later is expressed as
\begin{align}\label{Eq:Quantity DoF}
d^{\mathrm{Quan}}=\left\{ \begin{array}{ll} \min\left\{\frac{M}{K+C_{n}^{\mathrm{A}}},\frac{N}{1+\frac{K}{C_{n-1}^{\mathrm{A}}}}\right\}, &~\forall M/N\in\mathcal{R}^{\mathrm{A}}_{n} \\ \min\left\{\frac{M}{K+C_{n-1}^{\mathrm{B}}},\frac{N}{1+\frac{K}{C_{n}^{\mathrm{B}}}}\right\}, &~\forall M/N\in\mathcal{R}^{\mathrm{B}}_{n}
\end{array}
\right.
\end{align}
where $\mathcal{R}^{\mathrm{A}}_{n}\triangleq \left[C_{n}^{\mathrm{A}},C_{n-1}^{\mathrm{A}}\right)$, $\mathcal{R}^{\mathrm{B}}_{n}\triangleq \left(C_{n-1}^{\mathrm{B}},C_{n}^{\mathrm{B}}\right]$,
$C_{0}^{\mathrm{A}}=\infty$, $C_{0}^{\mathrm{B}}=0$, and
\begin{subequations}\label{Eq:Csequence_Recursive}
\begin{align}
\label{Eq:Csequence_RecursiveA} C_{n}^{\mathrm{A}}\triangleq &(G-1)K-{K}/{C_{n-1}^{\mathrm{A}}}
\\ \label{Eq:Csequence_RecursiveB}
C_{n}^{\mathrm{B}}\triangleq &\frac{1}{G-1-{C_{n-1}^{\mathrm{B}}}/{K}}
\end{align}
\end{subequations}
satisfying $C_{n}^{\mathrm{A}},C_{n}^{\mathrm{B}}\geq0,~\forall n \in \mathbb{Z}^{+}, n\leq n^{\max}$ with
\begin{align}\label{Eq:n_max}
n^{\max} = \left\{\begin{array}{ll}
 2, & \forall G=2,K=1 \\
 3, & \forall G=2,K=2 \\
 5, & \forall G=2,K=3 \\
 \infty, & \mathrm{otherwise.}\end{array}\right.
\end{align}

To understand the properties of the quantity DoF bound, we show several values of $\{C_{n}^{\mathrm{A}}\}$ and $\{C_{n}^{\mathrm{B}}\}$ in Table \ref{Table:CN}.
\begin{table}[htb!]\centering
\caption{Values of $\{C_{n}^{\mathrm{A}}\}$ and $\{C_{n}^{\mathrm{B}}\}$}\label{Table:CN}
\begin{tabular}{c|c|c}
 \hline
$n$ & $C_{n}^{\mathrm{A}}$ & $C_{n}^{\mathrm{B}}$ \\
 \hline
0 & $\infty$ & $0$\\
1 & $(G-1)K$ & $\cfrac{1}{G-1}$\\
2 & $(G-1)K-\cfrac{1}{G-1}$ & $\cfrac{1}{G-1-\cfrac{1}{(G-1)K}}$\\
3 & $(G-1)K-\cfrac{1}{G-1-\cfrac{1}{(G-1)K}}$ & $\cfrac{1}{G-1-\cfrac{1}{(G-1)K-\cfrac{1}{G-1}}}$\\
\vdots & \vdots & \vdots \\
 \hline
\end{tabular}
\end{table}

From Table \ref{Table:CN}, we can see that $C_{n}^{\mathrm{A}}$ and $C_{n}^{\mathrm{B}}$, $\forall n \in \mathbb{Z}^{+}, n\leq n^{\max}$ are the generalized continued fraction sequences as follows,
\begin{subequations}\label{Eq:Csequence_conti_frac}
\begin{align}
C_{n}^{\mathrm{A}}&=(G-1)\bar{K}_{0}-\cfrac{1}{(G-1)\bar{K}_{1}-\cfrac{1}{\cdots \cfrac{1}{(G-1)\bar{K}_{n-1}}}}\\
C_{n}^{\mathrm{B}}&=\cfrac{1}{(G-1)\bar{K}_{1}-\cfrac{1}{(G-1)\bar{K}_{2}- \cfrac{1}{\cdots \cfrac{1}{(G-1)\bar{K}_{n}}}}}
\end{align}
\end{subequations}
where
\begin{align}\label{Eq:K_Sequence}
\bar{K}_{n}\triangleq \left\{ \begin{array}{cc}
 K, &~\forall n\mathrm{~is~even.}\\
 1, &~\forall n\mathrm{~is~odd.} \end{array}\right.
\end{align}

According to Property~\ref{Property:Fibonacci_sequence}, $C_{n}^{\mathrm{A}}$ and $C_{n}^{\mathrm{B}}$, $\forall n \in \mathbb{Z}^{+}, n\leq n^{\max}$ can be rewritten as the ratios of generalized Fibonacci sequence-pairs, i.e.,
\begin{align}\label{Eq:C_pq}
 C_{n}^{\mathrm{A}}= {q_{n}^{\mathrm{A}}}/{p_{n}^{\mathrm{A}}}, ~C_{n}^{\mathrm{B}}= {q_{n}^{\mathrm{B}}}/{p_{n}^{\mathrm{B}}}
\end{align}
where $\left(p_{n}^{\mathrm{A}},q_{n}^{\mathrm{A}}\right)$ and $ \left(p_{n}^{\mathrm{B}},q_{n}^{\mathrm{B}}\right)$ are the generalized Fibonacci sequence-pairs satisfying $\left(p_{0}^{\mathrm{A}},q_{0}^{\mathrm{A}}\right)=(0,1)$, $\left(p_{-1}^{\mathrm{A}},q_{-1}^{\mathrm{A}}\right)=(-1,0)$, $\left(p_{0}^{\mathrm{B}},q_{0}^{\mathrm{B}}\right)=(1,0)$, $\left(p_{-1}^{\mathrm{B}},q_{-1}^{\mathrm{B}}\right)=(0,-1)$, and
\begin{subequations}\label{Eq:Fsequence_AB}
\begin{align}\label{Eq:Fsequence_A}
&\left\{\begin{array}{ll}
p_{n}^{\mathrm{A}}&\triangleq(G-1)\bar{K}_{n-1}p_{n-1}^{\mathrm{A}}-p_{n-2}^{\mathrm{A}}\\
q_{n}^{\mathrm{A}}&\triangleq(G-1)\bar{K}_{n-1}q_{n-1}^{\mathrm{A}}-q_{n-2}^{\mathrm{A}}\\
\end{array}\right.\\
&\label{Eq:Fsequence_B}
\left\{
\begin{array}{ll}
p_{n}^{\mathrm{B}}&\triangleq(G-1)\bar{K}_{n} p_{n-1}^{\mathrm{B}}-p_{n-2}^{\mathrm{B}}\\
q_{n}^{\mathrm{B}}&\triangleq(G-1)\bar{K}_{n} q_{n-1}^{\mathrm{B}}-q_{n-2}^{\mathrm{B}}\\
\end{array}\right.
\end{align}
\end{subequations}
satisfying $p_{n}^{\mathrm{A}},q_{n}^{\mathrm{A}},p_{n}^{\mathrm{B}},q_{n}^{\mathrm{A}}>0,~\forall n \in \mathbb{Z}^{+}, n\leq n^{\max}$.

According to Property~\ref{Property:Series}, $C_{n}^{\mathrm{A}}$ and $C_{n}^{\mathrm{B}}$, $\forall n \in \mathbb{Z}^{+}, n\leq n^{\max}$ can be rewritten as the series as follows,
\begin{subequations}\label{Eq:Csequence_series}
\begin{align}\label{Eq:Csequence_serieA}
C_{n}^{\mathrm{A}} &= (G-1)K - \sum_{m=1}^{n-1}\frac{1}{p_{m}^{\mathrm{A}}p_{m+1}^{\mathrm{A}}}\\
\label{Eq:Csequence_serieB}
C_{n}^{\mathrm{B}} &= \sum_{m=1}^{n-1}\frac{1}{p_{m}^{\mathrm{B}}p_{m+1}^{\mathrm{B}}}
\end{align}
\end{subequations}

Since $p_{m}^{\mathrm{A}},p_{m}^{\mathrm{B}}\geq 0,~\forall m\in \mathbb{Z}^{+},~m\leq n^{\max}$, $\left\{C_{n}^{\mathrm{A}}\right\}$ is a monotonically decreasing sequence, whereas $\left\{C_{n}^{\mathrm{B}}\right\}$ is a monotonically increasing sequence.

In \eqref{Eq:Quantity DoF}, $d^{\mathrm{Quan}}$ is a piecewise function defined in $M/N\in\mathcal{R}^{\mathrm{A}}\cup\mathcal{R}^{\mathrm{B}}$, where $\mathcal{R}^{\mathrm{A}}\triangleq \cup_{n=1}^{n^{\max}}\mathcal{R}^{\mathrm{A}}_{n}$ and $\mathcal{R}^{\mathrm{B}}\triangleq \cup_{n=1}^{n^{\max}}\mathcal{R}^{\mathrm{B}}_{n}$, respectively. According to the derivation in Appendices \ref{App_Region} and \ref{App_DoF_Comparison},
we obtain the feature of $d^{\mathrm{Quan}}$ in the following lemmas.

\begin{lemma}\label{Lemma_Region}
$\mathcal{R}^{\mathrm{A}}$ and $\mathcal{R}^{\mathrm{B}}$ are two subregions of Region II, i.e.,
\begin{align}\label{Eq:RegionIIAB}
 \mathcal{R}^{\mathrm{II}}=\mathcal{R}^{\mathrm{A}}\cup \mathcal{R}^{\mathrm{B}}
\end{align}
\end{lemma}

Lemma \ref{Lemma_Region} suggests that the quantity DoF bound $d^{\mathrm{Quan}}$ is a piecewise function defined in Region II.
We call the two subregions of of Region II as Region II-A and Region II-B and denote them by $\mathcal{R}^{\mathrm{II-A}}$ and $\mathcal{R}^{\mathrm{II-B}}$ in the following.

\begin{lemma}\label{Lemma_DoF_Comparison}
The quantity DoF bound $d^{\mathrm{Quan}}$ satisfies
\begin{align}\label{Eq:DoF_Comparison}
d^{\mathrm{Decom}}\leq d^{\mathrm{Quan}} \leq d^{\mathrm{Proper}},&~\forall M/N\in \mathcal{R}^{\mathrm{II}}
\end{align}
and
\begin{subequations}
\begin{align}
\label{Eq:DoF_Decom_Quan} d^{\mathrm{Quan}}&=d^{\mathrm{Decom}},~\forall M/N \in \{C_{n}^{\mathrm{A}},C_{n}^{\mathrm{B}}\}\\
\label{Eq:DoF_Proper_Quan} d^{\mathrm{Quan}}&=d^{\mathrm{Proper}},~\forall M/N \in \{D_{n}^{\mathrm{A}},D_{n}^{\mathrm{B}}\}
\end{align}
\end{subequations}
where
\begin{align}\label{Eq:Sequence_D}
 D_{n}^{\mathrm{A}}= \frac{K+C_{n+1}^{\mathrm{A}}}{1+K/C_{n}^{\mathrm{A}}},~
 D_{n}^{\mathrm{B}}= \frac{K+C_{n}^{\mathrm{B}}}{1+K/C_{n+1}^{\mathrm{B}}}
\end{align}%
\end{lemma}


As shown in Fig.~\ref{fig:DoF_Achievable}, in Region II, the quantity DoF bound is a zigzag curve between the proper DoF bound and the decomposition DoF bound.

\subsection{Information Theoretic Maximal DoF}
\begin{theorem}[DoF Upper-bound]\label{Theorem:DoF_Upperbound}
For the $G$-cell $K$-user $M \times N$ MIMO-IBC network, the DoF per user are bounded by
\begin{subequations}
\begin{align}\label{Eq:DoF_Bound_R1}
 d &\leq d^{\mathrm{Decom}},~\forall M/N \in \mathcal{R}^{\mathrm{I}}\cap\mathcal{Q}~\\
 \label{Eq:DoF_Bound_R2}
 d &\leq d^{\mathrm{Quan}},~\forall M/N \in \mathcal{R}^{\mathrm{II}}
\end{align}
\end{subequations}
where 
\begin{align}\label{Eq:Set_RegionI}
  \mathcal{Q}\triangleq\{\tilde{q}_{l,n}^{\alpha}/\tilde{p}_{l,n}^{\alpha}\},
~\alpha=\mathrm{A},~\mathrm{B},~n \in \mathbb{Z}^{+}
\end{align}
when $n=2m-1,~\alpha = A$ or $n=2m,~\alpha = B$, $\forall m\in \mathbb{Z}^{+}$, $l=j$,\footnote{It means that $\tilde{q}_{l,n}^{\alpha},~\tilde{p}_{l,n}^{\alpha}$  in \eqref{Eq:Set_RegionI} are the parameters of BS$_j$.} when $n=2m,~\alpha = A$ or $n=2m-1,~\alpha = B$, $l=i_k$.\footnote{It indicates that $\tilde{q}_{l,n}^{\alpha},~\tilde{p}_{l,n}^{\alpha}$ in \eqref{Eq:Set_RegionI} are the parameters of MS$_{i_k}$} Here, $(\tilde{p}_{l,n}^{\mathrm{A}},\tilde{q}_{l,n}^{\mathrm{A}})$ and $ (\tilde{p}_{l,n}^{\mathrm{B}},\tilde{q}_{l,n}^{\mathrm{B}})$ are the generalized Fibonacci sequence-pairs satisfying $(\tilde{p}_{l,0}^{\mathrm{A}},\tilde{q}_{l,0}^{\mathrm{A}})=(0,1)$, $(\tilde{p}_{l,-1}^{\mathrm{A}},\tilde{q}_{l,-1}^{\mathrm{A}})=(-1,0)$, $(\tilde{p}_{l,0}^{\mathrm{B}},\tilde{q}_{l,0}^{\mathrm{B}})=(1,0)$, $(\tilde{p}_{l,-1}^{\mathrm{B}},\tilde{q}_{l,-1}^{\mathrm{B}})=(0,-1)$, and
\begin{subequations}\label{Eq:Fsequence_abc}
\begin{align}
\label{Eq:Fsequence_A1}
\tilde{p}_{l,n}^{\alpha}&\triangleq \sum_{r\in \mathcal{I}_{l,n}^{\alpha}}\tilde{p}_{r,n-1}^{\alpha}-\tilde{p}_{l,n-2}^{\alpha}\\
\label{Eq:Fsequence_B1}
\tilde{q}_{l,n}^{\alpha}&\triangleq \sum_{r\in \mathcal{I}_{l,n}^{\alpha}}\tilde{q}_{r,n-1}^{\alpha}-\tilde{q}_{l,n-2}^{\alpha}
\end{align}
\end{subequations}
satisfying $\tilde{p}_{l,n}^{\alpha},\tilde{q}_{l,n}^{\alpha}>0$, $\forall n \in \mathbb{Z}^{+}$, and $\mathcal{I}_{l,n}^{\mathrm{A}},\mathcal{I}_{l,n}^{\mathrm{B}} \subseteq \mathcal{I}_{l}$ satisfy
\begin{subequations}
\begin{align}\label{Eq:SetA_Condition}
  &\frac{\sum_{r\in \mathcal{I}_{l,n}^{\mathrm{A}}}\tilde{q}_{r,n-1}^{\mathrm{A}}-2\tilde{q}_{l,n-2}^{\mathrm{A}}}
  {\sum_{r\in \mathcal{I}_{l,n}^{\mathrm{A}}}\tilde{p}_{r,n-1}^{\mathrm{A}}-2\tilde{p}_{l,n-2}^{\mathrm{A}}}
  \leq \frac{M}{N}\\
  \label{Eq:SetB_Condition}
  &\frac{\sum_{r\in \mathcal{I}_{l,n}^{\mathrm{B}}}\tilde{q}_{r,n-1}^{\mathrm{B}}-2\tilde{q}_{l,n-2}^{\mathrm{B}}}
  {\sum_{r\in \mathcal{I}_{l,n}^{\mathrm{B}}}\tilde{p}_{r,n-1}^{\mathrm{B}}-2\tilde{p}_{l,n-2}^{\mathrm{B}}}
  \geq \frac{M}{N}
\end{align}
\end{subequations}
$\mathcal{I}_{i_k}=\{j|j\neq i,~j\subseteq\{1,\cdots,G\}\}$ and $\mathcal{I}_{j}=\{i_k|i\neq j,~i\subseteq\{1,\cdots,G\},~k\subseteq\{1,\cdots,K\}\}$ denote the set of interfering BSs for MS$_{i_k}$ and the set of interfering users for BS$_{i_k}$, respectively.
\end{theorem}

In Theorem \ref{Theorem:DoF_Upperbound}, \eqref{Eq:DoF_Bound_R1} indicates that the decomposition DoF is the DoF upper-bound for most configurations in Region I,\footnote{We have studies the property of $\mathcal{Q}$ and found that $\mathcal{Q}$ includes the most antenna configurations in Region I. Due to the page limitation, we will provide the detailed proof in our future work.} and \eqref{Eq:DoF_Bound_R2} means that the quantity DoF is the DoF upper-bound for all configurations in Region II.

\begin{theorem}[Achievable DoF]\label{Theorem:Achievable_DoF}
For the $G$-cell $K$-user $M \times N$ MIMO-IBC network, the achievable DoF per user satisfy
\begin{subequations}
\begin{align}\label{Eq:DoF_lower_Bound_R1}
d&\geq d^{\mathrm{Decom}},~\forall M/N\in \mathcal{R}^{\mathrm{I}}\\
\label{Eq:DoF_lower_Bound_R2}
d&\geq d^{\mathrm{Quan}},~\forall M/N\in \mathcal{R}^{\mathrm{II}}
\end{align}
\end{subequations}
\end{theorem}
In Region I, the DoF of $d^{\mathrm{Decom}}$ are achievable by the asymptotic IA. In Region II, the DoF of $d^{\mathrm{Quan}}$ are achievable by the linear IA.

From Theorems~\ref{Theorem:DoF_Upperbound} and \ref{Theorem:Achievable_DoF}, we can obtain the information theoretic maximal DoF immediately.
\begin{corollary}[Information Theoretic Maximal DoF]\label{Corollary_Maximal_DoF}
For the $G$-cell $K$-user $M \times N$ MIMO-IBC network, the information theoretic maximal DoF per user are
\begin{subequations}
\begin{align}
 d^{\max}&= d^{\mathrm{Decom}},~\forall M/N\in \mathcal{R}^{\mathrm{I}}\cap\mathcal{Q}\\
 d^{\max}&= d^{\mathrm{Quan}},~\forall M/N\in \mathcal{R}^{\mathrm{II}}
\end{align}
\end{subequations}
\end{corollary}

Existing studies \cite{Jafar2010DoF,Ghasemi2010,Park2011,YuWei2013,Jafar_3cell,Jafar_IAchainKcell_all,Zhuguangxi2010} have investigated the information theoretic maximal DoF for the symmetric MIMO-IC and MIMO-IBC networks with some special configurations. Table \ref{Table:Existing_DoF} summarizes the considered cases in these studies, where $\surd$ denotes the networks whose information theoretic maximal DoF were obtained for all antenna configurations, and $\times$ denotes the remaining networks.

\begin{table}[htb!]
\begin{center}
\caption{Resolved cases in the literature}\label{Table:Existing_DoF}
\begin{tabular}{c|c|c|c|c}
 \hline
 $G$ & $K$ & Region I & Region II & \\
 \hline
 $G=2$ & $K=1$ & -- & $\mathcal{R}^{\mathrm{II}}$~\cite{Jafar2010DoF} & $\surd$\\
 \hline
 $G=2$ & $K=2,3$ & -- & $\mathcal{R}^{\mathrm{II}}$~\cite{YuWei2013} & $\surd$\\
 \hline
 $G=2$ & $K\geq 4$ & $\textcircled{4},\textcircled{5}$ & $\mathcal{R}^{\mathrm{II}}_{1}$ & $\times$\\
 \hline
 $G=3$ & $K=1$ & -- & $\mathcal{R}^{\mathrm{II}}$~\cite{Jafar_3cell} & $\surd$\\
 \hline
 $G = 4$ & $K=1$ & $\textcircled{1},\textcircled{2},\textcircled{3}$ & $\mathcal{R}^{\mathrm{II}}_{1}\cup\mathcal{R}^{\mathrm{II}}_{2}$~\cite{Jafar_IAchainKcell_all} & $\times$ \\
 \hline
 $G \geq 5$ & $K=1$ & $\textcircled{1},\textcircled{2}$ & $\mathcal{R}^{\mathrm{II}}_{1}\cup\mathcal{R}^{\mathrm{II}}_{2}$~\cite{Jafar_IAchainKcell_all}
 & $\times$ \\ \hline
 $G \geq 3$ & $K>1$ & $\textcircled{4},\textcircled{5}$ & $\mathcal{R}^{\mathrm{II}}_{1}$ & $\times$ \\
 \hline
\end{tabular}
\end{center}
~\\ where $\mathcal{R}^{\mathrm{II}}_{n}\triangleq \mathcal{R}^{\mathrm{II-A}}_{n}\cup\mathcal{R}^{\mathrm{II-B}}_{n}$,
\begin{itemize}
 \item[$\textcircled{1}$] $K=1$ and $\xi\triangleq\max\{M,N\}/\min\{M,N\}$ is an integer less than $G$~\cite{Jafar2010DoF}.
 \item[$\textcircled{2}$] $K=1$ and $G>(M+N)/\mathrm{gcd}(M,N)$, where $\mathrm{gcd}(M,N)$ is the greatest common divisor of $M$ and $N$~\cite{Ghasemi2010}.
 \item[$\textcircled{3}$] $K=1$, $G=4$, $\xi \in \mathcal{P}$, where $\mathcal{P}=\mathcal{P}_1\cup\mathcal{P}_2\cup\mathcal{P}_3$, $\mathcal{P}_1=\{a/b|1\leq a/b\leq 2, a,b \in \mathbb{Z}^{+}, a\leq 20\}$, $\mathcal{P}_2=(2,5/2]$, and $\mathcal{P}_3=\{21/8\}\cup\{(5c-2)/(2c-1)|c\in \mathbb{Z}^{+}\}$~\cite{Jafar_IAchainKcell_all}.
 \item[$\textcircled{4}$] $K\geq1$, $N=1$, and $\max\{M,K\}/\min\{M,K\}$ is an integer less than $G$~\cite{Zhuguangxi2010,Park2011}.
 \item[$\textcircled{5}$] $K\geq1$ and $M/N \in \{{a}/{b}|a,b\in\mathbb{Z}^{+}, a\leq (G-1)K, b\leq (G-1), a+Kb \leq GK \}$~\cite{YuWei2013}.
 \end{itemize}
\end{table}

When $M/N\in \mathcal{R}^{\mathrm{I}}$, from Table~\ref{Table:Existing_DoF} and \eqref{Eq:Set_RegionI}, it is not hard to prove that $\mathcal{Q}$ contains all configurations in $\textcircled{1}\sim\textcircled{5}$ except the two special cases in $\textcircled{3}$ where $\xi=19/10,17/9$. Therefore, our results in Corollary~\ref{Corollary_Maximal_DoF} are consistent with the most existing results.
When $M/N\in \mathcal{R}^{\mathrm{II}}$, our results in Corollary~\ref{Corollary_Maximal_DoF} are consistent with all existing results and provide the information theoretic maximal DoF for the networks with all configurations.

\subsection{IA Feasible Condition}

From Corollary~\ref{Corollary_Maximal_DoF}, according to the derivation in Appendix \ref{App_NS_Condition}, we obtain the sufficient and necessary condition of IA feasibility as follows.

\begin{corollary}\label{Corollary_NS_Condition}
To support $d$ data streams to each user, the necessary condition of both linear IA and asymptotic IA feasibility is
\begin{align}
\label{Eq:Sufficient_Condition_Compress}
\max\{pM,qN\}\geq (pK+q)d,~\forall (p,q)\in \mathcal{A}\cup\mathcal{B}\cup\mathcal{C}
\end{align}
where $\mathcal{A}\triangleq \left\{\left(p_{n}^{\mathrm{A}},q_{n}^{\mathrm{A}}\right)\right\}$, $\mathcal{B}\triangleq \left\{\left(p_{n}^{\mathrm{B}},q_{n}^{\mathrm{B}}\right)\right\}$, and $\mathcal{C}\triangleq \left\{(N,M)|M/N\in\mathcal{R}^{\mathrm{I}}\cap\mathcal{Q}\right\}$.

When $M/N\in\mathcal{R}^{\mathrm{I}}\cap\mathcal{Q}$, \eqref{Eq:Sufficient_Condition_Compress} is the sufficient and necessary condition of the asymptotic IA feasibility. When $M/N\in\mathcal{R}^{\mathrm{II}}$, \eqref{Eq:Sufficient_Condition_Compress} is the sufficient and necessary condition of the linear IA feasibility.
\end{corollary}


%
\begin{figure}[htb!]
\centering
\includegraphics[width=1.0\linewidth]{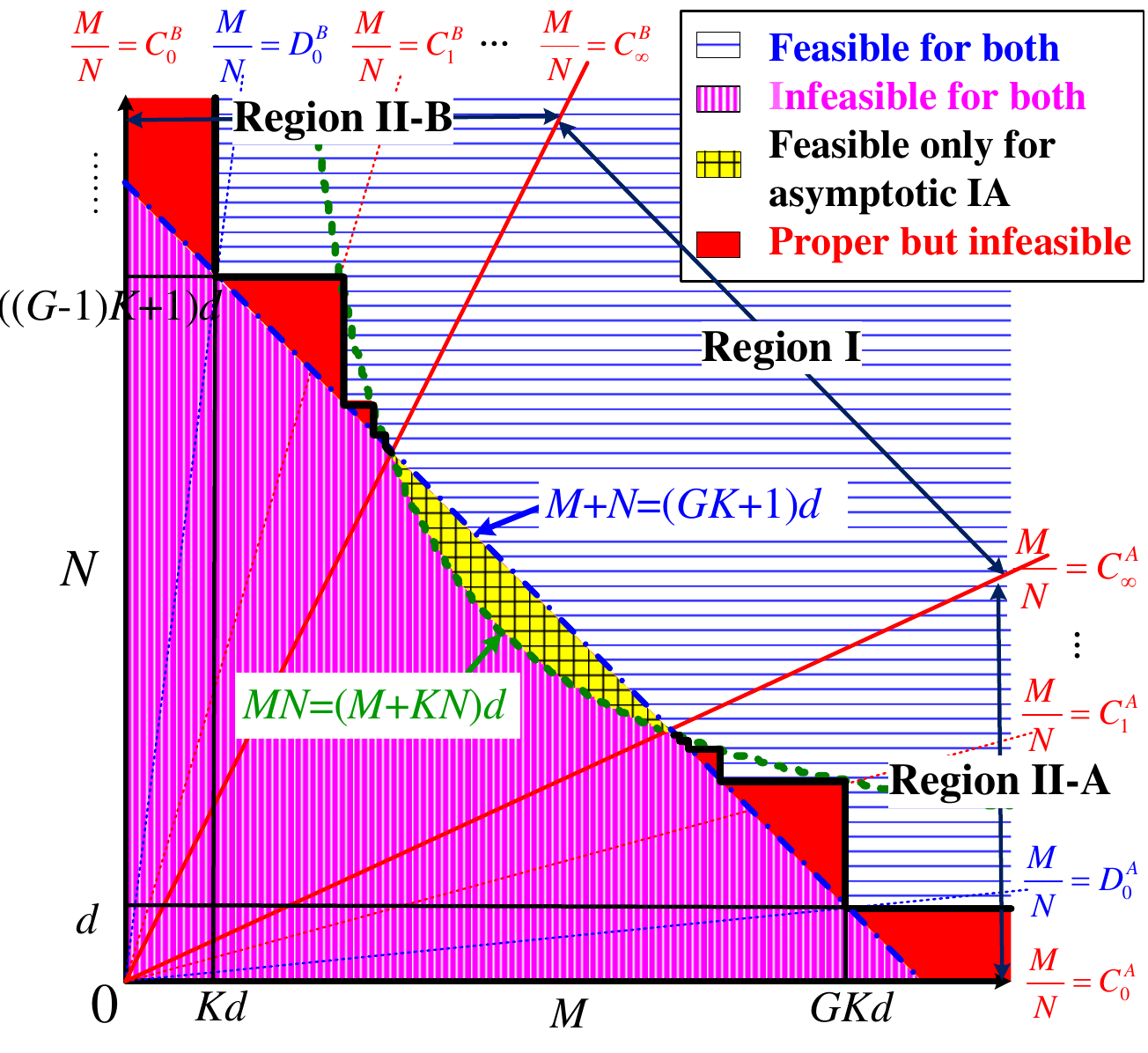}
\caption{IA feasible and infeasible regions for the symmetric
MIMO-IBC network with linear IA and asymptotic IA to support $d$ data streams per user.} \label{fig:Proper_Infeasible_IBC_G2}
\end{figure}

According to Corollary~\ref{Corollary_NS_Condition}, we show the feasible and infeasible regions for the symmetric MIMO-IBC networks with the linear IA and asymptotic IA in Fig.~\ref{fig:Proper_Infeasible_IBC_G2}. When the proper condition in \eqref{Eq:Proper_condition} is satisfied but \eqref{Eq:Sufficient_Condition_Compress} is not satisfied, the linear IA is still infeasible. The proper condition only indicates how many ICIs that can be jointly eliminated by the BS and user, while \eqref{Eq:Sufficient_Condition_Compress} indicates how many ICIs that must be solely eliminated by the BS or user, which explains why the proper condition is not sufficient for linear IA.

\section{Proof of DoF Upper-bound}\label{Sec:Proof1}
A usual approach to derive the DoF upper-bound is introducing a \emph{genie}, which is the side information introduced to a receiver to resolve all the ICIs and decode the desired signals~\cite{Jafar_3cell,Jafar_IAchainKcell_all}. However, how to construct the genie that can derive a tight DoF upper-bound is not easy.

In~\cite{Jafar_IAchainKcell_all}, four different ways of constructing genies were provided for MIMO-IC networks. Since different ways were required for different configurations, the tight DoF upper-bound was only obtained for some configurations. In this study, we provide a unified way to construct the genie and derive the closed-form expression of the tight DoF upper-bound for general configurations.

To provide a unified way to construct genies, we consider a genie-assisted ICI cancelation process where the BSs and users eliminate the ICIs alternately. In each step of the ICI cancelation, considering that each receiver cannot resolve and cancel all ICIs, we divide the ICIs into resolvable and irresolvable ICIs: the ICIs that can be canceled by the receiver are called \emph{resolvable ICIs} and the remaining ICIs are called \emph{irresolvable ICIs}.\footnote{In the forthcoming analysis, we will explain the meaning of the resolvable ICIs in detail.} To help the receiver resolve all ICIs, we construct the genie to resolve the irresolvable ICIs. Then, the tightest upper-bound is obtained by constructing genies to help each node eliminate the maximal number of irresolvable ICIs.

Our study finds that the genies with the same dimension can generate identical DoF upper-bound. Based on this fact, we can convert an information theoretic DoF upper-bound problem into a linear algebra problem and then derive the closed-form DoF upper-bound expression by solving the linear algebra inequality.

In the ICI cancelation process, multiple genies are constructed introduced to both BS and user sides and constitute a \emph{genie tree}. The genie tree can indicate how to introduce genies, i.e., introduce which nodes how many genies to derive the tight DoF upper-bound. Moreover, the genie tree can show how to align ICIs, i.e., align which ICIs at which nodes to achieve the maximal DoF. In this section, we use the genie tree to obtain a closed-form expression of the DoF upper-bound for the symmetric MIMO-IBC networks with most antenna configurations. In the next section, we use the genie tree to prove that the obtained DoF upper-bound is achievable, which means that the DoF upper-bound in Region II is tightest.

In the following, we first employ three representative examples in a three-cell two-user MIMO-IMAC network to show the basic idea to derive the DoF upper-bound in Subsection \ref{Sec:Proof1_Ex} and then provide the proof for general networks in Subsection \ref{Sec:Proof1_General}.

\subsection{Examples}\label{Sec:Proof1_Ex}
When $G=3$ and $K=2$, by substituting $C_{0}^{\mathrm{A}}=\infty,~C_{1}^{\mathrm{A}}=4,~C_{2}^{\mathrm{A}}={7}/{2}$, and $C_{3}^{\mathrm{A}}={24}/{7}$ from \eqref{Eq:Csequence_RecursiveA} into \eqref{Eq:Quantity DoF}, we obtain the quantity DoF bound as
\begin{align}\label{Eq:Quantity DoFG3K2}
d^{\mathrm{Quan}}=\left\{ \begin{array}{ll} \min\left\{\frac{M}{6},N\right\}, &~\forall 4 \leq M/N \\
\min\left\{\frac{2M}{11},\frac{2N}{3}\right\}, &~\forall \frac{7}{2} \leq M/N < 4 \\ \min\left\{\frac{7M}{38},\frac{7N}{11}\right\}, &~\forall \frac{7}{24} \leq M/N < \frac{7}{2} \\
\end{array}
\right.
\end{align}

To illustrate the basic idea to cancel ICIs with the help of the genie in general cases, we choose the examples whose antenna configurations fall in different regions in \eqref{Eq:Quantity DoFG3K2}. To simplify the description and the proof that the DoF upper-bounds are achievable in the next section, we consider the configurations where $d^{\mathrm{Quan}}$ is an integer. For these examples, the configurations and the DoF upper-bounds are listed in Table \ref{Table:CN}.

\begin{table}[htb!]\centering
\caption{Antenna configurations and DoF upper-bounds of examples, where $G=3$ and $K=2$.}\label{Table:CN}
\begin{tabular}{c|c|c|c}
 \hline
 & $M$ & $N$ & $d^{\mathrm{Quan}}$ \\
 \hline
 \hline
\textbf{Ex 1} & 6 & 1 & 1\\
 \hline
\textbf{Ex 2} & 11 & 3 & 2\\
 \hline
\textbf{Ex 3} & 38 & 11 & 7\\
 \hline
\end{tabular}
\end{table}

\begin{figure}[htb!]
\centering
\includegraphics[width=1.00\linewidth]{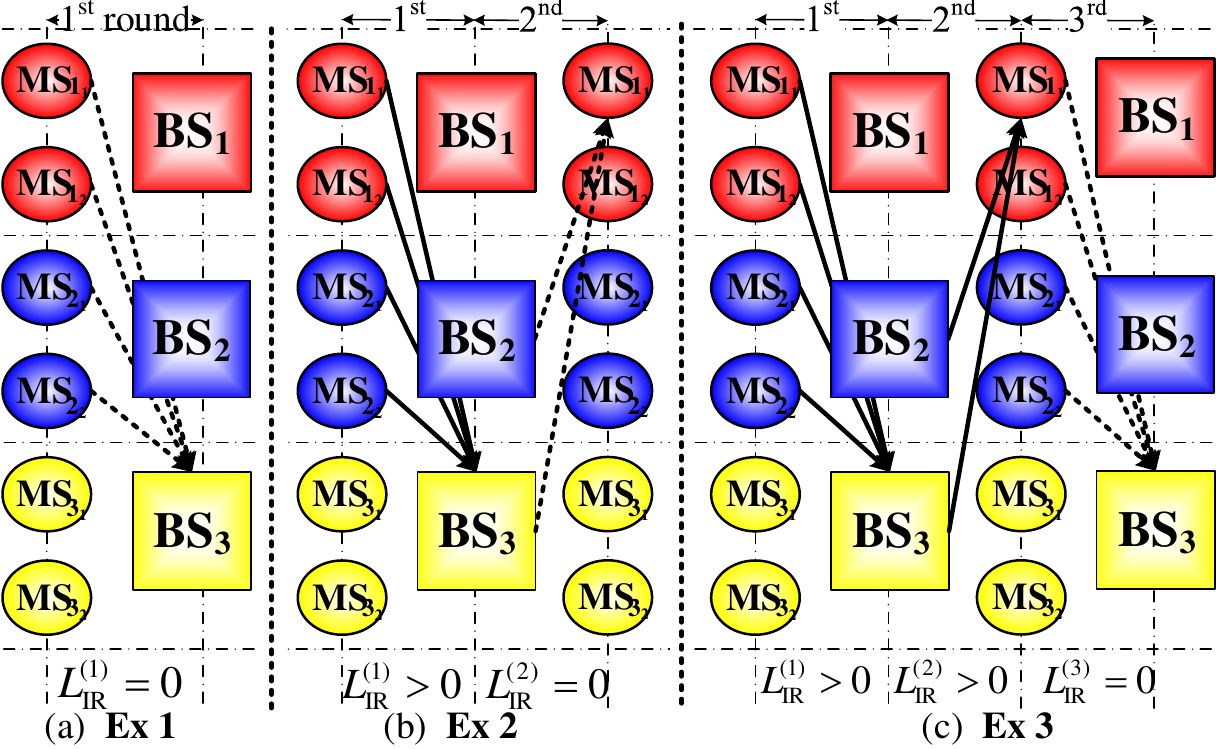}
\caption{ICI cancelation processes for \textbf{Ex 1}, \textbf{Ex 2}, and \textbf{Ex 3}.}
\label{fig:ICI_cancel_procedure}
\end{figure}

From the analysis in \cite{Jafar_IAchainKcell_all}, we know that the DoF upper-bound can be derived until all ICIs can be resolved with the assist of the genie. To obtain the DoF upper-bound for the examples, we consider a genie-assisted ICI cancelation process where the BSs and users eliminate the ICIs alternately.\footnote{It is worth to note that, the actual IA transceiver does not cancel the ICIs like this way. The genie-assisted ICI cancelation process is only used to derive the DoF upper-bound.} In Fig. \ref{fig:ICI_cancel_procedure}, we show the ICI cancelation processes for different examples. We can see that the ICIs need to be eliminated by multiple steps of cancelation, we call each step as one \emph{round}. All variables in terms of the $m$th round are denoted by superscript $(m)$.

Taking BS$_3$ as an example, from \eqref{Eq:Received_signal_IMAC} the received signals are
\begin{align}\label{Eq:Received_signal_IMAC1}
\pmb{y}_3
=&\underbrace{\sum_{k=1}^{2}\pmb{H}_{3,3_k}\pmb{x}_{3_k}}_{\mathrm{Desired~Signals}}+
\underbrace{\sum_{k=1}^{2}\pmb{H}_{3,1_k}\pmb{x}_{1_k}+
\pmb{H}_{3,2_k}\pmb{x}_{2_k}
}_{\mathrm{Recieved~ICIs}}
+\pmb{n}_3
\end{align}
The study in \cite{Jafar_IAchainKcell_all} indicates that the desired signals can be assumed to be decodable and only the ICIs need to be considered. In the following, we focus on investigating the received ICIs.

In the 1st round, the received ICIs of BS$_3$ can be expressed as
\begin{align}\label{Eq:Received_ICI_BS3_r1}
\pmb{z}_3^{(1)} &=
\pmb{H}_{3,1_1}\pmb{x}_{1_1}+\pmb{H}_{3,1_2}\pmb{x}_{1_2}+
\pmb{H}_{3,2_1}\pmb{x}_{2_1}+\pmb{H}_{3,2_2}\pmb{x}_{2_2}
 \\
&= \pmb{H}_3^{(1)}\pmb{x}_3^{(1)}\nonumber
\end{align}
where $\pmb{x}_3^{(1)} = \left[ \pmb{x}_{1_1}^{T},\pmb{x}_{1_2}^{T},\pmb{x}_{2_1}^{T},\pmb{x}_{2_2}^{T} \right]^{T} \in \mathbb{C}{^{4N \times 1}}$ and $\pmb{H}_3^{(1)}=\left[ \pmb{H}_{3,1_1},\pmb{H}_{3,1_2}, \pmb{H}_{3,2_1},\pmb{H}_{3,2_2}\right]\in\mathbb{C}{^{M \times 4N}}$.

As shown in \eqref{Eq:Received_ICI_BS3_r1}, the received ICIs are the projection of transmit signals of interfering users into the interference subspace. Since the received ICIs can be expressed as a linear combination of the column vectors of $\pmb{H}_3^{(1)}$, we can use $\mathrm{span}\{\pmb{H}_3^{(1)}\}$ to denote the interference subspace of BS$_3$.

Considering that $\pmb{H}_3^{(1)}$ is of size $M \times 4N$, the projection is from a $4N$-dimensional subspace into a $M$-dimensional subspace.
When $4N \leq M$, the projection is from a low-dimensional subspace into a high-dimensional subspace. Hence, BS$_3$ is able to resolve and cancel all ICIs. We call the ICIs \emph{resolvable ICIs}. By contrast, if $4N >M$, BS$_3$ cannot resolve all ICIs. The analysis in \cite{Jafar_IAchainKcell_all} indicates that, when the genie provides the side information to help the BS cancel ICIs in $(4N-M)$-dimensional subspace, the BS can resolve and cancel the remaining ICIs. Therefore, we call the ICIs that are canceled by the BS \emph{resolvable ICIs} and the remaining ICIs \emph{irresolvable ICIs} in the 1st round.

\subsubsection{\textbf{Ex 1}}
We first consider \textbf{Ex 1} where $M=6$ and $N=1$.
Since $\pmb{H}_3^{(1)}$ is of size $6 \times 4$, BS$_3$ is able to resolve and cancel all ICIs in the 1st round. Thus, we have
\begin{subequations}\label{Eq:Entropy_Ex1_R1_BS3}
\begin{align}\label{Eq:Entropy_Ex1_R1_BS3_a}
 4\ell R - \ell{\varepsilon _\ell} & \le I\left( W_{1_1},W_{1_2},W_{2_1},W_{2_2};{\pmb{z}}_3^{(1)\ell} \right)\\
 \label{Eq:Entropy_Ex1_R1_BS3_b}
 &\le \hbar\left({{\pmb{z}}_3^{(1)\ell}} \right)\\
 \label{Eq:Entropy_Ex1_R1_BS3_c}
 &\le M\ell\left(\log\gamma + o(\log\gamma)\right)-2\ell R
\end{align}
\end{subequations}
where $W_{{i_k}}\in [1,2^{\ell R}]$ denotes the message transmitted by $\pmb{x}_{i_k}^{\ell}$, $\hbar\left(\pmb{x}\right)\triangleq h\left(\pmb{x}+\pmb{n}\right)$ is the differential entropy defined in \cite{Jafar_IAchainKcell_all}, $h(\cdot)$ is the stand differential entropy, $\pmb{n}\sim \mathcal{CN}(\pmb{0},\pmb{I})$ is a circularly symmetrically additive white Gaussian noise vector that is independent of $\pmb{x}$, $\varepsilon_\ell$ is an arbitrary small numbers of $\ell$, \eqref{Eq:Entropy_Ex1_R1_BS3_a} follows from Fano's inequality~\cite{Information_Theory2006}, \eqref{Eq:Entropy_Ex1_R1_BS3_b} follows because all ICIs can be resolved from ${\pmb{z}}_3^{(1)}$, and \eqref{Eq:Entropy_Ex1_R1_BS3_c} follows from $\hbar\left({{\pmb{z}}_3^{(1)\ell}} \right)=\hbar\left({{\pmb{y}}_3^{\ell}}\right)- \hbar\left(\sum_{k=1}^{2}\pmb{H}_{3,3_k}^{\ell}\pmb{x}_{3_k}^{\ell}\right)
=M\ell\left(\log\gamma + o(\log\gamma)\right)-2\ell R$, which means that the BS needs to reserve enough antennas to receive the desired signals and use the remaining antennas to cancel ICIs.

By dividing $\ell\log\gamma $ on both sides of \eqref{Eq:Entropy_Ex1_R1_BS3_c} and letting $\gamma\rightarrow \infty$ and $\ell\rightarrow \infty$, we obtain
\begin{align}
 d\leq M/6 = 1
\end{align}
which leads to the DoF upper-bound $d\leq d^{\mathrm{Quan}}$ for \textbf{Ex 1}.

\subsubsection{\textbf{Ex 2}}
We then consider \textbf{Ex 2} where $M=11$ and $N=3$.
In \eqref{Eq:Received_ICI_BS3_r1}, $\pmb{H}_3^{(1)}$ is of size $11 \times 12$. As a result, BS$_3$ cannot resolve all ICIs and it is necessary to introduce a genie in the $12-11=1$-dimensional interference subspace to help BS$_3$.

To find an appropriate genie for BS$_3$, we divide the received ICIs into two parts as

\begin{figure}[htb!]
\centering
\includegraphics[width=0.95\linewidth]{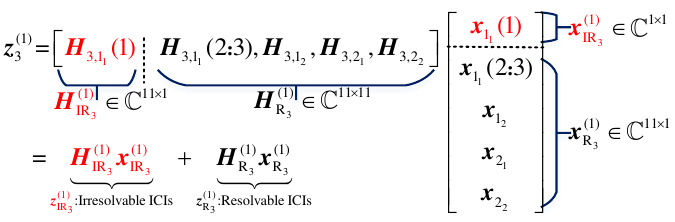}
\end{figure}

When introducing a genie ${\pmb{G}}_{3}^{(1)}= \pmb{x}_{\mathrm{IR}_{3}}^{(1)} =\pmb{x}_{1_1}(1)$ to help BS$_3$ resolve and cancel $\pmb{z}_{\mathrm{IR}_{3}}^{(1)}$, BS$_3$ itself can resolve and cancel $\pmb{z}_{\mathrm{R}_{3}}^{(1)}$. Therefore, for BS$_3$, $\pmb{z}_{\mathrm{R}_{3}}^{(1)}$ are the \emph{resolvable ICIs}, while $\pmb{z}_{\mathrm{IR}_{3}}^{(1)}$ are the \emph{irresolvable ICIs} in the 1st round.

Thus, we have
\begin{subequations}\label{Eq:Entropy_Ex2_R1_BS3}
\begin{align}
\label{Eq:Entropy_Ex2_R1_BS3_a}
 4\ell R - \ell{\varepsilon _\ell} &\le I\left( W_{1_1},W_{1_2},W_{2_1},W_{2_2};{\pmb{z}}_3^{(1)\ell},
 {\pmb{G}}_{3}^{(1)\ell}\right) \\
\label{Eq:Entropy_Ex2_R1_BS3_b}
 &\le \hbar\left( \pmb{z}_{{3}}^{(1)\ell}|\pmb{x}_{\mathrm{IR}_{3}}^{(1)\ell} \right)
 +\hbar\left( \pmb{x}_{\mathrm{IR}_{3}}^{(1)\ell} \right) \\
\label{Eq:Entropy_Ex2_R1_BS3_c}
 &\le \hbar\left( \pmb{z}_{\mathrm{R}_{3}}^{(1)\ell}\right) + \hbar\left( {{\pmb{x}}_{\mathrm{IR}_3}^{(1)\ell}} \right)\\
\label{Eq:Entropy_Ex2_R1_BS3_d}
 &= M\ell\left(\log\gamma + o(\log\gamma)\right)-2\ell R + \hbar\left( \pmb{x}_{1_1}^{\ell}(1) \right)
\end{align}
\end{subequations}
where \eqref{Eq:Entropy_Ex2_R1_BS3_a} follows from Fano's inequality, \eqref{Eq:Entropy_Ex2_R1_BS3_b} follows from the chain rule, and \eqref{Eq:Entropy_Ex2_R1_BS3_c} follows from ${\pmb{z}}_{\mathrm{R}_3}^{(1)}={\pmb{z}}_{3}^{(1)}-{\pmb{z}}_{\mathrm{IR}_3}^{(1)}$.

In \eqref{Eq:Entropy_Ex2_R1_BS3}, the introduced genie $\pmb{x}_{1_1}(1)$ is the transmit signal from MS$_{1_1}$'s first antenna.  In fact, there are more than one signal can be chosen as the genie.  For example, we can consider $\pmb{x}_{1_1}(2)$, the signal from MS$_{1_1}$'s second antenna. Moreover, we can also obtain the similar inequality as \eqref{Eq:Entropy_Ex1_R1_BS3} by introducing the genie to other BSs. If we introduce BS$_2$ a genie  ${\pmb{G}}_{2}^{(1)}= \pmb{x}_{\mathrm{IR}_{2}}^{(1)}= \pmb{x}_{1_1}(2)$, following the similar derivation in \eqref{Eq:Entropy_Ex1_R1_BS3}, we have
\begin{align}\label{Eq:Entropy_Ex2_R1_BS31}
  4\ell R - \ell{\varepsilon _\ell} \le M\ell\left(\log\gamma + o(\log\gamma)\right)-2\ell R + \hbar\left(\pmb{x}_{1_1}^{\ell}(2) \right)
\end{align}

As shown in \eqref{Eq:Entropy_Ex2_R1_BS3} and \eqref{Eq:Entropy_Ex2_R1_BS31}, to obtain the DoF upper-bound, we need to derive $\hbar\left( \pmb{x}_{1_1}^{\ell}(1) \right)$ or $\hbar\left(\pmb{x}_{1_1}^{\ell}(2) \right)$, i.e., resolve $\pmb{x}_{1_1}(1)$ or $\pmb{x}_{1_1}(2)$. Since MS$_{1_1}$ can help BS$_3$ and BS$_2$ resolve their irresolvable ICIs,
these two BSs can return their irresolvable ICIs to MS$_{1_1}$ in the 2nd round. As a result, the received ICIs of MS$_{1_1}$ can be expressed as
\begin{align}\label{Eq:Received_ICI_MS11_r2}
\pmb{z}_{1_1}^{(2)} &=
\pmb{H}_{3,1_1}^{T}\pmb{z}_{\mathrm{IR}_{3}}^{(1)}+
\pmb{H}_{2,1_1}^{T}\pmb{z}_{\mathrm{IR}_{2}}^{(1)}
\\
&=
{\pmb{H}_{3,1_1}^{T}\pmb{H}_{\mathrm{IR}_{3}}^{(1)}}
\pmb{x}_{\mathrm{IR}_{3}}^{(1)}+
{\pmb{H}_{2,1_1}^{T}\pmb{H}_{\mathrm{IR}_{2}}^{(1)}}
\pmb{x}_{\mathrm{IR}_{2}}^{(1)} \nonumber \\ &=\pmb{H}_{1_1}^{(2)}\pmb{x}_{1_1}^{(2)}\nonumber
\end{align}
where $\pmb{x}_{1_1}^{(2)} =\left[\pmb{x}_{\mathrm{IR}_{3}}^{(1)T},\pmb{x}_{\mathrm{IR}_{2}}^{(1)T}
\right]^{T}=\pmb{x}_{1_1}(1:2) \in \mathbb{C}{^{2 \times 1}}$, a two-dimensional vector obtained from the 1st and 2nd variables of $\pmb{x}_{1_1}$, $\pmb{H}_{1_1}^{(2)}
= \left[\pmb{H}_{1_1,3}^{(2)},\pmb{H}_{1_1,2}^{(2)}\right]
\in \mathbb{C}{^{3 \times 2}}$, and $\pmb{H}_{1_1,i}^{(2)}=\pmb{H}_{i,1_1}^{T}\pmb{H}_{\mathrm{IR}_{i}}^{(1)}\in \mathbb{C}^{3 \times 1}$ is the equivalent channel matrix from BS$_i$ to MS$_{1_1}$ in the 2nd round.

In \eqref{Eq:Received_ICI_MS11_r2}, $\pmb{H}_{1_1}^{(2)}$ is of size $3 \times 2$, which means that MS$_{1_1}$ can resolve all returned ICIs without any genie. Then, we have
\begin{subequations}\label{Eq:Entropy_Ex2_R2_MS11}
\begin{align}\label{Eq:Entropy_Ex2_R2_MS11a}
\hbar\left(\pmb{x}_{1_1}^{\ell}(1)\right) +\hbar\left(\pmb{x}_{1_1}^{\ell}(2)\right)
&\le
\hbar\left( \pmb{z}_{1_1}^{(2)\ell} \right) = \hbar\left( \pmb{x}_{1_1}^{\ell}(1:2) \right)\\
&\le \hbar\left( {{\pmb{x}}_{1_1}^{\ell}} \right)= \ell R
\end{align}
\end{subequations}

By dividing $\ell\log\gamma $ on both sides of \eqref{Eq:Entropy_Ex2_R1_BS3}, \eqref{Eq:Entropy_Ex2_R1_BS31}, and \eqref{Eq:Entropy_Ex2_R2_MS11}, and letting $\gamma\rightarrow \infty$ and $\ell\rightarrow \infty$, we can obtain
\begin{subequations}\label{Eq:DoF_Ex1_RALL}
\begin{align}\label{Eq:DoF_Ex1_R1_BS1}
4d&\le M-2d + \omega\left(\pmb{x}_{1_1}^{\ell}(1) \right)\\
\label{Eq:DoF_Ex1_R1_BS2}
4d&\le M-2d + \omega\left(\pmb{x}_{1_1}^{\ell}(2) \right)\\
\label{Eq:DoF_Ex1_R2_MS11}
\omega\left(\pmb{x}_{1_1}^{\ell}(1)\right) +\omega\left(\pmb{x}_{1_1}^{\ell}(2)\right)
&\le d
\end{align}
\end{subequations}
where $\omega\left(\pmb{x}\right)\triangleq\lim_{\ell,\gamma\rightarrow \infty}\hbar\left(\pmb{x} \right)/\left(\ell\log\gamma\right)$.

By adding up the inequalities \eqref{Eq:DoF_Ex1_R1_BS1}, \eqref{Eq:DoF_Ex1_R1_BS2}, and \eqref{Eq:DoF_Ex1_R2_MS11}, we have
\begin{align}\label{Eq:DoF_bound_Ex2}
8d \le 2M- 3d \Rightarrow d\leq 2M/11 = 2
\end{align}
which leads to the DoF upper-bound $d\leq d^{\mathrm{Quan}}$ for \textbf{Ex 2}.

\begin{figure}[htb!]
\centering
\includegraphics[width=0.50\linewidth]{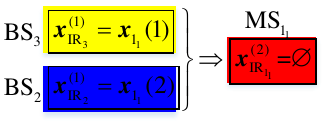}
\caption{Required genies for \textbf{Ex 2}.}
\label{fig:Genie_Tree_Ex2}
\end{figure}

In Fig. \ref{fig:Genie_Tree_Ex2}, we show all the required genies to derive the DoF upper-bound for \textbf{Ex 2}. In the following, we will discuss the key factors to design the genies.

As shown in \eqref{Eq:Received_ICI_MS11_r2}, when introducing BS$_2$ and BS$_3$ different genies, i.e., ${\pmb{x}}_{\mathrm{IR}_3}^{(1)}\neq{\pmb{x}}_{\mathrm{IR}_2}^{(1)}$,
MS$_{1_1}$ can resolve two different ICIs.
If introducing BS$_2$ and BS$_3$ the same genie, i.e., ${\pmb{x}}_{\mathrm{IR}_3}^{(1)}={\pmb{x}}_{\mathrm{IR}_2}^{(1)}=\pmb{x}_{1_1}(1)$, \eqref{Eq:Received_ICI_MS11_r2} becomes
\begin{align}\label{Eq:Received_ICI_MS11_r2cc}
\pmb{z}_{1_1}^{(2)} =\left(\pmb{H}_{3,1_1}^{T}\pmb{H}_{\mathrm{IR}_{3}}^{(1)}+\pmb{H}_{2,1_1}^{T}\pmb{H}_{\mathrm{IR}_{2}}^{(1)}
\right)\pmb{x}_{1_1}(1)
\end{align}
so that MS$_{1_1}$ can resolve one ICI, which cannot exploit MS$_{1_1}$'s ability to cancel interference sufficiently. Consequently, we obtain $\hbar\left(\pmb{x}_{1_1}^{\ell}(1)\right) \le \ell R$ and $\omega\left(\pmb{x}_{1_1}^{\ell}(1)\right) \le d$, which leads to a looser DoF upper-bound $d\leq M/5=2.2$ than \eqref{Eq:DoF_bound_Ex2}. Then, we can obtain a following conclusion.

\begin{remark}\label{Remark_Max}
To obtain the tight DoF upper-bound, each node (either BS or user) needs to help the nodes in the last round resolve the ICIs as much as possible. Therefore, the introduced genies for different nodes should be different. 
\end{remark}

In \eqref{Eq:DoF_Ex1_R1_BS1} and \eqref{Eq:DoF_Ex1_R1_BS2}, $\omega\left(\pmb{x}_{1_1}^{\ell}(1)\right)$ and $\omega\left(\pmb{x}_{1_1}^{\ell}(2)\right)$ denote how many ICIs the two genies can resolve, respectively. 
Combining \eqref{Eq:DoF_Ex1_R1_BS1} and \eqref{Eq:DoF_Ex1_R1_BS2}, we obtain
\begin{align}\label{Eq:DoF_Ex1_R1_BS3}
 4d\le M-2d + \min\left\{\omega\left(\pmb{x}_{1_1}^{\ell}(1)\right), \omega\left(\pmb{x}_{1_1}^{\ell}(2)\right)\right\}
\end{align}
If $\omega\left(\pmb{x}_{1_1}^{\ell}(1)\right) \neq\omega\left(\pmb{x}_{1_1}^{\ell}(2)\right)$, we can find a DoF upper-bound tighter than  \eqref{Eq:DoF_bound_Ex2}, since $\min\left\{\omega\left(\pmb{x}_{1_1}^{\ell}(1)\right), \omega\left(\pmb{x}_{1_1}^{\ell}(2)\right)\right\}< d/2$ from \eqref{Eq:DoF_Ex1_R2_MS11}. As a result, it brings a new question: Does there exist other upper-bound tighter than \eqref{Eq:DoF_bound_Ex2}? In Subsection \ref{Sec:Proof2_Ex}, we will prove that $d^{\mathrm{Quan}}=2$ is achievable,  which means that \eqref{Eq:DoF_bound_Ex2} is the tightest DoF upper-bound.
Therefore, $\omega\left(\pmb{x}_{1_1}^{\ell}(1)\right) =\omega\left(\pmb{x}_{1_1}^{\ell}(2)\right)$ always holds. Then, we can obtain a conclusion as follows.

\begin{remark}\label{Remark_Same}
For two arbitrary genies, if their dimensions are identical, they can resolve the same number of irresolvable ICIs and generate identical DoF upper-bound.
\end{remark}

\subsubsection{\textbf{Ex 3}}
We finally consider \textbf{Ex 3} where $M=38$ and $N=11$.
In the 1st round, since $\pmb{H}_j^{(1)}$ is of size $38 \times 44$, each BS cannot resolve all ICIs and it is necessary to  introduce a genie in the $44-38=6$-dimensional interference subspace.

As shown in Fig. \ref{fig:ICI_cancel_procedure}(c), MS$_{1_1}$ can help both BS$_3$ and BS$_2$ cancel their irresolvable ICIs. According to Remark \ref{Remark_Max}, to ensure MS$_{1_1}$ to cancel the maximal number of ICIs,  $\pmb{x}_{\mathrm{IR}_{3}}^{(1)}$ and $\pmb{x}_{\mathrm{IR}_{2}}^{(1)}$ should contain different variables of $\pmb{x}_{1_1}$.
Considering that $\pmb{x}_{\mathrm{IR}_{3}}^{(1)}$ and $\pmb{x}_{\mathrm{IR}_{2}}^{(1)}$ have six entries, whereas $\pmb{x}_{1_1}$  only has eleven entries, we choose $\pmb{x}_{\mathrm{IR}_{3}}^{(1)}=\pmb{x}_{1_1}(1:6)$ and $\pmb{x}_{\mathrm{IR}_{2}}^{(1)}=[\pmb{x}_{{1_1}}^{T}(7:11),\pmb{x}_{{3_1}}^{T}(1)]^{T}$.

Therefore, the received ICIs of BS$_3$ and BS$_2$ are divided into
\begin{figure}[htb!]
\centering
\includegraphics[width=0.95\linewidth]{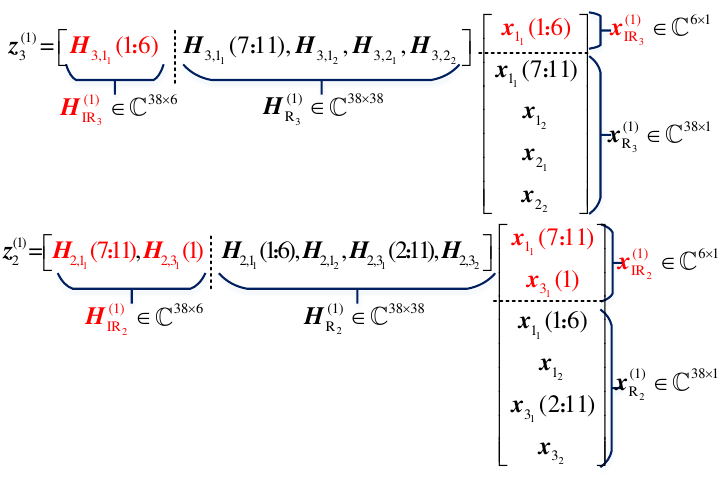}
\end{figure}

When introducing ${\pmb{G}}_{3}^{(1)}= \pmb{x}_{\mathrm{IR}_{3}}^{(1)}$ and ${\pmb{G}}_{2}^{(1)}= \pmb{x}_{\mathrm{IR}_{2}}^{(1)}$ to BS$_3$ and BS$_2$, respectively, each BS of these two BSs can resolve all ICIs. Following the similarly derivation in \eqref{Eq:Entropy_Ex2_R1_BS3}, we have
\begin{subequations}\label{Eq:Entropy_Ex3_R1}
\begin{align}\label{Eq:Entropy_Ex3_R1_BS3}
 4\ell R - \ell{\varepsilon _\ell} \le & M\ell\left(\log\gamma + o(\log\gamma)\right)-2\ell R+ \hbar\left( \pmb{x}_{{1_1}}^{\ell}(1:6)\right) \\
 \label{Eq:Entropy_Ex3_R1_BS2}
 4\ell R - \ell{\varepsilon _\ell} \le & M\ell\left(\log\gamma + o(\log\gamma)\right)-2\ell R + \hbar\left( \pmb{x}_{{1_1}}^{\ell}(7:11),\pmb{x}^{\ell}_{{3_1}}(1) \right)
\end{align}
\end{subequations}

As shown in Fig. \ref{fig:ICI_cancel_procedure}(c),
BS$_2$ and BS$_3$ return their irresolvable ICIs to MS$_{1_1}$ in the 2nd round, then the received ICIs of MS$_{1_1}$ can be expressed as
\begin{align}\label{Eq:Received_ICI_MS11_r2_Ex3}
\pmb{z}_{1_1}^{(2)} =&
\pmb{H}_{3,1_1}^{T}\pmb{z}_{\mathrm{IR}_{3}}^{(1)}+
\pmb{H}_{2,1_1}^{T}\pmb{z}_{\mathrm{IR}_{2}}^{(1)}
= \pmb{H}_{1_1}^{(2)}\pmb{x}_{1_1}^{(2)}
\end{align}
where $\pmb{x}_{1_1}^{(2)} = [\pmb{x}_{{3_1}}^{T}(1), \pmb{x}_{{1_1}}^{T}]^{T} \in \mathbb{C}{^{12 \times 1}}$ and
$\pmb{H}_{1_1}^{(2)} = [\pmb{H}_{2,1_1}^{T}\pmb{H}_{2,{3_1}}(1),\pmb{H}_{3,1_1}^{T}\pmb{H}_{3,{1_1}}(1:6), \pmb{H}_{2,1_1}^{T}\pmb{H}_{2,{1_1}}(7:11)\pmb{x}_{{1_1}}(7:11)] \in \mathbb{C}{^{11 \times 12}}$.

In \eqref{Eq:Received_ICI_MS11_r2_Ex3}, $\pmb{H}_{1_1}^{(2)}$ is of size $11 \times 12$, so that MS$_{1_1}$ cannot resolve all received ICIs and it is necessary to introduce a genie in the $12-11=1$-dimensional interference subspace to help MS$_{1_1}$.

Since MS$_{1_1}$ can use $N=11$ antennas to resolve all ICIs related to $\pmb{x}_{{1_1}}$, the irresolvable ICIs of MS$_{1_1}$ are $\pmb{z}_{\mathrm{IR}_{1_1}}^{(2)} = \pmb{H}_{2,1_1}^{T}\pmb{H}_{2,{3_1}}(1)\pmb{x}_{{3_1}}(1)$ as shown in the sequel. Therefore, \eqref{Eq:Received_ICI_MS11_r2_Ex3} is divided into
\begin{figure}[htb!]
\centering
\includegraphics[width=0.95\linewidth]{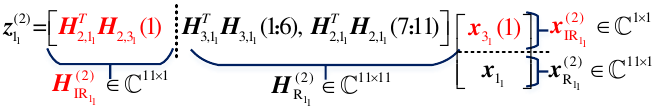}
\end{figure}

When introducing ${\pmb{G}}_{1_1}^{(2)}= \pmb{x}_{\mathrm{IR}_{1_1}}^{(2)}=\pmb{x}_{{3_1}}(1)$ to MS$_{1_1}$, MS$_{1_1}$ can resolve all ICIs in the 2nd round. Thus, we have
\begin{subequations}\label{Eq:Entropy_Ex3_R2_MS11}
\begin{align}
\label{Eq:Entropy_Ex3_R2_MS11a}
&\hbar\left( \pmb{x}_{{1_1}}^{\ell}(1:6)\right)+\hbar\left( \pmb{x}_{{1_1}}^{\ell}(7:11),\pmb{x}^{\ell}_{{3_1}}(1) \right)\nonumber\\
\le& \hbar\left( \pmb{z}_{{1_1}}^{(2)\ell}|\pmb{x}_{\mathrm{IR}_{1_1}}^{(2)\ell} \right)+\hbar\left( \pmb{x}_{\mathrm{IR}_{1_1}}^{(2)\ell} \right) \\
\label{Eq:Entropy_Ex3_R2_MS11b}
\le& \hbar\left( \pmb{z}_{\mathrm{R}_{1_1}}^{(2)\ell} \right)+\hbar\left( \pmb{x}_{\mathrm{IR}_{1_1}}^{(2)\ell} \right) \\
\label{Eq:Entropy_Ex3_R2_MS11c}
\le& \hbar\left( {{\pmb{x}}_{1_1}^{\ell}} \right)+ \hbar\left( {{\pmb{x}}_{3_1}^{\ell}(1)} \right)\\
\label{Eq:Entropy_Ex3_R2_MS11d}
\le& \ell R+\hbar\left( {{\pmb{x}}_{3_1}^{\ell}(1)} \right)
\end{align}
\end{subequations}
where \eqref{Eq:Entropy_Ex3_R2_MS11a} follows from the chain rule, \eqref{Eq:Entropy_Ex3_R2_MS11b} follows from $\hbar\left( \pmb{z}_{{1_1}}^{(2)\ell}|\pmb{x}_{\mathrm{IR}_{1_1}}^{(2)\ell} \right)=\hbar\left(\pmb{z}_{\mathrm{R}_{1_1}}^{(2)\ell} \right)$, and \eqref{Eq:Entropy_Ex3_R2_MS11c} follows because we can resolve $ {{\pmb{x}}_{1_1}}$ from $\pmb{z}_{\mathrm{R}_{1_1}}^{(2)}$ and resolve ${{\pmb{x}}_{1_1}(1)}$ from $\pmb{z}_{\mathrm{IR}_{1_1}}^{(2)}$ due to $\pmb{H}_{\mathrm{R}_{1_1}}^{(2)}\in \mathbb{C}^{11\times 11}$ and $\pmb{H}_{\mathrm{IR}_{1_1}}^{(2)}\in \mathbb{C}^{11\times 1}$.

\begin{figure}[htb!]
\centering
\includegraphics[width=0.90\linewidth]{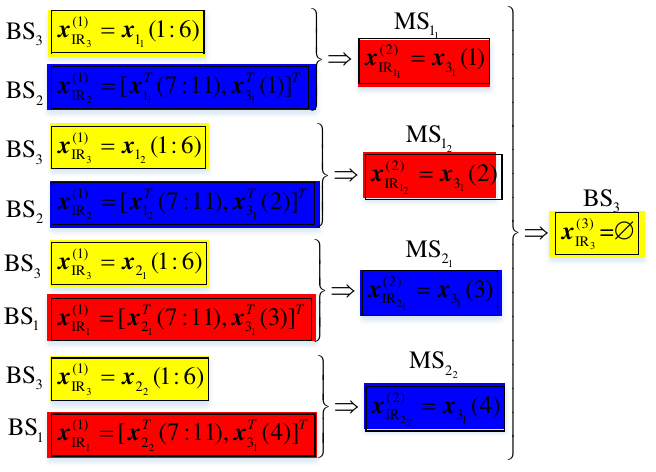}
\caption{Introduced genies for \textbf{Ex 3}.}
\label{fig:Genie_Tree_Ex3}
\end{figure}

As shown in Fig. \ref{fig:ICI_cancel_procedure}(c), BS$_{3}$ can help MS$_{1_1}$, MS$_{1_2}$, MS$_{2_1}$ and MS$_{2_2}$ resolve ICIs simultaneously. According to Remark \ref{Remark_Max}, as shown in Fig. \ref{fig:Genie_Tree_Ex3}, we choose the genies as  $\pmb{x}_{\mathrm{IR}_{1_2}}^{(2)} =\pmb{x}_{{3_1}}(2)$, $\pmb{x}_{\mathrm{IR}_{2_1}}^{(2)} = \pmb{x}_{{3_1}}(3)$, and $\pmb{x}_{\mathrm{IR}_{2_2}}^{(2)} = \pmb{x}_{{3_1}}(4)$.

As a result, the received ICIs of BS$_{3}$ in the 3rd round can be expressed as
\begin{align}\label{Eq:Received_ICI_BS3_r3_Ex3}
\pmb{z}_{3}^{(3)} =&
\sum_{k=1}^{2}\pmb{H}_{1_k,3}^{(2)T}\pmb{z}_{\mathrm{IR}_{1_k}}^{(2)}
+\pmb{H}_{2_k,3}^{(2)T}\pmb{z}_{\mathrm{IR}_{2_k}}^{(2)}\\
=& \sum_{k=1}^{2}{\pmb{H}_{1_k,3}^{(2)T} \pmb{H}_{\mathrm{IR}_{1_k}}^{(2)}}\pmb{x}_{\mathrm{IR}_{1_k}}^{(2)}+
{\pmb{H}_{2_k,3}^{(2)T}\pmb{H}_{\mathrm{IR}_{2_k}}^{(2)}} \pmb{x}_{\mathrm{IR}_{2_k}}^{(2)}\nonumber\\
=&\pmb{H}_{3}^{(3)}\pmb{x}_{3}^{(3)}\nonumber
\end{align}
where $\pmb{x}_{3}^{(3)} = \left[ \pmb{x}_{\mathrm{IR}_{1_1}}^{(2)T},\pmb{x}_{\mathrm{IR}_{1_2}}^{(2)T}, \pmb{x}_{\mathrm{IR}_{2_1}}^{(2)T},\pmb{x}_{\mathrm{IR}_{2_2}}^{(2)T} \right]^T =\pmb{x}_{3_1}(1:4)\in \mathbb{C}^{4 \times 1}$, $\pmb{H}_3^{(3)}=\left[\pmb{H}_{3,1_1}^{(3)},\pmb{H}_{3,1_2}^{(3)}, \pmb{H}_{3,2_1}^{(3)},\pmb{H}_{3,2_2}^{(3)}\right]\in \mathbb{C}^{6 \times 4}$, and $\pmb{H}_{3,i_k}^{(3)}=\pmb{H}_{{i_k,3}}^{(2)T} \pmb{H}_{\mathrm{IR}_{i_k}}^{(2)}\in \mathbb{C}^{6\times 1}$ is the equivalent channel matrix from MS$_{i_k}$ to BS$_3$ in the 3rd round.

In \eqref{Eq:Received_ICI_BS3_r3_Ex3},
$\pmb{H}_{{3}}^{(3)}$ is of size $6\times 4$, hence BS$_3$ can resolve all received ICIs without any genie. Thus, we have
\begin{subequations}\label{Eq:Entropy_Ex3_R3_BS3}
\begin{align}
\label{Eq:Entropy_Ex3_R3_BS3a}
 &\hbar\left( {{\pmb{x}}_{3_1}^{\ell}(1)} \right)+
  \hbar\left( {{\pmb{x}}_{3_1}^{\ell}(2)} \right)+
  \hbar\left( {{\pmb{x}}_{3_1}^{\ell}(3)} \right)+
  \hbar\left( {{\pmb{x}}_{3_1}^{\ell}(4)} \right)\nonumber\\
  \le& \hbar\left(\pmb{z}_{3}^{(3)\ell}\right)= \hbar\left( \pmb{x}_{3_1}^{\ell}(1:4) \right)\\
\label{Eq:Entropy_Ex3_R3_BS3c}
\le & \hbar\left( {{\pmb{x}}_{3_1}^{\ell}(1:6)} \right)
\end{align}
\end{subequations}

By dividing $\ell\log\gamma $ on both sides of \eqref{Eq:Entropy_Ex3_R1},
\eqref{Eq:Entropy_Ex3_R2_MS11},
and \eqref{Eq:Entropy_Ex3_R3_BS3}, and letting $\gamma\rightarrow \infty$ and $\ell\rightarrow \infty$, we obtain
\begin{subequations}\label{Eq:DoF_Ex2}
\begin{align}\label{Eq:DoF_Ex2_R1}
4d&\le M-2d + \omega\left( \pmb{x}_{{1_1}}^{\ell}(1:6)\right) \\
\label{Eq:DoF_Ex2_R2}
2\omega\left( \pmb{x}_{{1_1}}^{\ell}(1:6)\right) &\le d + \omega\left( {{\pmb{x}}_{3_1}^{\ell}(1)} \right)\\
\label{Eq:DoF_Ex2_R3}
4\omega\left( {{\pmb{x}}_{3_1}^{\ell}(1)} \right)
&\le \omega\left( \pmb{x}_{{3_1}}^{\ell}(1:6)\right)
\end{align}
\end{subequations}
where \eqref{Eq:DoF_Ex2_R2} follows from $\omega\left( \pmb{x}_{{1_1}}^{\ell}(1:6)\right)=\omega\left( \pmb{x}_{{1_1}}^{\ell}(7:11),\pmb{x}^{\ell}_{{3_1}}(1) \right)$, \eqref{Eq:DoF_Ex2_R3} follows from $\omega\left( {{\pmb{x}}_{3_1}^{\ell}(1)} \right)=\omega\left( {{\pmb{x}}_{3_1}^{\ell}(2)} \right)=
\omega\left( {{\pmb{x}}_{3_1}^{\ell}(3)} \right)= \omega\left( {{\pmb{x}}_{3_1}^{\ell}(4)} \right)$.

From \eqref{Eq:DoF_Ex2}, we can see that the DoF upper-bound is not available since $\omega\left( \pmb{x}_{{3_1}}^{\ell}(1:6)\right)$ is unknown. Fortunately, we have $\omega\left( {{\pmb{x}}_{3_1}^{\ell}(1:6)} \right)$ = $\omega\left( {{\pmb{x}}_{1_1}^{\ell}(1:6)} \right)$ according to Remark \ref{Remark_Same}. Consequently, by substituting $\omega\left( {{\pmb{x}}_{3_1}^{\ell}(1:6)} \right)$ = $\omega\left( {{\pmb{x}}_{1_1}^{\ell}(1:6)} \right)$ into \eqref{Eq:DoF_Ex2_R3}, we obtain $\omega\left( {{\pmb{x}}_{3_1}^{\ell}(1)} \right)\leq \omega\left( \pmb{x}_{{3_1}}^{\ell}(1:6)\right)/4$. Then, upon substituting into \eqref{Eq:DoF_Ex2_R2}, we have $\omega\left( \pmb{x}_{{3_1}}^{\ell}(1:6)\right)\leq 4d/7$. Finally, upon substituting into \eqref{Eq:DoF_Ex2_R1}, we obtain
\begin{align}\label{Eq:DoF_bound_Ex3}
 d\leq 7M/38 = 7
\end{align}
which leads to the DoF upper-bound $d\leq d^{\mathrm{Quan}}$ for \textbf{Ex 3}.

\begin{remark}\label{Remark_Simple}
From \textbf{Ex 3}, we can see that according to Remark \ref{Remark_Same}, the information theoretic DoF upper-bound problem can be converted into a linear algebra problem. Therefore, we can derive the DoF upper-bound by
solving a series of algebra inequalities, which is make it easy to derive the closed-form DoF upper-bound expression for general configurations.
\end{remark}

\subsection{Proof of DoF Upper-bound}\label{Sec:Proof1_General}
From the three examples, we know that to derive the DoF upper-bound, it is necessary to establish the algebra inequalities obtained from the sum rate inequalities. To this end, we first introduce a notion of \emph{genie tree} to show which nodes require genies in Subsection \ref{Sec:Proof1_Tree}, provide a unified way to construct genies in Subsection \ref{Sec:Proof1_Genie}, then establish the sum rate inequalities based on the constructed genies in Subsection \ref{Sec:Proof1_Inequality}, and finally derive the DoF upper-bound in Region II and Region I from the inequalities in Subsections \ref{Sec:Proof1_Bound1} and \ref{Sec:Proof1_Bound2}, respectively.

\subsubsection{Genie Tree}\label{Sec:Proof1_Tree}
From the above examples, we can see that the nodes who have irresolvable ICIs will require genies. Moreover, the irresolvable ICIs of each node (either BS or user) in the $m$th round come from the irresolvable ICIs of multiple nodes in the $(m-1)$th round, $\forall m \in \mathbb{Z}^{+}$. All nodes who have the irresolvable ICIs (i.e., require genies) in different rounds constitute a tree, called \emph{genie tree}.

\begin{definition}
\emph{Genie tree} is a rooted tree that is comprised of one node who has no irresolvable ICIs and multiple nodes who have irresolvable ICIs.
\end{definition}

\begin{figure}[htb!]
\centering
\includegraphics[width=0.90\linewidth]{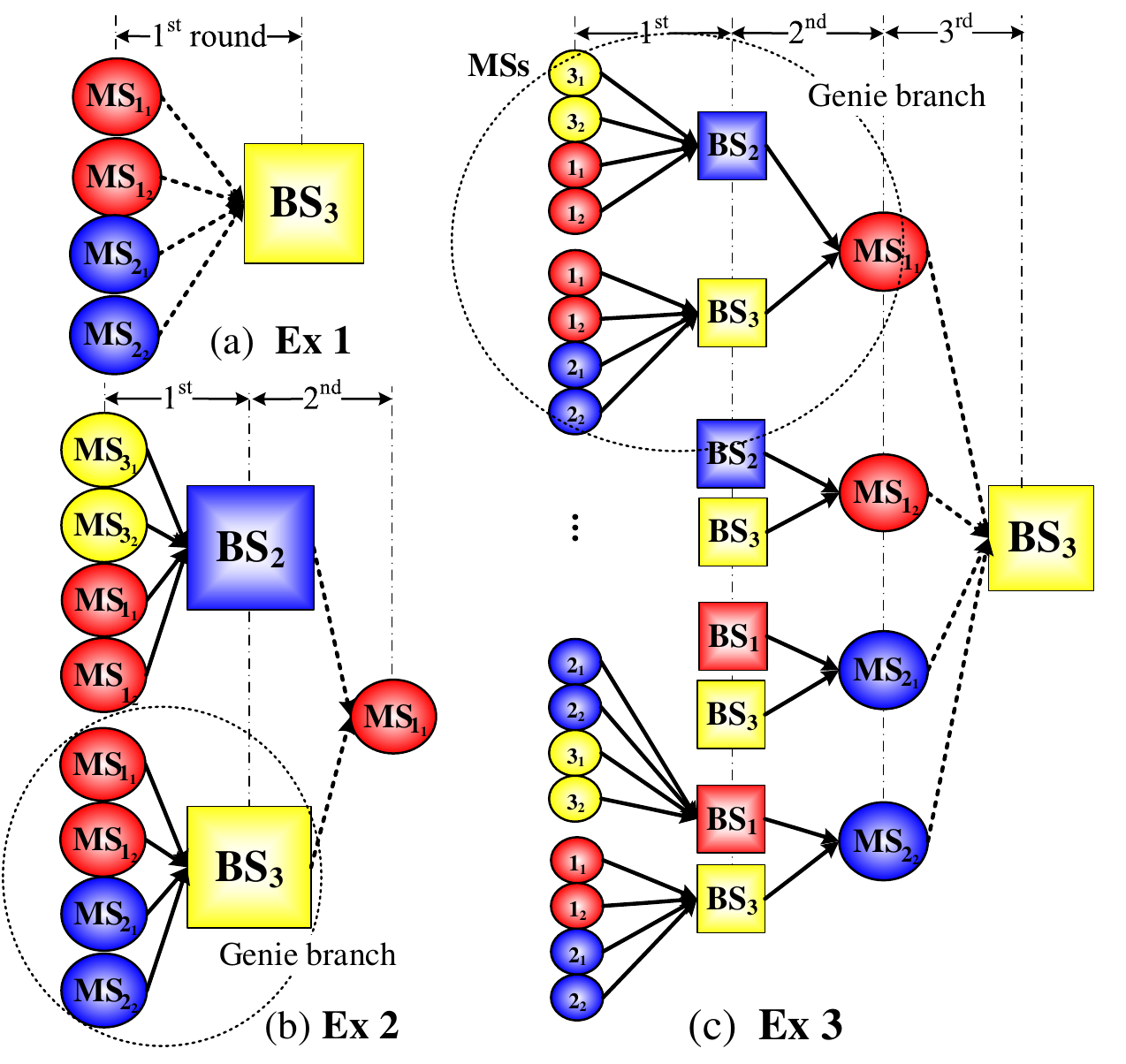}
\caption{Genie trees for \textbf{Ex 1}, \textbf{Ex 2}, and \textbf{Ex 3}.} \label{fig:Genie_Tree}
\end{figure}

From the genie-assisted ICI cancelation processes in Fig. \ref{fig:ICI_cancel_procedure} and the required genies shown in Figs. \ref{fig:Genie_Tree_Ex2} and \ref{fig:Genie_Tree_Ex3}, we obtain the genie trees for \textbf{Ex 1}, \textbf{Ex 2}, and \textbf{Ex 3} in Fig. \ref{fig:Genie_Tree}. To illustrate where the irresolvable ICIs of the nodes in the 1st round come from, as shown in Fig. \ref{fig:Genie_Tree}, the genie trees also contain the nodes in the initial round, where $\pmb{x}_{\mathrm{IR}_{l}}^{(0)}=\pmb{x}_{{l}}$.

In a genie tree, each vertex except the root denotes a node (either BS or user) who cannot resolve the received ICIs. The root is the only one node who can resolve all ICIs, which denotes the end of the genie tree. Every edge is directed from the node in the $(m-1)$th round to the node in the $m$th round, which denotes where the irresolvable ICIs of the node in the $m$th round come from, $\forall m \in \mathbb{Z}^{+},~m \leq n$. All edges directing the root are dash lines, which means that there are no irresolvable ICIs in the last round. The height of the genie tree is the maximal number of rounds in the genie-assist ICI cancelation process.

A maximal branch of genie tree is called \emph{maximal genie branch}, and is called \emph{genie branch} for short. In Fig. \ref{fig:Genie_Tree}(b) and \ref{fig:Genie_Tree}(c), we show the genie branches for \textbf{Ex 2} and \textbf{Ex 3}. We can see that the genie branch is also a rooted tree and its height is one less than the height of genie tree. In a genie branch, all nodes have irresolvable ICIs.

In Figs.~\ref{fig:Genie_Tree}(a) and \ref{fig:Genie_Tree}(c),
we show the genie tree whose root is  BS$_3$. If the root becomes one of other BSs, we can obtain other genie trees. When the maximal number of rounds is $n$, there are $G\bar{K}_{n}$ genie trees with height of $n$. From these genie trees, we can obtain $G\bar{K}_{n-1}$ genie branches with height of $n-1$.

In the genie tree, when the ICIs of one node are generated from all interfering nodes, the tree is a \emph{full tree}, otherwise, it is a \emph{partial tree}. In a partially tree, only some interference links are considered. In the forthcoming analysis, we use different genie trees to derive the DoF upper-bounds in different regions.

\subsubsection{Genie Construction}\label{Sec:Proof1_Genie}
As shown in Fig. \ref{fig:Genie_Tree}, the received ICIs of node $l$ in the $m$th round are from its children nodes (the nodes in the $(m-1)$th round), which can be expressed as
\begin{align}\label{Eq:Received_ICI_general0}
\pmb{z}_{l}^{(m)}&=\sum_{j\in \mathcal{J}_{l}^{(m)}}\pmb{H}_{l,j}^{(m-1)T}\pmb{z}_{\mathrm{IR}_{j}}^{(m-1)},~\forall m \in \mathbb{Z}^{+}\\
&=\sum_{j\in \mathcal{J}_{l}^{(m)}}
\pmb{H}_{l,j}^{(m)}\pmb{x}_{\mathrm{IR}_{j}}^{(m-1)}\nonumber\\
&=\pmb{H}_{l}^{(m)}\pmb{x}_{l}^{(m)}\nonumber
\end{align}
where $l=j$ or $l=i_k$ denotes BS$_j$ or MS$_{i_k}$, $\mathcal{J}_{l}^{(m)}\subseteq \mathcal{I}_{l}$ is the set of children of node $l$ in the $m$th round, 
\begin{align}\label{Eq:channel_matrix}
\pmb{H}_{l,j}^{(m)}=\pmb{H}_{l,j}^{(m-1)T} \pmb{H}_{\mathrm{IR}_{l}}^{(m-1)}
\end{align}
is the equivalent channel matrix from node $j$ to node $l$, $\pmb{H}_{l,j}^{(0)}=\pmb{H}_{l,j}$, $\pmb{H}_{\mathrm{IR}_{l}}^{(0)}=\pmb{I}$, and $\pmb{x}_{\mathrm{IR}_{j}}^{(0)}=\pmb{x}_{{j}}$.

The analysis in Subsection \ref{Sec:Proof1_Ex} shows that
node $l$ in the $m$th round can help its children nodes resolve their irresolvable ICIs. According to Remark \ref{Remark_Max}, we know that to derive the tight DoF upper-bound, node $l$ needs to help them resolve the maximal number of ICIs. Therefore, it is necessary to introduce different 
children nodes different genies, i.e.,
\begin{align}\label{Eq:Genie_rule11}
 &\{\pmb{x}_{\mathrm{IR}_i}^{(m-1)}\}\cap\{\pmb{x}_{\mathrm{IR}_j}^{(m-1)}\} =\varnothing,~\forall i,j\in\mathcal{J}_{l}^{(m)},~i\neq j
\end{align}
where $\{\pmb{x}\}$ denotes a set whose elements are the variables in $\pmb{x}$ and $\varnothing$ denotes an empty set.

Then, we can divide the received ICIs in \eqref{Eq:Received_ICI_general0} into
\begin{align}\label{Eq:Received_ICI_general}
 \pmb{z}_{l}^{(m)}=& \pmb{H}_{l}^{(m)}\pmb{x}_{l}^{(m)} \\
 =& \underbrace{\pmb{H}_{\mathrm{IR}_{l}}^{(m)}\pmb{x}_{\mathrm{IR}_{l}}^{(m)}} _{\pmb{z}_{\mathrm{IR}_{l}}^{(m)}:\mathrm{Irresolvable~ICIs}}
 + \underbrace{\pmb{H}_{\mathrm{R}_{l}}^{(m)}\pmb{x}_{\mathrm{R}_{l}}^{(m)}} _{\pmb{z}_{\mathrm{R}_{l}}^{(m)}:\mathrm{Resolvable~ICIs}}\nonumber
\end{align}
where $\pmb{H}_{{l}}^{(m)}\in \mathbb{C}^{L_{\mathrm{R}_l}^{(m)}\times L_{l}^{(m)} }$,
$\pmb{H}_{\mathrm{IR}_{l}}^{(m)}\in \mathbb{C}^{L_{\mathrm{R}_l}^{(m)}\times L_{\mathrm{IR}_l}^{(m)}}$,
$\pmb{H}_{\mathrm{R}_{l}}^{(m)}\in \mathbb{C}^{L_{\mathrm{R}_l}^{(m)}\times L_{\mathrm{R}_l}^{(m)}}$, $\pmb{x}_{{l}}^{(m)}\in \mathbb{C}^{L_{l}^{(m)}\times 1}$, $\pmb{x}_{\mathrm{IR}_{l}}^{(m)}\in \mathbb{C}^{L_{\mathrm{IR}_l}^{(m)}\times 1}$, and
$\pmb{x}_{\mathrm{R}_{l}}^{(m)}\in \mathbb{C}^{L_{\mathrm{R}_l}^{(m)}\times 1}$.

In \eqref{Eq:Received_ICI_general}, $L_l^{(m)}$ is the number of columns in $\pmb{H}_{l}^{(m)}$, while $L_{\mathrm{R}_l}^{(m)}$ is the number of rows in $\pmb{H}_{l}^{(m)}$, and $L_{\mathrm{IR}_l}^{(m)}$ satisfies
\begin{align}
\label{Eq:dim_irr_space0}
L_{\mathrm{IR}_l}^{(m)}&=\left(L_l^{(m)}-L_{\mathrm{R}_l}^{(m)}\right)^{+}
\end{align}
where $x^{+}=\max\{x,0\}$.

When the received ICIs are divided into the resolvable and irresolvable ICIs, the interference subspace is also divided into the \emph{resolvable subspace} (denoted by $\mathrm{span}\{\pmb{H}_{\mathrm{R}_l}^{(m)}\}$) and \emph{irresolvable subspace} (denoted by $\mathrm{span}\{\pmb{H}_{\mathrm{IR}_l}^{(m)}\}$). Therefore, $L_l^{(m)}$, $L_{\mathrm{R}_l}^{(m)}$, and $L_{\mathrm{IR}_l}^{(m)}$ denote the dimension of interference, resolvable, and irresolvable subspaces, respectively.

To introduce the genie to help node $l$  cancel its irresolvable ICIs, we construct the genie as
\begin{align}\label{Eq:Genie_general}
 \pmb{G}_{l}^{(m)}=\pmb{x}_{\mathrm{IR}_l}^{(m)}
\end{align}

Moreover, for node $l$ in the $m$th round, to make its parent node (i.e., the node in the $(m+1)$th round) help its children nodes resolve the maximal number of ICIs, $\pmb{x}_{\mathrm{IR}_l}^{(m)}$ in \eqref{Eq:Received_ICI_general} also needs to satisfy $\{\pmb{x}_{\mathrm{IR}_l}^{(m)}\}\cap\{\pmb{x}_{\mathrm{IR}_j}^{(m)}\} =\varnothing,~\forall~l\neq j,~l,j\in\mathcal{J}_{i}^{(m+1)}$.

As shown in \eqref{Eq:Genie_general}, we only use one way rather than four different ways to construct genies for arbitrary antenna configurations, which is helpful to derive the DoF upper-bound for general cases.

\subsubsection{Sum Rate Inequality}\label{Sec:Proof1_Inequality}
When introducing the genie according to \eqref{Eq:Genie_general}, node $l$ can resolve all received ICIs. Thus, we obtain the sum rate inequality in the $m$th round as
\begin{align}\label{Eq:Entropy_All}
\sum_{j\in \mathcal{J}_{l}^{(m)}}\hbar\left(\pmb{z}_{\mathrm{IR}_j}^{(m-1)\ell}\right)
\leq & \hbar\left(\pmb{z}_{\mathrm{R}_l}^{(m)\ell}\right)+\hbar\left(\pmb{x}_{\mathrm{IR}_l}^{(m)\ell} \right)
\end{align}
 
From the analysis in \cite{Jafar_IAchainKcell_all}, we have
\begin{subequations}
\begin{align}\label{Eq:Entropy_Equ1}
\hbar\left(\pmb{x}_{\mathrm{R}_l}^{(m)\ell}\right)&= \hbar\left(\pmb{H}_{\mathrm{R}_l}^{(m)\ell}\pmb{x}_{\mathrm{R}_l}^{(m)\ell}\right)
  = \hbar\left(\pmb{z}_{\mathrm{R}_l}^{(m)\ell}\right)\\
\label{Eq:Entropy_Equ2}
\hbar\left(\pmb{x}_{\mathrm{IR}_l}^{(m)\ell}\right)&= \hbar\left(\pmb{H}_{\mathrm{IR}_l}^{(m)\ell}\pmb{x}_{\mathrm{IR}_l}^{(m)\ell}\right)
  = \hbar\left(\pmb{z}_{\mathrm{IR}_l}^{(m)\ell}\right)
\end{align}
\end{subequations}
where \eqref{Eq:Entropy_Equ1} always holds since $\pmb{H}_{\mathrm{R}_{l}}^{(m)}\in \mathbb{C}^{L_{\mathrm{R}_l}^{(m)}\times L_{\mathrm{R}_l}^{(m)}}$ and \eqref{Eq:Entropy_Equ2} holds when $L_{\mathrm{IR}_l}^{(m)} \leq L_{\mathrm{R}_l}^{(m)}$,\footnote{In the forthcoming analysis, we provide the condition that can ensure $L_{\mathrm{IR}_l}^{(m)} \leq L_{\mathrm{R}_l}^{(m)}$.} since $\pmb{H}_{\mathrm{IR}_{l}}^{(m)}\in \mathbb{C}^{L_{\mathrm{R}_l}^{(m)}\times L_{\mathrm{IR}_l}^{(m)}}$.

By substituting \eqref{Eq:Entropy_Equ1} and \eqref{Eq:Entropy_Equ2} into \eqref{Eq:Entropy_All}, dividing $\ell\log\gamma $ on both sides of \eqref{Eq:Entropy_All}, and letting $\gamma\rightarrow \infty$ and $\ell\rightarrow \infty$, we obtain
\begin{align}\label{Eq:inequation_general1}
\sum_{j\in \mathcal{J}_{l}^{(m)}}\Delta_{\mathrm{IR}_j}^{(m-1)}\leq \Delta_{\mathrm{R}_l}^{(m)}+\Delta_{\mathrm{IR}_l}^{(m)},~\forall m \in \mathbb{Z}^{+}
\end{align}
where
\begin{subequations}\label{Eq:Delta_general1}
\begin{align}
\Delta_{\mathrm{IR}_l}^{(m)}&\triangleq \omega\left(\pmb{x}_{\mathrm{IR}_l}^{(m)\ell}\right)=
\lim_{\ell,\gamma\rightarrow \infty}{\hbar\left(\pmb{x}_{\mathrm{R}_l}^{(m)\ell} \right)}/ {(\ell\log\gamma)}
\\
\Delta_{\mathrm{R}_l}^{(m)}&\triangleq \omega\left(\pmb{x}_{\mathrm{R}_l}^{(m)\ell}\right)=
\lim_{\ell,\gamma\rightarrow \infty}{\hbar\left(\pmb{x}_{\mathrm{R}_l}^{(m)\ell}\right)}/ {(\ell\log\gamma)}
\end{align}
\end{subequations}
with $\Delta_{\mathrm{IR}_l}^{(0)}=d$ and $\Delta_{\mathrm{R}_l}^{(1)}=M-Kd$ follow because $\hbar\left(\pmb{x}_{\mathrm{IR}_l}^{(0)\ell}\right)=\ell R$ and $\hbar\left(\pmb{x}_{\mathrm{R}_l}^{(1)\ell}\right)=M\ell\left(\log\gamma + o(\log\gamma)\right)-K\ell R$, respectively.

In \eqref{Eq:inequation_general1}, $\Delta_{\mathrm{R}_l}^{(m)}$ and $\Delta_{\mathrm{IR}_l}^{(m)}$ are equal to the numbers of resolvable and irresolvable ICIs in the $m$th round, which reflect how many ICIs can be resolved by node $l$ and the genie $\pmb{G}_{l}^{(m)}$, respectively. To derive the DoF upper-bound, it is necessary to find the upper-bound of $\Delta_{\mathrm{IR}_l}^{(m)}$.

From \eqref{Eq:Entropy_Equ1}, \eqref{Eq:Entropy_Equ2}  and \eqref{Eq:Delta_general1}, we have
\begin{subequations}\label{Eq:Delta34}
\begin{align}
\label{Eq:Delta3}
\Delta_{\mathrm{IR}_l}^{(m)}=&\omega\left(\pmb{H}_{\mathrm{IR}_l}^{(m)\ell}\pmb{x}_{\mathrm{IR}_l}^{(m)\ell}\right)\leq L_{\mathrm{IR}_l}^{(m)},~\forall m \in \mathbb{Z}^{+}\\
\label{Eq:Delta4}
\Delta_{\mathrm{R}_l}^{(m)}=&\omega\left(\pmb{H}_{\mathrm{R}_l}^{(m)\ell}\pmb{x}_{\mathrm{R}_l}^{(m)\ell}\right)\leq L_{\mathrm{R}_l}^{(m)},~\forall m \in \mathbb{Z}^{+}
\end{align}
\end{subequations}

From \eqref{Eq:Delta3}, we can take $L_{\mathrm{IR}_l}^{(m)}$ as the upper-bound of $\Delta_{\mathrm{IR}_l}^{(m)}$. In the following, we derive the expression of $L_{\mathrm{IR}_l}^{(m)}$.

Since \eqref{Eq:Received_ICI_general0} shows that the received ICIs in the $m$th round are from the irresolvable ICIs of nodes in the $(m-1)$th round,
the dimension of interference subspace is
\begin{align}
\label{Eq:dim_total_space}
L_l^{(m)}&=\sum_{j\in \mathcal{J}_{l}^{(m)}}L_{\mathrm{IR}_j}^{(m-1)},~\forall m \in \mathbb{Z}^{+}
\end{align}
where $L_{\mathrm{IR}}^{(0)}=N$.

In \eqref{Eq:Received_ICI_general}, $\pmb{H}_{l,j}^{(m)}$ has $L_{\mathrm{R}_j}^{(m)}$ rows. By substituting $\pmb{H}_{\mathrm{IR}_{l}}^{(m-1)} \in \mathbb{C}^{ L_{\mathrm{R}_j}^{(m-1)}\times L_{\mathrm{IR}_j}^{(m-1)}}$ and $\pmb{H}_{l,j}^{(m-1)T} \in \mathbb{C}^{ L_{\mathrm{IR}_j}^{(m-2)}\times L_{\mathrm{R}_j}^{(m-1)}}$ into \eqref{Eq:channel_matrix}, we know that $\pmb{H}_{l,j}^{(m)}$ has $L_{\mathrm{IR}_j}^{(m-2)}$ rows, $\forall m\geq 2$. Therefore, we have $L_{\mathrm{R}_l}^{(m)}=L_{\mathrm{IR}_l}^{(m-2)},~\forall m\geq 2$. By letting $L_{\mathrm{IR}}^{(-1)}=L_{\mathrm{R}_l}^{(1)}=M$,
we obtain a unified expression of $L_{\mathrm{R}_l}^{(m)}$ as follows,
\begin{align}
\label{Eq:dim_ir_space}
L_{\mathrm{R}_l}^{(m)}&=L_{\mathrm{IR}_l}^{(m-2)},~\forall m \in \mathbb{Z}^{+}
\end{align}
It indicates that the irresolvable subspace in the $(m-2)$th round $\mathrm{span}\{\pmb{H}_{\mathrm{IR}_{l}}^{(m-2)}\}$ becomes the resolvable subspace in the $m$th round $\mathrm{span}\{\pmb{H}_{\mathrm{R}_{l}}^{(m)}\}$. It is because in the $(m-2)$th round, node $l$ has canceled the ICIs in $\mathrm{span}\{\pmb{H}_{\mathrm{R}_{l}}^{(m-2)}\}$. As a result, in the $m$th round, node $l$ cannot receive the ICIs in $\mathrm{span}\{\pmb{H}_{\mathrm{R}_{l}}^{(m-2)}\}$ any more but can receive the ICIs in $\mathrm{span}\{\pmb{H}_{\mathrm{IR}_{l}}^{(m-2)}\}$.

By substituting \eqref{Eq:dim_total_space} and \eqref{Eq:dim_ir_space} into \eqref{Eq:dim_irr_space0}, we obtain the recursive expression of $L_{\mathrm{IR}_l}^{(m)}$ as
\begin{align}\label{Eq:recursive_L}
L_{\mathrm{IR}_l}^{(m)}&=\left(\sum_{j\in \mathcal{J}_{l}^{(m)}}L_{\mathrm{IR}_j}^{(m-1)}- L_{\mathrm{IR}_l}^{(m-2)}\right)^{+},~\forall m \in \mathbb{Z}^{+}
\end{align}
where $L_{\mathrm{IR}_l}^{(-1)}=M$ and $L_{\mathrm{IR}_l}^{(0)}=N$.

The previous analysis indicates that to ensure \eqref{Eq:Entropy_Equ2}, it is necessary to meet $L_{\mathrm{IR}_l}^{(m)} \leq  L_{\mathrm{R}_l}^{(m)}$. Therefore, from \eqref{Eq:dim_ir_space} and \eqref{Eq:recursive_L}, 
$\mathcal{J}_{l}^{(m)}$ should satisfy
\begin{align}\label{Eq:Condition_full_tree}
  \sum_{j\in \mathcal{J}_{l}^{(m)}}L_{\mathrm{IR}_j}^{(m-1)} \leq 2L_{\mathrm{IR}_l}^{(m-2)}
\end{align}

From \eqref{Eq:dim_ir_space}, we know that $\pmb{x}_{\mathrm{IR}_{l}}^{(m-2)}$ and $\pmb{x}_{\mathrm{R}_{l}}^{(m)}$ have the same dimensions. According to Remark \ref{Remark_Same}, we have
\begin{align}\label{Eq:inequation_general2}
 \Delta_{\mathrm{R}_l}^{(m)}=\Delta_{\mathrm{IR}_l}^{(m-2)}
\end{align}
where $\Delta_{\mathrm{IR}_l}^{(-1)}=\Delta_{\mathrm{R}_l}^{(1)}=M-Kd$.

By substituting \eqref{Eq:inequation_general2} into \eqref{Eq:inequation_general1}, we obtain the recursive inequality of $\Delta_{\mathrm{IR}_l}^{(m)}$ as
\begin{align}\label{Eq:recursive_D}
\Delta_{\mathrm{IR}_l}^{(m)} \geq \left(\sum_{j\in \mathcal{J}_{l}^{(m)}}\Delta_{\mathrm{IR}_j}^{(m-1)}- \Delta_{\mathrm{IR}_l}^{(m-2)}\right)^{+}
\end{align}
where $\Delta_{\mathrm{IR}_l}^{(-1)}=M-Kd$ and $\Delta_{\mathrm{IR}_l}^{(0)}=d$.

As shown in \eqref{Eq:recursive_L}, $L_{\mathrm{IR}_l}^{(m)}$ is independent of the unknown variable $d$ (i.e., the DoF). Therefore, we can derive the DoF upper-bound by substituting \eqref{Eq:recursive_L} and \eqref{Eq:recursive_D} into \eqref{Eq:Delta3}.

From \eqref{Eq:Condition_full_tree}, we obtain $L_{\mathrm{IR}_l}^{(m)}\leq  L_{\mathrm{IR}_l}^{(m-2)}$, which means that $L_{\mathrm{IR}_l}^{(m)}$ decreases when $m$ increases as $m=2n$ or $m=2n+1$, $n=1,2,\cdots$. As a result, for each node (either BS or user), its dimension of irresolvable subspace is decreasing. Consequently, all the irresolvable ICIs can be resolved and canceled until $L_{\mathrm{IR}_l}^{(m)}=0$. Therefore, for a genie tree satisfying \eqref{Eq:Condition_full_tree}, its height is finite, otherwise, it may be infinite. Since the number of DoF inequalities depends on the height of the genie tree, we can derive the DoF upper-bound from the genie tree with a finite height. It is not hard to prove that when $M/N\in \mathcal{R}^{\mathrm{II}}$, full genie trees satisfy \eqref{Eq:Condition_full_tree}, while when $M/N\in \mathcal{R}^{\mathrm{I}}$, only partial genie trees satisfy \eqref{Eq:Condition_full_tree}. In the sequel, we derive the DoF upper-bound in Region II from full trees and the DoF upper-bound in Region I from partial trees.

\subsubsection{DoF Upper-bound in Region II}\label{Sec:Proof1_Bound1}

In this subsection, we prove that the quantity DoF bound is the DoF upper-bound in Region II from full genie trees.

As shown in Fig. \ref{fig:Genie_Tree}, all BSs suffer from the ICIs from $(G-1)K$ interfering users and all users suffer from the ICIs from $(G-1)$ interfering BSs. Consequently, all nodes in the same round have the identical dimension of subspace and number of ICIs. Therefore, by omitting the subscript $l$, we can use $L^{(m)}$, $L_{\mathrm{R}}^{(m)}$, and $L_{\mathrm{IR}}^{(m)}$ to denote the dimension of interference, resolvable, and irresolvable subspaces, $\Delta^{(m)}$, $\Delta_{\mathrm{R}}^{(m)}$, and $\Delta_{\mathrm{IR}}^{(m)}$ to denote the number of received, resolvable, and irresolvable ICIs, respectively, $\forall m \in \mathbb{Z}^{+}$.

From \eqref{Eq:recursive_L} and \eqref{Eq:recursive_D}, we have
\begin{subequations}\label{Eq:recursive_LD1}
\begin{align}\label{Eq:recursive_L1}
L_{\mathrm{IR}}^{(m)}&=\left((G-1)\bar{K}_{m-1}L_{\mathrm{IR}}^{(m-1)} -L_{\mathrm{IR}}^{(m-2)}\right)^{+}\\
\label{Eq:recursive_D1}
\Delta_{\mathrm{IR}}^{(m)} &\geq \left((G-1)\bar{K}_{m-1}\Delta_{\mathrm{IR}}^{(m-1)}- \Delta_{\mathrm{IR}}^{(m-2)}\right)^{+}
\end{align}
\end{subequations}
where $\Delta_{\mathrm{IR}}^{(-1)}=M-Kd$, $\Delta_{\mathrm{IR}}^{(0)}=d$, $L_{\mathrm{IR}}=M$, and $L_{\mathrm{IR}}=N$.

Using mathematical induction, from \eqref{Eq:recursive_LD1} we obtain
the closed-form expressions of $L_{\mathrm{IR}}^{(m)}$ and $\Delta_{\mathrm{IR}}^{(m)}$ in the following lemma.
\begin{lemma}\label{Lemma_Dim_space}
When $M/N \in \mathcal{R}^{\mathrm{II-A}}_{n},~\forall n \in \mathbb{Z}^{+},~n\leq n^{\max}$, from a full genie tree whose leaf vertexes are users, the dimension of irresolvable subspace and the number of irresolvable ICIs in the $m$th round are, respectively,
\begin{subequations}\label{Eq:dim_irr_spaceDim}
\begin{align}
\label{Eq:dim_irr_space}
L_{\mathrm{IR}}^{(m)}&=\left(q_{m}^{\mathrm{A}}N-p_{m}^{\mathrm{A}}M\right)^{+},~\forall m \in \mathbb{Z}^{+}\\
\label{Eq:dim_irr_interference}
\Delta_{\mathrm{IR}}^{(m)}&\geq\left(q_{m}^{\mathrm{A}}d-p_{m}^{\mathrm{A}}(M-Kd)\right)^{+},~\forall m \in \mathbb{Z}^{+}
\end{align}
\end{subequations}
where $\{q_{m}^{\mathrm{A}},p_{m}^{\mathrm{A}}\}$ is the generalized Fibonacci sequence-pair defined in \eqref{Eq:Fsequence_A}.
\end{lemma}

According to the value of $L_{\mathrm{IR}}^{(m)}$, we can judge whether the ICIs are resolvable or not in the $m$th round. If $L_{\mathrm{IR}}^{(m)}>0$, some ICIs in $\pmb{z}_{l}^{(m)}$ are irresolvable. When $L_{\mathrm{IR}}^{(m)}=0$, all ICIs in $\pmb{z}_{l}^{(m)}$ are resolvable.
From Lemma \ref{Lemma_Dim_space}, we can obtain the maximal number of rounds for different antenna configurations.
\begin{lemma}\label{Lemma_height_space}
When $M/N \in \mathcal{R}^{\mathrm{II-A}}_{n},~\forall n \in \mathbb{Z}^{+},~n\leq n^{\max}$, we have
\begin{align}\label{Eq:Len_subspace_chain}
L_{\mathrm{IR}}^{(m)}\left\{
\begin{array}{ll}
 >0, &~\forall m < n \\
 =0, &~\forall m=n
\end{array}
\right.
\end{align}
It indicates that the maximal number of rounds is $n,~\forall M/N \in \mathcal{R}^{\mathrm{II-A}}_{n}$.
\end{lemma}

\begin{proof}
When $M/N \in \mathcal{R}^{\mathrm{II-A}}_{n}$, i.e., $C_{n}^{\mathrm{A}} \leq M/N < C_{n-1}^{\mathrm{A}}$, from \eqref{Eq:C_pq} we have $q_{n}^{\mathrm{A}}N-p_{n}^{\mathrm{A}}M \leq 0$ and $q_{n-1}^{\mathrm{A}}N-p_{n-1}^{\mathrm{A}}M > 0$. Upon substituting into \eqref{Eq:dim_irr_space}, we have $L_{\mathrm{IR}}^{(m)}>0,~\forall m<n$ and $L_{\mathrm{IR}}^{(n)}=0$.
\end{proof}

When $M/N \in \mathcal{R}^{\mathrm{II-A}}_{n}$, according to Lemma \ref{Lemma_Dim_space}, \eqref{Eq:Delta3} and Lemma \ref{Lemma_height_space}, we have $0 \leq \Delta_{\mathrm{IR}}^{(m)}\leq L_{\mathrm{IR}}^{(m)},~\forall m\leq n-1$ and $\Delta_{\mathrm{IR}}^{(n)}\leq 0$. By substituting \eqref{Eq:dim_irr_interference} into $\Delta_{\mathrm{IR}}^{(n)}\leq 0$, we obtain $q_{n}^{\mathrm{A}}d-p_{n}^{\mathrm{A}}(M-Kd)\leq0$, i.e.,
\begin{align}\label{Eq:upper-bound_M}
d \leq \frac{p_{n}^{\mathrm{A}}M}{p_{n}^{\mathrm{A}}K+q_{n}^{\mathrm{A}}}=\frac{M}{K+C_{n}^{\mathrm{A}}} 
\end{align}
By substituting \eqref{Eq:dim_irr_spaceDim} into $0 \leq \Delta_{\mathrm{IR}}^{(m)}\leq L_{\mathrm{IR}}^{(m)},~\forall m\leq n-1$, we obtain $q_{m}^{\mathrm{A}}d-p_{m}^{\mathrm{A}}(M-Kd)\leq q_{m}^{\mathrm{A}}N-p_{m}^{\mathrm{A}}M$, i.e.,
\begin{align}\label{Eq:upper-bound_N}
d \leq \frac{q_{m}^{\mathrm{A}}N}{q_{m}^{\mathrm{A}}+Kp_{m}^{\mathrm{A}}} =\frac{N}{1+K/C_{m}^{\mathrm{A}}},~\forall m\leq n-1
\end{align}

Since $C_{m}^{\mathrm{A}}$ is a monotonically decreasing sequence, the tightest bound in \eqref{Eq:upper-bound_N} is $d\leq N/({1+K/C_{n-1}^{\mathrm{A}}})$.
As a result, the upper-bound of DoF per user is
\begin{align}\label{Eq:DoF_Bound_A}
d \leq \min\left\{\frac{M}{K+C_{n}^{\mathrm{A}}}, \frac{N}{1+\frac{K}{C_{n-1}^{\mathrm{A}}}}\right\},~\forall \frac{M}{N}\in\mathcal{R}^{\mathrm{II-A}}_{n}
\end{align}
where $\forall n \in \mathbb{Z}^{+}, n\leq n^{\max}$. It indicates that the quantity DoF bound is the DoF upper-bound in $\mathcal{R}^{\mathrm{II-A}}= \cup_{n=1}^{n^{\max}}\mathcal{R}^{\mathrm{II-A}}_{n}$, i.e., Region II-A.

Based on the signal model of the MIMO-IMAC network in \eqref{Eq:Received_signal_IMAC}, we have obtained the full genie trees whose leaf vertexes are users and proved that quantity DoF bound is the DoF upper-bound in Region II-A. Similarly, based on the signal model of the MIMO-IBC network in \eqref{Eq:Received_signal_IBC}, we can obtain the full genie trees whose leaf vertexes are BSs and prove that the quantity DoF bound is the DoF upper-bound in Region II-B.

This completes the proof that the quantity DoF bound is the DoF upper-bound in Region II, i.e., \eqref{Eq:DoF_Bound_R2}.

\subsubsection{DoF Upper-bound in Region I}\label{Sec:Proof1_Bound2}

In this subsection, we prove that when $\forall M/N \in \mathcal{R}^{\mathrm{I}}\cap\mathcal{Q}$, the decomposition DoF bound is the DoF upper-bound. 

\begin{lemma}\label{Lemma_Dim_space_partial}
When $M/N\in\{\tilde{q}_{l,n}^{\mathrm{A}}/\tilde{p}_{l,n}^{\mathrm{A}}\}$, from a partial tree whose leaf vertexes are users and the sets of children nodes are $\mathcal{J}_{l}^{(m)}=\mathcal{I}_{l,m}^{\mathrm{A}}$ satisfying \eqref{Eq:SetA_Condition},\footnote{The constraint in \eqref{Eq:SetA_Condition} ensures \eqref{Eq:Condition_full_tree}.} $\forall m\leq n$,  the dimension of irresolvable subspace and the number of irresolvable ICIs are, respectively,
\begin{subequations}
\begin{align}
\label{Eq:dim_irr_space_0a}
L_{\mathrm{IR}_l}^{(m)}&=\left(\tilde{q}_{l,m}^{\mathrm{A}}N-\tilde{p}_{l,m}^{\mathrm{A}}M\right)^{+},~\forall m \in \mathbb{Z}^{+}\\
\label{Eq:dim_irr_singal_0a}
\Delta_{\mathrm{IR}_l}^{(m)}&\geq\left(\tilde{q}_{l,m}^{\mathrm{A}}d-\tilde{p}_{l,m}^{\mathrm{A}}(M-Kd)\right)^{+},~\forall m \in \mathbb{Z}^{+}
\end{align}
\end{subequations}
where $\{\tilde{q}_{l,m}^{\mathrm{A}},\tilde{p}_{l,m}^{\mathrm{A}}\}$ is the generalized Fibonacci sequence-pair defined in \eqref{Eq:Fsequence_A1}.
\end{lemma}

Since $M/N=\tilde{q}_{l,n}^{\mathrm{A}}/\tilde{p}_{l,n}^{\mathrm{A}}$, from \eqref{Eq:dim_irr_space_0a} we have $L_{\mathrm{IR}_{l}}^{(n)}=\tilde{q}_{l,n}^{\mathrm{A}}N-\tilde{p}_{l,n}^{\mathrm{A}}M =0$, so that $\Delta_{\mathrm{IR}_{l}}^{(n)}\leq L_{\mathrm{IR}_{l}}^{(n)} = 0$. Upon substituting into \eqref{Eq:dim_irr_singal_0a}, we have
\begin{align}\label{Eq:DoF_Bound_RI}
 d\leq \frac{M}{K+\tilde{q}_{l,n}^{\mathrm{A}}/\tilde{p}_{l,n}^{\mathrm{A}}}=\frac{MN}{M+KN}
\end{align}
It means that the decomposition bound is the DoF upper-bound.
 
Similarly, when $M/N\in\{\tilde{q}_{l,n}^{\mathrm{B}}/\tilde{p}_{l,n}^{\mathrm{B}}\}$, from a partial tree whose leaf vertexes are BSs and $\mathcal{J}_{l}^{(m)}=\mathcal{I}_{l,m}^{\mathrm{B}}$ satisfying \eqref{Eq:SetB_Condition}, $\forall m\leq n$, we can prove that the decomposition bound is the DoF upper-bound.

This completes the proof of Theorem \ref{Theorem:DoF_Upperbound}.

\subsection{Applications}\label{Sec:Proof1_Applications}
In this subsection, we take several examples to show how to use the genie tree to derive the DoF upper-bound for the networks with configurations in both Region I and Region II. To understand the relationship of our genie tree and the genie chain in \cite{Jafar_IAchainKcell_all}, we consider five examples in a four-cell one-user MIMO-IC network, whose configurations and the corresponding DoF upper-bounds are listed in Table \ref{Table:CN1}.

\begin{table}[htb!]\centering
\caption{Antenna configurations and DoF upper-bounds of examples, where $G=4$ and $K=1$.}\label{Table:CN1}
\begin{tabular}{c|c|c|c|c}
 \hline
 & $M$ & $N$ & Upper-bound & \\
 \hline
 \hline
\textbf{Ex 4} & 29 & 11 & $d^{\mathrm{Quan}}=d^{\mathrm{Proper}}=8$ & Unsolved in \cite{Jafar_IAchainKcell_all}\\
 \hline
\textbf{Ex 5} & 31 & 12 & $d^{\mathrm{Decom}}=372/43$ & Unsolved in \cite{Jafar_IAchainKcell_all}\\
 \hline
\textbf{Ex 6} & 5 & 2 & $d^{\mathrm{Decom}}=10/7$ & \textbf{Example 1} in \cite{Jafar_IAchainKcell_all}\\
 \hline
\textbf{Ex 7} & 7 & 3 & $d^{\mathrm{Decom}}=21/10$ & \textbf{Example 2} in \cite{Jafar_IAchainKcell_all}\\
 \hline
\textbf{Ex 8} & 8 & 3 & $d^{\mathrm{Quan}}=d^{\mathrm{Decom}}=24/11$ & \textbf{Example 3} in \cite{Jafar_IAchainKcell_all}\\
 \hline
\end{tabular}
\end{table}

\subsubsection{\textbf{Ex 4}}
We first consider \textbf{Ex 4} where $M=29$ and $N=11$.
Since $M/N \in \mathcal{R}_{3}^{\mathrm{II-A}} = [21/8,8/3)$ from \eqref{Eq:Quantity DoF}, the DoF upper-bound is $d^{\mathrm{Quan}}=8$ and it can be derived from a full genie tree with height of three.

From \eqref{Eq:dim_irr_space}, we have
\begin{subequations}\label{Eq:Dim_Ex4}
\begin{align}
L_{\mathrm{IR}}^{(1)} & = (3L_{\mathrm{IR}}^{(0)}-L_{\mathrm{IR}}^{(-1)})^{+} =33-29 = 4\\
L_{\mathrm{IR}}^{(2)} & = (3L_{\mathrm{IR}}^{(1)}-L_{\mathrm{IR}}^{(0)})^{+} = 12-11= 1\\
L_{\mathrm{IR}}^{(3)} & = (3L_{\mathrm{IR}}^{(2)}-L_{\mathrm{IR}}^{(1)})^{+} = (3-4)^{+} = 0
\end{align}
\end{subequations}
where $L_{\mathrm{IR}}^{(-1)}=M=29$ and $L_{\mathrm{IR}}^{(0)}=N=11$.

It means that to resolve all ICIs, we need to introduce BSs $L_{\mathrm{IR}}^{(1)}=4$-dimensional genies in the 1st round, and introduce users $L_{\mathrm{IR}}^{(2)}=1$-dimensional genies in the 2nd round, and do not introduce BSs any genie in the 3rd round, since they can resolve all ICIs.

\begin{figure}[htb!]
\centering
\includegraphics[width=0.90\linewidth]{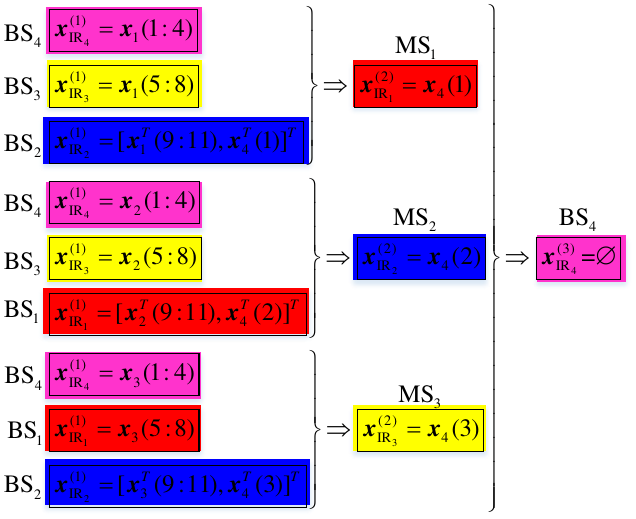}
\caption{Introduced genies for \textbf{Ex 4}.}
\label{fig:Genie_Tree_Ex4}
\end{figure}

According to the required genie dimensions in \eqref{Eq:Dim_Ex4} and the basic principle to construct genies \eqref{Eq:Genie_rule11}, we design the genies for different nodes shown in Fig. \ref{fig:Genie_Tree_Ex4}. To derive the DoF upper-bound, we need to introduce the same node different genies to obtain different sum rate inequalities.

All the sum rate inequalities obtained from the genie tree can be converted into the following algebra inequalities
\begin{subequations}\label{Eq:DoF_Ex4}
\begin{align}
\label{Eq:DoF_Ex4_R1}
3d&\le M-d +  \Delta_{\mathrm{IR}}^{(1)}\\
\label{Eq:DoF_Ex4_R2}
3\Delta_{\mathrm{IR}}^{(1)} &\le d + \Delta_{\mathrm{IR}}^{(2)}\\
\label{Eq:DoF_Ex4_R3}
3\Delta_{\mathrm{IR}}^{(2)}
&\le \Delta_{\mathrm{IR}}^{(1)}
\end{align}
\end{subequations}
where $\Delta_{\mathrm{IR}}^{(m)}=\omega\left( \pmb{x}_{\mathrm{IR}_{l}}^{(m)\ell}\right)$, $m=1,2$.

By solving \eqref{Eq:DoF_Ex4}, we obtain
\begin{align}\label{Eq:DoF_bound_Ex4}
 d\leq 8M/29 = 8
\end{align}
which leads to the DoF upper-bound $d\leq d^{\mathrm{Quan}}=d^{\mathrm{Proper}}$ for \textbf{Ex 4}.

\subsubsection{\textbf{Ex 5}}
We first consider \textbf{Ex 5} where $M=31$ and $N=12$.
Since $M/N \in \mathcal{R}^{\mathrm{I}}=((3-\sqrt{5})/2,(3+\sqrt{5})/2)$ from \eqref{Eq:Region},
the DoF upper-bound is $d^{\mathrm{Decom}}=MN/(M+N)=372/43$ and it can be derived from a partial genie tree.

According to the constraint in \eqref{Eq:Condition_full_tree}, we design the genie tree where each BS helps three users resolve ICIs in the 1st round, then each user helps three BSs resolve ICIs in the 2nd round, each BS cannot help three BSs but can help two BSs resolve ICIs in the 3rd round, until each user can help three users resolve ICIs without any genies in the 4st round. Consequently, from \eqref{Eq:dim_irr_space_0a}, we have
\begin{subequations}
\label{Eq:Dim_Ex4}
\begin{align}
\label{Eq:Dim_Ex4_R1}
L_{\mathrm{IR}}^{(1)} &=  (3L_{\mathrm{IR}}^{(0)}-L_{\mathrm{IR}}^{(-1)})^{+} = 36-31 = 5\\
\label{Eq:Dim_Ex4_R2}
L_{\mathrm{IR}}^{(2)} & = (3L_{\mathrm{IR}}^{(1)}-L_{\mathrm{IR}}^{(0)})^{+} = 15-12 = 3\\
\label{Eq:Dim_Ex4_R3}
L_{\mathrm{IR}}^{(3)} & = (2L_{\mathrm{IR}}^{(2)}-L_{\mathrm{IR}}^{(1)})^{+} = 6-5=1\\
\label{Eq:Dim_Ex4_R4}
L_{\mathrm{IR}}^{(4)} & = (3L_{\mathrm{IR}}^{(3)}-L_{\mathrm{IR}}^{(2)})^{+} = 3-3=0
\end{align}
\end{subequations}

Therefore, from \eqref{Eq:Dim_Ex4}, we know that
to resolve all ICIs, we need to introduce BSs $L_{\mathrm{IR}}^{(1)}=5$-dimensional genie in the 1st round, and introduce users $L_{\mathrm{IR}}^{(2)}=3$-dimensional genie in the 2nd round and introduce BSs $L_{\mathrm{IR}}^{(3)}=1$-dimensional genie in the 3rd round.

\begin{figure}[htb!]
\centering
\includegraphics[width=0.99\linewidth]{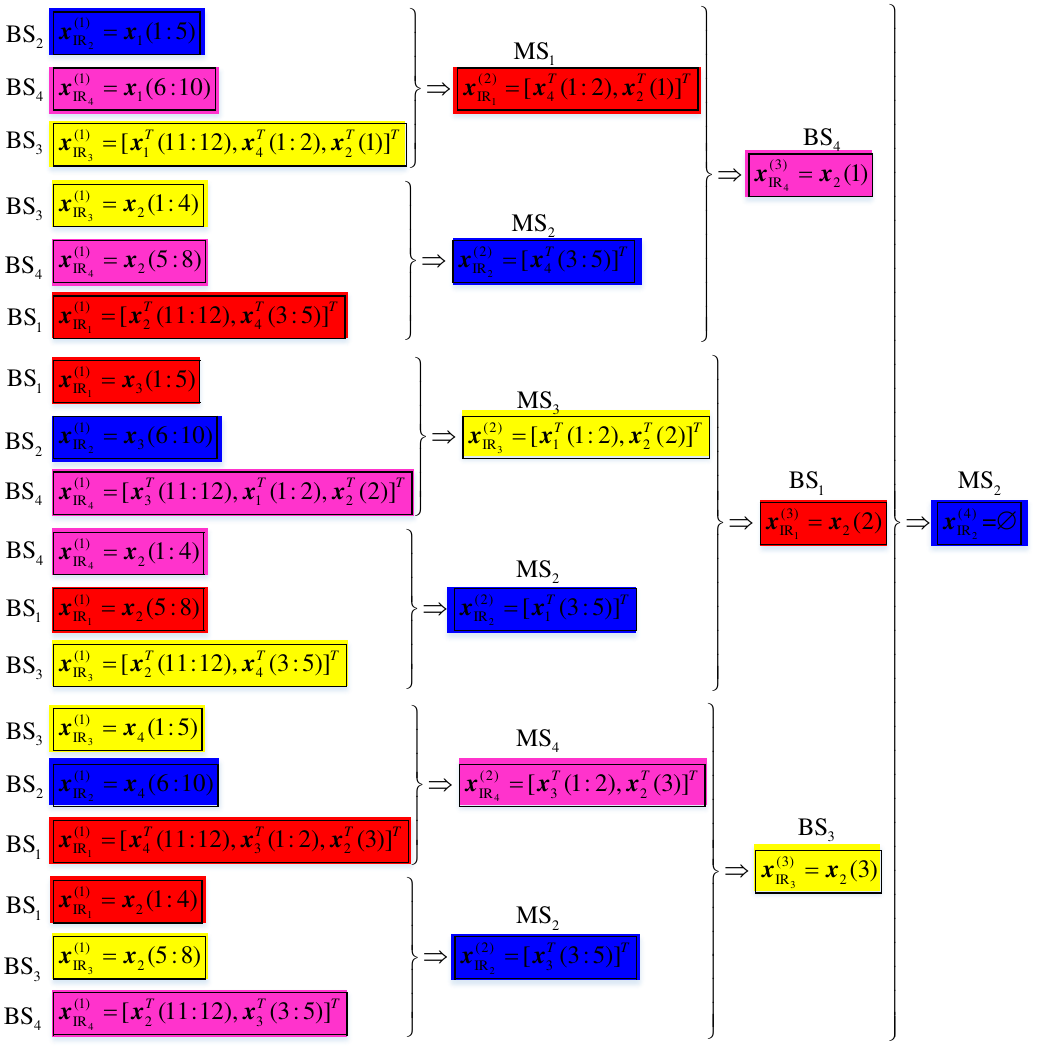}
\caption{Introduced genies for \textbf{Ex 5}.}
\label{fig:Genie_Tree_Ex5}
\end{figure}

According to \eqref{Eq:Dim_Ex4} and the principle in \eqref{Eq:Genie_rule11}, we design the genies as shown in Fig. \ref{fig:Genie_Tree_Ex5} and obtain the inequalities as follows,
\begin{subequations}\label{Eq:DoF_Ex5}
\begin{align}
\label{Eq:DoF_Ex5_R1}
3d&\le M-d + \Delta_{\mathrm{IR}}^{(1)} \\
\label{Eq:DoF_Ex5_R2}
3\Delta_{\mathrm{IR}}^{(1)} &\le d + \Delta_{\mathrm{IR}}^{(2)}\\
\label{Eq:DoF_Ex5_R3}
2\Delta_{\mathrm{IR}}^{(2)}
&\le \Delta_{\mathrm{IR}}^{(1)}+\Delta_{\mathrm{IR}}^{(3)}\\
\label{Eq:DoF_Ex5_R4}
3\Delta_{\mathrm{IR}}^{(3)}
&\le \Delta_{\mathrm{IR}}^{(2)}
\end{align}
\end{subequations}
where $\Delta_{\mathrm{IR}}^{(m)}=\omega\left( \pmb{x}_{\mathrm{IR}_{l}}^{(m)\ell}\right)$, $m=1,2,3$.

By solving \eqref{Eq:DoF_Ex5}, we have
\begin{align}
  d\leq 12M/43 = 372/43
\end{align}
which leads to the DoF upper-bound $d\leq d^{\mathrm{Decom}}$ for \textbf{Ex 5}.

\subsubsection{\textbf{Ex 6}}
We consider \textbf{Ex 6} where $M=5$ and $N=2$.
Since $M/N \in \mathcal{R}^{\mathrm{I}}=((3-\sqrt{5})/2,(3+\sqrt{5})/2)$, the DoF upper-bound is $d^{\mathrm{Decom}}=MN/(M+N)=10/7$ and it can be derived from a partial genie tree.

Following the similar analysis in \textbf{Ex 5}, we have
\begin{subequations}\label{Eq:Dim_Ex6}
\begin{align}
L_{\mathrm{IR}}^{(1)} & = (3L_{\mathrm{IR}}^{(0)}-L_{\mathrm{IR}}^{(-1)})^{+} =6-5 = 1\\
L_{\mathrm{IR}}^{(2)} & = (2L_{\mathrm{IR}}^{(1)}-L_{\mathrm{IR}}^{(0)})^{+} = 2-2= 0
\end{align}
\end{subequations}

\begin{figure}[htb!]
\centering
\includegraphics[width=0.60\linewidth]{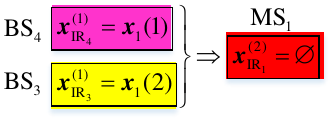}
\caption{Introduced genies for \textbf{Ex 6}.}
\label{fig:Genie_Tree_Ex6}
\end{figure}

According to \eqref{Eq:Dim_Ex6} and \eqref{Eq:Genie_rule11}, we design the genies as shown in Fig. \ref{fig:Genie_Tree_Ex6} and obtain the inequalities as follows,
\begin{subequations}\label{Eq:DoF_Ex6}
\begin{align}
\label{Eq:DoF_Ex6_R1}
3d&\le M-d + \Delta_{\mathrm{IR}}^{(1)} \\
\label{Eq:DoF_Ex6_R2}
2\Delta_{\mathrm{IR}}^{(1)} &\le d
\end{align}
\end{subequations}
where $\Delta_{\mathrm{IR}}^{(1)}=\omega\left( \pmb{x}_{\mathrm{IR}_{l}}^{(1)\ell}\right)$.

By solving \eqref{Eq:DoF_Ex6}, we obtain
\begin{align}\label{Eq:DoF_bound_Ex6}
 d\leq 2M/7 = 10/7
\end{align}
which leads to the DoF upper-bound $d\leq d^{\mathrm{Decom}}$ for \textbf{Ex 6}.

\subsubsection{\textbf{Ex 7}}
We consider \textbf{Ex 7} where $M=7$ and $N=3$.
Since $M/N \in \mathcal{R}^{\mathrm{I}}=((3-\sqrt{5})/2,(3+\sqrt{5})/2)$, the DoF upper-bound is $d^{\mathrm{Decom}}=MN/(M+N)=21/10$ and it can be derived from a partial genie tree.

Following the similar analysis in \textbf{Ex 5}, we have
\begin{subequations}\label{Eq:Dim_Ex7}
\begin{align}
L_{\mathrm{IR}}^{(1)} & = (3L_{\mathrm{IR}}^{(0)}-L_{\mathrm{IR}}^{(-1)})^{+} =9-7 = 2\\
L_{\mathrm{IR}}^{(2)} & = (2L_{\mathrm{IR}}^{(1)}-L_{\mathrm{IR}}^{(0)})^{+} = 4-3= 1\\
L_{\mathrm{IR}}^{(3)} & = (2L_{\mathrm{IR}}^{(2)}-L_{\mathrm{IR}}^{(1)})^{+} = 2-2 = 0
\end{align}
\end{subequations}

\begin{figure}[htb!]
\centering
\includegraphics[width=0.90\linewidth]{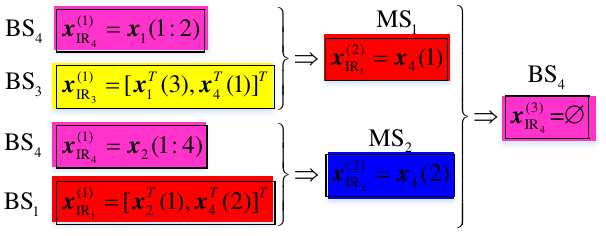}
\caption{Introduced genies for \textbf{Ex 7}.}
\label{fig:Genie_Tree_Ex7}
\end{figure}

According to \eqref{Eq:Dim_Ex7} and \eqref{Eq:Genie_rule11}, we design the genies as shown in Fig. \ref{fig:Genie_Tree_Ex7} and obtain the inequalities as follows,
\begin{subequations}\label{Eq:DoF_Ex7}
\begin{align}
\label{Eq:DoF_Ex7_R1}
3d&\le M-d + \Delta_{\mathrm{IR}}^{(1)} \\
\label{Eq:DoF_Ex7_R2}
2\Delta_{\mathrm{IR}}^{(1)} &\le d + \Delta_{\mathrm{IR}}^{(2)}\\
\label{Eq:DoF_Ex7_R3}
2\Delta_{\mathrm{IR}}^{(2)}
&\le \Delta_{\mathrm{IR}}^{(1)}
\end{align}
\end{subequations}
where $\Delta_{\mathrm{IR}}^{(m)}=\omega\left( \pmb{x}_{\mathrm{IR}_{l}}^{(m)\ell}\right)$, $m=1,2$.

By solving \eqref{Eq:DoF_Ex7}, we obtain
\begin{align}\label{Eq:DoF_bound_Ex7}
 d\leq 3M/11 = 21/11
\end{align}
which leads to the DoF upper-bound $d\leq d^{\mathrm{Decom}}$ for \textbf{Ex 7}.

\subsubsection{\textbf{Ex 8}}
We consider \textbf{Ex 8} where $M=8$ and $N=3$.
Since $M/N \in \mathcal{R}_{2}^{\mathrm{II-A}} = [8/3,3)$ from \eqref{Eq:Quantity DoF}, the DoF upper-bound is $d^{\mathrm{Quan}}=\min\{3M/11,3N/4\}=24/11$ and it can be derived from a full genie tree.

From \eqref{Eq:dim_irr_space}, we have
\begin{subequations}\label{Eq:Dim_Ex8}
\begin{align}
L_{\mathrm{IR}}^{(1)} & = (3L_{\mathrm{IR}}^{(0)}-L_{\mathrm{IR}}^{(-1)})^{+} =9-8 = 1\\
L_{\mathrm{IR}}^{(2)} & = (3L_{\mathrm{IR}}^{(1)}-L_{\mathrm{IR}}^{(0)})^{+} = 3-3 = 0
\end{align}
\end{subequations}

\begin{figure}[htb!]
\centering
\includegraphics[width=0.60\linewidth]{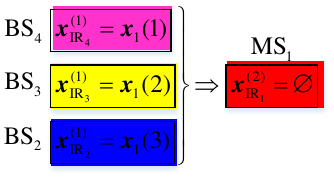}
\caption{Introduced genies for \textbf{Ex 8}.}
\label{fig:Genie_Tree_Ex8}
\end{figure}

According to \eqref{Eq:Dim_Ex8} and \eqref{Eq:Genie_rule11}, we can design the genies as shown in Fig. \ref{fig:Genie_Tree_Ex8} and obtain the inequalities as follows,
\begin{subequations}\label{Eq:DoF_Ex8}
\begin{align}
\label{Eq:DoF_Ex8_R1}
3d&\le M-d + \Delta_{\mathrm{IR}}^{(1)}\\
\label{Eq:DoF_Ex8_R2}
3\Delta_{\mathrm{IR}}^{(1)} &\le d
\end{align}
\end{subequations}
where $\Delta_{\mathrm{IR}}^{(1)}=\omega\left( \pmb{x}_{\mathrm{IR}_{l}}^{(1)\ell}\right)$.

By solving \eqref{Eq:DoF_Ex8}, we obtain
\begin{align}\label{Eq:DoF_bound_Ex8}
 d\leq 3M/11 = 24/11
\end{align}
which leads to the DoF upper-bound $d\leq d^{\mathrm{Quan}}=d^{\mathrm{Decom}}$ for \textbf{Ex 8}.

From the above examples, we show how to use the genie trees to derive the DoF upper-bounds in both Region I and Region II. We can see the differences between our genie tree and the genie chain in \cite{Jafar_IAchainKcell_all}. Firstly, \textbf{Ex 4} and \textbf{Ex 5} are the unsolved cases in \cite{Jafar_IAchainKcell_all}, which means that genie trees can provide the DoF upper-bound that cannot be derived from the genie chain. Moreover, the genie chain considers the four different ways to construct the genies to derive the DoF upper-bound for \textbf{Ex 6}, \textbf{Ex 7}, and \textbf{Ex 8}, where different ways need to be selected carefully based on different antenna configurations. By contrast, our genie tree can use a unified way to solve all cases that require different ways by using genie chain. In other words, our genie tree can be applied to more general cases.

\section{Proof of Achievable DoF}\label{Sec:Proof2}
In Section III.A, we have shown that the decomposition DoF bound is achievable by the asymptotic IA in Region I, which indicates that \eqref{Eq:DoF_lower_Bound_R1} in Theorem \ref{Theorem:Achievable_DoF} is true. In the following, we prove that the quantity DoF bound is achievable by the linear IA in Region II, i.e., \eqref{Eq:DoF_lower_Bound_R2}.

\subsection{Feasible Condition of Linear IA}
The basic idea of the proof is to find a linear IA transceiver that can support $d=d^{\mathrm{Quan}}$ data streams for each user. To prove the feasibility of linear IA, we need to check the IA conditions in~\cite{TTTSP2013}, i.e.,
\begin{subequations}\label{Eq:IA_Conditions}
\begin{align}
\label{Eq:Constraint_MUI_free}
\mathrm{rank} \left(
\pmb{V}_i^{H}\left[\begin{array}{c}
\pmb{H}_{i,i_{1}} \\
 \vdots \\
 \pmb{H}_{i,i_{K}}
 \end{array}\right]
\left[\begin{array}{ccc}
 \pmb{U}_{i_1} & {} & {\pmb{0}} \\
 {} & \ddots & {} \\
 {\pmb{0}} & {} & \pmb{U}_{i_{K}}
 \end{array}
\right]
\right)&=Kd\\
\label{Eq:Constraint_ICI_free}
\pmb{V}_j^{H}\pmb{H}_{j,i_k}\pmb{U}_{i_k}&=\pmb{0},~\forall i\neq j
\end{align}
\end{subequations}
where $\pmb{V}_{j}$ and $\pmb{U}_{i_k}$ denote the receive matrix of BS$_j$ and the transmit matrix of MS$_{i_k}$. \eqref{Eq:Constraint_MUI_free} is a rank constraint to convey the desired signals for each BS and each user, which is equivalent to $\mathrm{rank}(\pmb{V}_j)=Kd$ and $\mathrm{rank}(\pmb{U}_{i_k})=d$, and \eqref{Eq:Constraint_ICI_free} is the ICI-free constraint.

Since Region II is comprised of Regions II-A and II-B, and the proof in Region II-B is similar, we only provide the proof in Region II-A for conciseness.

When $M/N \in \mathcal{R}^{\mathrm{II-A}}_{n}$, i.e., $C_{n}^{\mathrm{A}} \leq M/N < C_{n-1}^{\mathrm{A}}$, Corollary \ref{Corollary_NS_Condition} indicates that it is necessary to satisfy $M\geq M_{\min}$ and $N\geq N_{\min}$ to support $d$ data steams per user, where $M_{\min}=(K+C_{n}^{\mathrm{A}})d$ and $N_{\min}=(1+K/C_{n-1}^{\mathrm{A}})d$ denote the minimal antenna configuration in Subregion $\mathcal{R}^{\mathrm{II-A}}_{n}$. When the linear IA is feasible with the minimal antenna configuration, i.e., $\forall M/N \in \{D_{n}^{\mathrm{A}}\}$, the linear IA must be feasible with other configurations in Subregion $\mathcal{R}^{\mathrm{II-A}}_{n}$ by removing the redundant antennas and considering the finite spatial extensions. In the following, we only need to design the feasible IA transceiver with the minimal antenna configuration.

\subsection{Basic Idea of Transceiver Design}

To obtain the feasible IA transceiver, we first design the transmit matrices of users to align some ICIs and then design the receive matrices of BSs to cancel the remaining ICIs. Once the transmit matrices are obtained, the receive matrices can be designed as follows,
\begin{align}\label{Eq:V1}
\pmb{V}_{j}
&=\mathcal{V}\left\{\pmb{Q}_{j}^{H}\pmb{Q}_{j}\right\},~j=1,\cdots,G
\end{align}
where $\mathcal{V}\left\{\pmb{X}\right\}$ is a matrix whose columns are the eigenvectors corresponding to zero eigenvalues of $\pmb{X}$
and satisfy $\pmb{X}\mathcal{V}\left\{\pmb{X}\right\}=\pmb{0}$, $\pmb{Q}_{j} = [\pmb{Q}_{j,1},\cdots,\pmb{Q}_{j,j-1},\pmb{Q}_{j,j+1},\cdots,\pmb{Q}_{j,G}]\in \mathbb{C}^{(G-1)Kd\times M}$ contains all the received ICIs of the $j$th BS, and $\pmb{Q}_{j,i} = [\pmb{U}_{i_1}^{H}\pmb{H}_{j,i_1}^{H},\cdots,\pmb{U}_{i_K}^{H}\pmb{H}_{j,i_K}^{H}]\in \mathbb{C}^{Kd\times M}$ denotes the received ICIs of the $j$th BS from all the users in the $i$th cell.

Therefore, the key factor of devising the feasible IA transceiver is to establish the alignment equations to design the transmit matrices. The subspace alignment chains proposed in \cite{Jafar_3cell} can only describe which two ICIs can be aligned together. To describe which ICIs can be aligned for general cases, we introduce a notion of \emph{alignment graph}.
\begin{definition}
\emph{Alignment graph} is a bipartite graph $\mathcal{G}=\{\mathcal{X},\mathcal{Y};\mathcal{E}\}$ to describe how to align the ICIs from the data streams of the nodes in $\mathcal{X}$ at the nodes in $\mathcal{Y}$. If $(x_i,y_j)\in \mathcal{E},~\forall x_i \in \mathcal{Z}_j\subseteq \mathcal{X},~y_j \in \mathcal{Y}$, the ICIs from the data streams of nodes in the set $\mathcal{Z}_j$ can be aligned at node $y_j$.
\end{definition}

To achieve the information theoretical maximal DoF, we need to find out which ICIs can be aligned together. Since the aligned ICIs cannot be resolved by the receiver, we know that the irresolvable ICIs are the aligned ICIs. Considering that the genie branch shows which nodes have the irresolvable ICIs, it also indicates where the ICIs can be aligned together. Therefore, we can establish the alignment equations from the genie branch to obtain transmit matrices.

\subsection{Examples}\label{Sec:Proof2_Ex}
In the following, we employ \textbf{Ex 1}, \textbf{Ex 2} and \textbf{Ex 3} listed in Table \ref{Table:CN} to illustrate the basic idea to establish alignment equations from genie branches.

\subsubsection{\textbf{Ex 1}}
We first consider \textbf{Ex 1} where $M=6$, $N=1$, and $d=1$. As shown in Fig. \ref{fig:ICI_cancel_procedure}(a), there are no irresolvable ICIs, i.e., BSs are capable of canceling all ICIs. Therefore, the transmit matrices of users do not need to align ICIs and can be designed to satisfy $\mathrm{rank}(\pmb{U}_{i_k})\geq 1$.

\subsubsection{\textbf{Ex 2}}
We then consider \textbf{Ex 2} where $M=11$, $N=3$, and $d=2$. Since BSs cannot cancel all ICIs, it is necessary to design the transmit matrices to align some ICIs.

From the genie branch in Fig. \ref{fig:Genie_Tree}(b), we can see that BS$_2$ and BS$_3$ have irresolvable ICIs. It indicates that the ICIs generated from the signals of MS$_{1_1}$, MS$_{1_2}$, MS$_{3_1}$, and MS$_{3_2}$ can be aligned at BS$_2$, and the ICIs from the signals of MS$_{1_1}$, MS$_{1_2}$, MS$_{2_1}$, and MS$_{2_2}$ can be aligned at BS$_3$.

Let $\pmb{z}_{\mathrm{IR}_{2}}^{(1)}$ and $\pmb{z}_{\mathrm{IR}_{3}}^{(1)}$ represent the aligned ICIs (i.e., irresolvable ICIs) at BS$_2$ and BS$_3$ in the 1st round. As shown in Fig. \ref{fig:Genie_Tree}(b), $\pmb{z}_{\mathrm{IR}_{2}}^{(1)}$ and $\pmb{z}_{\mathrm{IR}_{3}}^{(1)}$ contain the ICIs from MS$_{1_1}$. From \eqref{Eq:Received_ICI_MS11_r2cc}, we can see that when $\pmb{z}_{\mathrm{IR}_{2}}^{(1)}$ and $\pmb{z}_{\mathrm{IR}_{3}}^{(1)}$ contain the data stream from MS$_{1_1}$, they can be aligned at MS$_{1_1}$ in the 2nd round, otherwise, they cannot be aligned together. In Fig. \ref{fig:Genie_Tree}(b), $\pmb{z}_{\mathrm{IR}_{2}}^{(1)}$ and $\pmb{z}_{\mathrm{IR}_{3}}^{(1)}$ are not aligned, which indicates that the data streams of MS$_{1_1}$ in $\pmb{z}_{\mathrm{IR}_{2}}^{(1)}$ and $\pmb{z}_{\mathrm{IR}_{3}}^{(1)}$ are different. Let $s_{1_1,1}$ and $s_{1_1,2}$ represent two different data streams of MS$_{1_1}$, without loss of generality, we consider that $\pmb{z}_{\mathrm{IR}_{2}}^{(1)}$ contains the ICI from $s_{1_1,1}$, while $\pmb{z}_{\mathrm{IR}_{3}}^{(1)}$ contains the ICI from $s_{1_1,2}$, where $s_{i_k,j}$ denotes the $j$th data stream of MS$_{i_k}$,
\begin{figure}[htb!]
\centering
\includegraphics[width=0.7\linewidth]{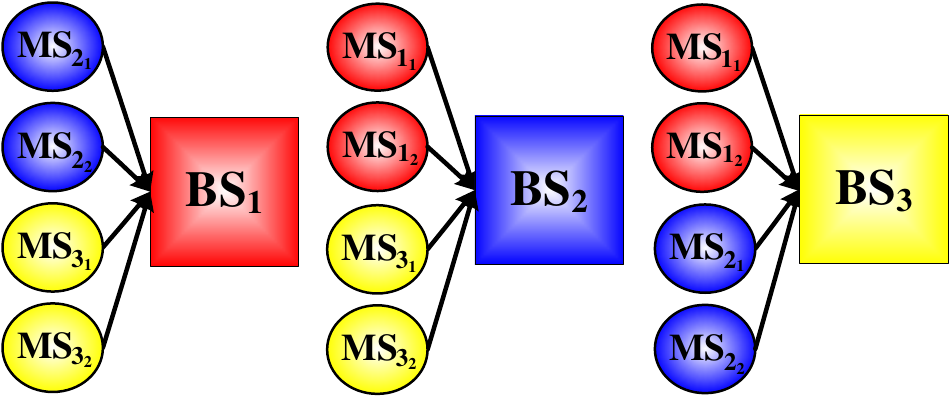}
\caption{Alignment graphs for \textbf{Ex 2}.} \label{fig:Aligned_matrix1}
\end{figure}

The genie branch shows where the ICIs can be aligned, we construct the alignment graph from the genie branch. From the genie branch in Fig. \ref{fig:Genie_Tree}(b), we obtain the alignment graph in Fig. \ref{fig:Aligned_matrix1}. In the alignment graph, all vertexes with the same label represent different data streams from the same node. According to the alignment graph in Fig. \ref{fig:Aligned_matrix1}, we can align the ICIs from $s_{2_1,1}$, $s_{2_2,1}$, $s_{3_1,1}$, $s_{3_2,1}$ at BS$_1$, the ICIs from $s_{1_1,1}$, $s_{1_2,1}$, $s_{3_1,2}$, $s_{3_2,2}$ at BS$_2$, and the ICIs from $s_{1_1,2}$, $s_{1_2,2}$, $s_{2_1,2}$, $s_{2_2,2}$ at BS$_3$. Then, the alignment equations are obtained as follows,
\begin{align*}
 &\pmb{H}_{1,2_1}\pmb{u}_{2_1,1}+\pmb{H}_{1,2_2}\pmb{u}_{2_2,1}+ \pmb{H}_{1,3_1}\pmb{u}_{3_1,1}+\pmb{H}_{1,3_2}\pmb{u}_{3_2,1}=\pmb{0} \\
 &\pmb{H}_{2,1_1}\pmb{u}_{1_1,1}+\pmb{H}_{2,1_2}\pmb{u}_{1_2,1}+ \pmb{H}_{2,3_1}\pmb{u}_{3_1,2}+\pmb{H}_{2,3_2}\pmb{u}_{3_2,2}=\pmb{0} \\
 &\pmb{H}_{3,1_1}\pmb{u}_{1_1,2}+\pmb{H}_{3,1_2}\pmb{u}_{1_2,2}+ \pmb{H}_{3,2_1}\pmb{u}_{2_1,2}+\pmb{H}_{3,2_2}\pmb{u}_{2_2,2}=\pmb{0}
\end{align*}
where $\pmb{u}_{i_k,\ell}$ denotes the transmit vector of $s_{i_k,\ell}$.

The alignment equations can be expressed as
\begin{align*}
\pmb{A}_{\ell}\pmb{W}^{U}_{\ell}=\pmb{0},~\ell=1,2,3
\end{align*}
where $\pmb{A}_{1}=[\pmb{H}_{1,2_1},\pmb{H}_{1,2_2}, \pmb{H}_{1,3_1},\pmb{H}_{1,3_2}]$, $\pmb{A}_{2}=[\pmb{H}_{2,1_1},$ $\pmb{H}_{2,1_2}, \pmb{H}_{2,3_1},\pmb{H}_{2,3_2}]$, $\pmb{A}_{3}=[\pmb{H}_{3,1_1},\pmb{H}_{3,1_2}, \pmb{H}_{3,2_1},\pmb{H}_{3,2_2}]$, $\pmb{W}^{U}_{1} = [\pmb{u}_{2_1,1}^{T},\pmb{u}_{2_2,1}^{T}, \pmb{u}_{3_1,1}^{T},\pmb{u}_{3_2,1}^{T}]^{T}$, $\pmb{W}^{U}_{2} = [\pmb{u}_{1_1,1}^{T},\pmb{u}_{1_2,2}^{T},$ $\pmb{u}_{3_1,2}^{T},\pmb{u}_{3_2,2}^{T}]^{T}$, and $\pmb{W}^{U}_{3} = [\pmb{u}_{1_1,2}^{T},\pmb{u}_{1_2,2}^{T}, \pmb{u}_{2_1,2}^{T},\pmb{u}_{2_2,2}^{T}
]^{T}$. $\pmb{A}_{\ell}\in \mathbb{C}^{M\times 4N=11\times 12}$ is called \emph{alignment matrix},
and $\pmb{W}^{U}_{\ell}\in \mathbb{C}^{4N\times 1=12\times 1}$ is comprised of the transmit vectors.

For each BS, the number of received ICIs is $(G-1)Kd=8$. There are three alignment equations for three BSs and each equation can align one ICI. Therefore, the users can help each BS cancel one ICI. Meanwhile, each BS can cancel the remaining $M-Kd = 11-4=7$ ICIs. Therefore, all the ICIs are jointly eliminated by the users and BSs, and the designed IA transceiver is feasible.

\subsubsection{\textbf{Ex 3}}
We next consider \textbf{Ex 3} where $M=38$, $N=11$, and $d=7$. In this case, BSs cannot cancel all ICIs, it is still necessary to design the transmit matrices to align some ICIs.

\begin{figure}[htb!]
\centering
\includegraphics[width=0.6\linewidth]{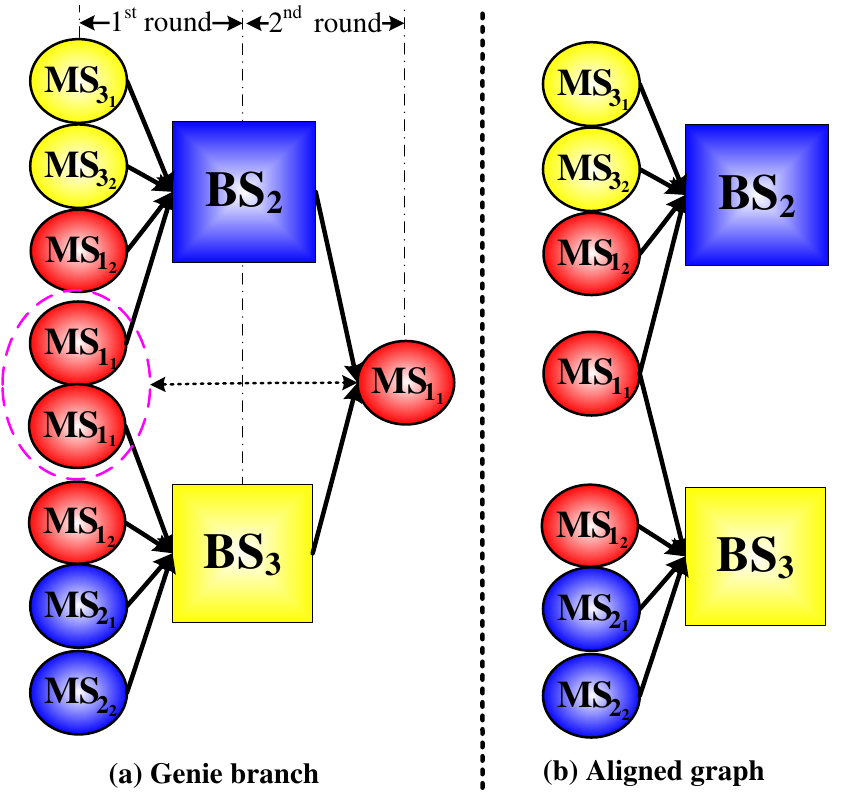}
\caption{Genie branch and alignment graph for \textbf{Ex 3}.} \label{fig:Aligned_matrix2}
\end{figure}
Take the genie branch in Fig. \ref{fig:Aligned_matrix2}(a) as an example, similar to \textbf{Ex 2}, the ICIs generated from the signals of MS$_{1_1}$, MS$_{1_2}$, MS$_{3_1}$, and MS$_{3_2}$ can be aligned at BS$_2$, the ICIs from the signals of MS$_{1_1}$, MS$_{1_2}$, MS$_{2_1}$, and MS$_{2_2}$ can be aligned at BS$_2$. Different from \textbf{Ex 2}, $\pmb{z}_{\mathrm{IR}_{2}}^{(1)}$ and $\pmb{z}_{\mathrm{IR}_{3}}^{(1)}$ can be further aligned at
MS$_{1_1}$ in the 2nd round. It means that $\pmb{z}_{\mathrm{IR}_{1}}^{(1)}$ and $\pmb{z}_{\mathrm{IR}_{2}}^{(1)}$ are comprised of the ICIs from the same data stream of MS$_{1_1}$. Therefore, in Fig. \ref{fig:Aligned_matrix2}(a), the two leaf vertexes labeled by MS$_{1_1}$ denote the same data stream of MS$_{1_1}$.
Without loss of generality, we assume that the data stream is $s_{1_1,1}$, the 1st data stream of MS$_{1_1}$. To avoid establishing redundant alignment equations, it is necessary to merge one redundant vertex of the genie branch  in Fig. \ref{fig:Aligned_matrix2}(a) to obtain the alignment graph in Fig. \ref{fig:Aligned_matrix2}(b).

According to the alignment graph, we can align the ICIs from $s_{1_1,1}$, $s_{1_2,1}$, $s_{3_1,1}$, and $s_{1_1,1}$ at BS$_2$, the ICIs from
$s_{1_1,1}$, $s_{1_2,2}$, $s_{2_1,1}$, and $s_{2_2,1}$ at BS$_3$. The ICI from $s_{1_1,1}$ is aligned at both BS$_2$ and BS$_3$ simultaneously. Then, the alignment equations are obtained as follows,
\begin{align*}
\left\{
\begin{array}{l}
\pmb{H}_{2,1_1}\pmb{u}_{1_1,1}+\pmb{H}_{2,1_2}\pmb{u}_{1_2,1}+ \pmb{H}_{2,3_1}\pmb{u}_{3_1,1}+\pmb{H}_{2,3_2}\pmb{u}_{3_2,1}=\pmb{0} \\
\pmb{H}_{3,1_1}\pmb{u}_{1_1,1}+\pmb{H}_{3,1_2}\pmb{u}_{1_2,2}+ \pmb{H}_{3,2_1}\pmb{u}_{2_1,1}+\pmb{H}_{3,2_2}\pmb{u}_{2_2,1}=\pmb{0}
\end{array}
\right.
\end{align*}

The alignment equations are from one genie branch whose root is MS$_{1_1}$. When considering other five different genie branches, we can obtain twelve alignment equations.

For each BS, the number of received ICIs is $(G-1)Kd=28$ ICIs. There are twelve alignment equations for three BSs. Therefore, the users can help each BS cancel four ICIs. Meanwhile, each BS can cancel $M-Kd = 38-14=24$ ICIs. As a result, all the ICIs can be jointly eliminated by the users and BSs, and the designed IA transceiver is feasible.

\subsection{IA Transceiver Design}
From these examples, we can summarize the basic steps to design the linear IA transceiver for general cases as:
\begin{enumerate}
 \item Show the genie tree according to the configuration;
 \item Obtain the genie branches from the genie tree;
 \item Construct the alignment graphs from the genie branches;
 \item Establish the alignment equations according to the alignment graphs;
 \item Design the transmit matrices by solving the alignment equations and design the receive matrices based on the transmit matrices.
\end{enumerate}

\subsubsection{Genie Branches}
When $M/N \in \mathcal{R}^{\mathrm{II-A}}_{n}$, from Lemma \ref{Lemma_height_space}, we know that the genie tree is a full tree with height of $n$. According to the genie tree, we can obtain $G\bar{K}_{n-1}$ genie branches that are also full trees with height of $n-1$.

We use $\mathcal{T}(\ell)=\{\mathcal{O};\mathcal{D}\}$ to denote the genie branch whose root is the $\ell$th vertex in the $(n-1)$ round, $\ell=1,\cdots,G\bar{K}_{n-1}$, where $\mathcal{O}=\cup_{m=0}^{n-1}\mathcal{O}^{(m)}$, $\mathcal{D}=\cup_{m=1}^{n-1}\mathcal{D}^{(m)}$, $\mathcal{O}^{(m)}$ is the set of vertexes in the $m$th round, and $\mathcal{D}^{(m)}$ is the set of edges from the vertexes in the $(m-1)$th round to those in the $m$th round.

Let $\mathcal{I}_{j}^{(m)}$ denote the set of children of node $j$ in the $m$th round, $\forall 1 \leq m \leq n-1$, we have
\begin{subequations}
\begin{align}
 &\mathcal{O}^{(m-1)}=\cup_{j\in \mathcal{O}^{(m)} } \mathcal{I}_{j}^{(m)} \\
 &\mathcal{D}^{(m)}=\left\{(i,j)|\forall j \in \mathcal{I}_{i}^{(m)},~i \in \mathcal{O}^{(m)},~j \in \mathcal{O}^{(m-1)}\right\}
\end{align}
\end{subequations}

\subsubsection{Alignment Graphs}
According to the genie branch $\mathcal{T}(\ell)$, we can construct an alignment graph, denoted by $\mathcal{G}(\ell)$. Based on the previous examples, we know that when $n=1$, it is not necessary to align ICIs. When $n=2$, the genie branch can be taken as the alignment graph directly. When $n>2$, in the genie branch, the vertex in the $m$th round indicates which vertexes in the $(m-2)$th round contain a redundant vertex. We need to first merge the redundant vertexes to obtain the alignment graph.

\begin{figure}[htb!]
\centering
\includegraphics[width=0.9\linewidth]{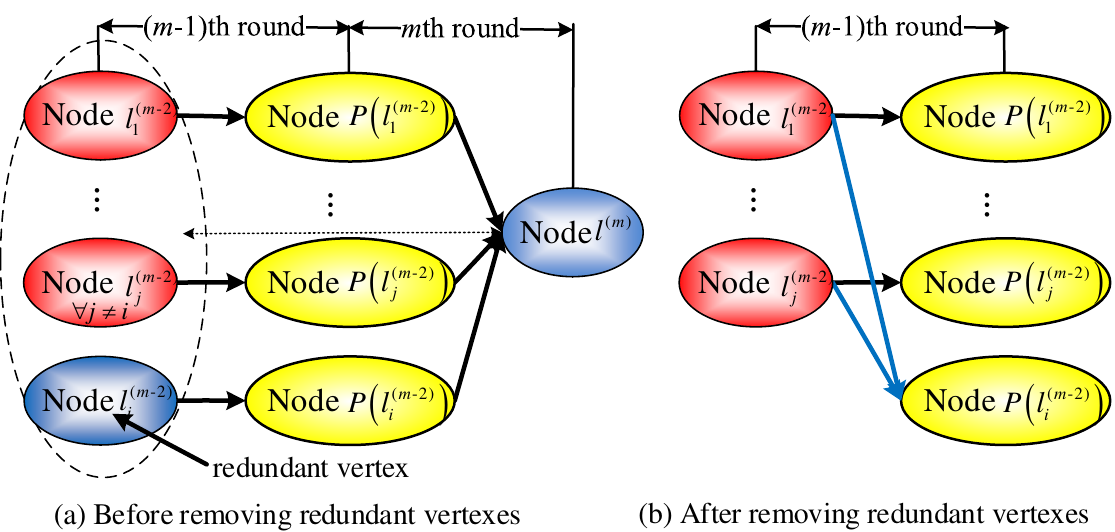}
\caption{Genie branches before and after merging redundant vertexes, where the blue vertexes in (a) are removed but the blue edges in (b) are added.} \label{fig:Redundant}
\end{figure}

In Fig. \ref{fig:Redundant}, we show how to merge the redundant vertexes in the genie branch. For a vertex labeled by node $l$ (either BS or user) in the $m$th round (denoted by $l^{(m)}$), when $m\geq 2$ we know that there exists a redundant vertex among multiple vertexes labeled by node $l$ in the $(m-2)$th round (denoted by $l_j^{(m-2)}$\footnote{As shown in Fig.\ref{fig:Aligned_matrix2}, there are two vertexes who have the same label (i.e., MS$_{1_1}$) in the first round of the genie branch. To distinguish different vertexes with the same label, here we use the subscript $l_j^{(m-2)}$ to denote the $j$th vertex who is labeled node $l$ in the $(m-2)$th round.}), $\forall P(l_j^{(m-2)})\in \mathcal{J}_{l}^{(m)}$, where $P(x)$ is the parent of vertex $x$.
If the redundant vertex is $l_i^{(m-2)}$, it means that since $l_j^{(m-2)},~\forall j\neq i$ can align their ICIs at $P(l_i^{(m-2)})$ in the $(m-1)$th round, the ICIs of $P(l_i^{(m-2)})$ are the linear combinations of the ICIs of $l_j^{(m-2)},~\forall j\neq i$, so that $P(l_j^{(m-2)})\in \mathcal{I}_{l,m}^{\mathrm{A}}$ can align their ICIs at $l^{(m)}$ in the $m$th round. Therefore, we need to remove the redundant vertex $l_i^{(m-2)}$ and add the edges from $l_j^{(m-2)},~\forall j\neq i$ to $P(l_i^{(m-2)})$.

To obtain the alignment graph from the genie branch, we need to delete the vertex in the $m$th round and merge the redundant vertex in the $(m-2)$th round in the following way.
\begin{subequations}
\begin{align}
 \mathcal{O}^{(m)}=&\mathcal{O}^{(m)}\setminus\{l^{(m)}\},~\forall P(l_i^{(m-2)})\in \mathcal{J}_{l}^{(m)}\\
 \mathcal{O}^{(m-1)}=&\mathcal{O}^{(m-1)}\setminus\{l_i^{(m-2)}\}\\
 \mathcal{D}^{(m)}=&\mathcal{D}^{(m)}\cup\left\{\left(l_j^{(m-2)},P(l_i^{(m-2)})\right)\right\},\nonumber\\
 &\forall j\neq i, P(l_j^{(m-2)})\in \mathcal{J}_{l}^{(m)}
\end{align}
\end{subequations}
After deleting all vertexes in the $m$th round, the height of the genie branch is reduced to $m-1$.

By repeating the above merging steps for $m=n,\cdots,2$, we can obtain the final alignment graph $\mathcal{G}(\ell)=\{\mathcal{X},\mathcal{Y};\mathcal{E}\}$ by letting $\mathcal{X}=\mathcal{O}^{(0)}$, $\mathcal{Y}=\mathcal{O}^{(1)}$, and $\mathcal{E}=\mathcal{D}^{(1)}$, where $\mathcal{O}^{(0)}$, $\mathcal{O}^{(1)}$, and $\mathcal{D}^{(1)}$ are obtained from the genie branch with height of two. Using mathematical induction, it is not difficult to prove that $\mathcal{X}$ contains $q_{n-1}^{\mathrm{A}}$ vertexes and
$\mathcal{Y}$ contains $p_{n-1}^{\mathrm{A}}$ vertexes. As a result, the alignment graph indicates how the ICIs from $q_{n-1}^{\mathrm{A}}$ data streams of users are aligned at $p_{n-1}^{\mathrm{A}}$ BSs.

\subsubsection{Alignment Equations}
From alignment graphs $\mathcal{G}(\ell),~\ell=1,\cdots,G\bar{K}_{n-1}$, we can obtain
$G\bar{K}_{n-1}$ groups of alignment equations
\begin{align}\label{Eq:Alignment_Equation}
 \pmb{A}_{\ell}\pmb{W}^{U}_{\ell}=\pmb{0},~\ell=1,\cdots,G\bar{K}_{n-1}
\end{align}
where $\pmb{A}_{\ell}\in \mathbb{C}^{p_{n-1}^{\mathrm{A}}M \times q_{n-1}^{\mathrm{A}}N}$ is the alignment matrix obtained from the alignment graph $\mathcal{G}(\ell)$ and $\pmb{W}^{U}_{\ell}$ is the rearranged matrix comprised of transmit vectors.

From the analysis in Subsection \ref{Sec:Proof2_Ex}, we know that the different vertexes labeled by the same node in the alignment graph denote the different data streams from the node. Therefore, we can obtain different transmit vectors from different alignment equations.
Since there are $GKd$ transmit vectors
and $G\bar{K}_{n-1}$ rearranged matrices, each rearranged matrix $\pmb{W}^{U}_{\ell}$ needs to contain $Kd/\bar{K}_{n-1}$ transmit vectors. Since $\pmb{A}^{V}_{(\ell)}$ has $q_{n-1}^{\mathrm{A}}N$ columns, $\pmb{W}^{U}_{\ell}$ has
$q_{n-1}^{\mathrm{A}}N$ rows and each column contains $q_{n-1}^{\mathrm{A}}$ transmit vectors. When $\pmb{W}^{U}_{\ell}$ has $Kd/(\bar{K}_{n-1}q_{n-1}^{\mathrm{A}})$ columns,
$\pmb{W}^{U}_{\ell},~\ell=1,\cdots,G\bar{K}_{n-1}$ can contain all transmit vectors.
As a result, $\pmb{W}^{U}_{\ell}\in \mathbb{C}^{q_{n-1}^{\mathrm{A}}N \times Kd/(\bar{K}_{n-1}q_{n-1}^{\mathrm{A}})}$.

\subsubsection{Transmit and Receive Matrices}
From \eqref{Eq:Alignment_Equation}, we can design the transmit matrices as follows,
\begin{align}\label{Eq:U1}
\pmb{W}^{U}_{\ell}&=\mathcal{V}\left\{\pmb{A}_{\ell}^{H}\pmb{A}_{\ell}\right\},~\ell=1,\cdots,G\bar{K}_{n-1}
\end{align}
By substituting the transmit matrices solved by \eqref{Eq:U1} into \eqref{Eq:V1}, we obtain the receive matrices. 

Since the genie branch indicates which ICIs can be aligned together, it is undoubted that the alignment equations obtained from genie branch are always solvable. Therefore, to prove the achievability, it is only necessary to prove that the IA transceiver obtained from the genie tree can satisfy \eqref{Eq:IA_Conditions}.

\subsection{IA Transceiver Feasibility}
To satisfy $\mathrm{rank}(\pmb{U}_{i_k}) = d$, it is necessary to satisfy $\mathrm{rank}(\pmb{W}^{U}_{\ell})=Kd/(\bar{K}_{n-1}q_{n-1}^{\mathrm{A}})$, i.e., $\pmb{A}_{\ell}^{H}\pmb{A}_{\ell}$ has $Kd/(\bar{K}_{n-1}q_{n-1}^{\mathrm{A}})$ independent eigenvectors corresponding to zero eigenvalues. Since $\pmb{A}_{\ell}\in\mathbb{C}^{p_{n-1}^{\mathrm{A}}M\times q_{n-1}^{\mathrm{A}}N}$, from \cite{Rank_nullity} we have
\begin{subequations}\label{Eq:Dim_Null_U0_ALL}
\begin{align}
 &\mathrm{dim}\left( \mathcal{V}\left\{\pmb{A}_{\ell}^{H}\pmb{A}_{\ell}\right\}\right)\geq q_{n-1}^{\mathrm{A}}N-p_{n-1}^{\mathrm{A}}M\nonumber \\
\label{Eq:Dim_Null_U3}
             =&q_{n-1}^{\mathrm{A}}\left(1+K/{C_{n-1}^{\mathrm{A}}}\right)d -p_{n-1}^{\mathrm{A}}\left(K+C_{n}^{\mathrm{A}}\right)d\\
\label{Eq:Dim_Null_U4}
             =&p_{n-1}^{\mathrm{A}}\left( C_{n-1}^{\mathrm{A}} - C_{n}^{\mathrm{A}}\right)d \\
\label{Eq:Dim_Null_U5}
  =& 1/{p_{n}^{\mathrm{A}}}d \\
\label{Eq:Dim_Null_U6}
  =& Kd/(\bar{K}_{n-1}q_{n-1}^{\mathrm{A}})
\end{align}
\end{subequations}
where \eqref{Eq:Dim_Null_U3} is from $M= \left(K+C_{n}^{\mathrm{A}}\right)d$ and $N= \left(1+K/C_{n-1}^{\mathrm{A}}\right)d$, \eqref{Eq:Dim_Null_U4} follows from $C_{n-1}^{\mathrm{A}}=q_{n-1}^{\mathrm{A}}/p_{n-1}^{\mathrm{A}}$ in \eqref{Eq:C_pq}, \eqref{Eq:Dim_Null_U5} comes from $C_{n-1}^{\mathrm{A}} - C_{n}^{\mathrm{A}} = 1/(p_{n-1}^{\mathrm{A}}p_{n}^{\mathrm{A}})$ in \eqref{Eq:Csequence_serieA}, and \eqref{Eq:Dim_Null_U6} follows from ${q_{n-1}^{\mathrm{A}}}/{p_{n}^{\mathrm{A}}}=K/\bar{K}_{n-1}$,
which is obtained by using the mathematical induction.

Since $\pmb{W}^{U}_{\ell}$ has $ Kd/(\bar{K}_{n-1}q_{n-1}^{\mathrm{A}})$ column vectors, \eqref{Eq:Dim_Null_U6} ensures that the column vectors of $\pmb{W}^{U}_{\ell}$ are independent. Therefore, the transmit vectors are independent, i.e., $\mathrm{rank}(\pmb{U}_{i_k}) = d$.

To satisfy \eqref{Eq:Constraint_ICI_free}, the BSs and users need to cancel all ICIs, i.e., $I^{N}+I^{M}\geq(G-1)Kd$, where $I^{N}$ and $I^{M}$ are the numbers of ICIs that can be canceled by the BS and the user, respectively.

In \eqref{Eq:U1}, since
$\pmb{A}_{\ell}$ has $p_{n-1}^{\mathrm{A}}M$ rows and $\pmb{W}^{U}_{\ell}$ has
$Kd/(\bar{K}_{n-1} q_{n-1}^{\mathrm{A}})$ columns, we can obtain $p_{n-1}^{\mathrm{A}}Kd/(\bar{K}_{n-1} q_{n-1}^{\mathrm{A}})$ alignment equations from each alignment matrix. There are $G\bar{K}_{n-1}$ alignment matrices and $G$ BSs, then $I^{N}$ is obtained as
\begin{align}
I^{N}=\frac{p_{n-1}^{\mathrm{A}}Kd}{\bar{K}_{n-1}q_{n-1}^{\mathrm{A}}}
\frac{G\bar{K}_{n-1}}{G}=\frac{Kd}{C_{n-1}^{\mathrm{A}}}
\end{align}

To reserve $Kd$-dimensional subspace to receive the desired signals, i.e., ensure $\mathrm{rank}(\pmb{V}_{j}) = Kd$, $I^{N}$ is obtained as
\begin{align}
I^{M}=M-Kd
=C_{n}^{\mathrm{A}}d
\end{align}

From \eqref{Eq:Csequence_RecursiveA}, we have $I^{N}+I^{M}=(G-1)Kd$. It means that the total ICIs are canceled by BSs and users thoroughly.

As a result, the jointly designed transmitters in \eqref{Eq:U1} and receivers in \eqref{Eq:V1} satisfy the IA conditions in \eqref{Eq:IA_Conditions}, which indicates that the linear IA is feasible and the quantity DoF bound is achievable.

This completes the proof of Theorem~\ref{Theorem:Achievable_DoF}.

\subsection{Applications}
In this subsection, we take several examples to show how to use the genie branches to establish the alignment equations. To understand the relationship of our alignment graph and alignment matrix with the subspace alignment chain in \cite{Jafar_3cell} and the block matrix in \cite{Tse_3cell_2011}, we consider three examples in a three-cell one-user MIMO-IC network, whose configurations and the corresponding achievable DoF are listed in Table \ref{Table:CN2}.

\begin{table}[htb!]\centering
\caption{Antenna configurations and achievable DoF of examples, where $G=3$ and $K=1$.}\label{Table:CN2}
\begin{tabular}{c|c|c|c}
 \hline
 & $M$ & $N$ & $d^{\mathrm{Quan}}$ \\
 \hline
 \hline
\textbf{Ex 9} & 5 & 3 & 2\\
 \hline
\textbf{Ex 10} & 7 & 5 & 3\\
 \hline
\textbf{Ex 11} & 9 & 7 & 4\\
 \hline
\end{tabular}
\end{table}

\subsubsection{\textbf{Ex 9}}
We consider \textbf{Ex 9} where $M=5$, $N=3$, and $d=2$.
Since $M/N \in \mathcal{R}^{\mathrm{II-A}}_{2}=[3/2,2)$, we know that the genie tree is a full tree with height of two. According to the genie tree, we can obtain three genie branches with height of one.

\begin{figure}[htb!]
\centering
\includegraphics[width=1.00\linewidth]{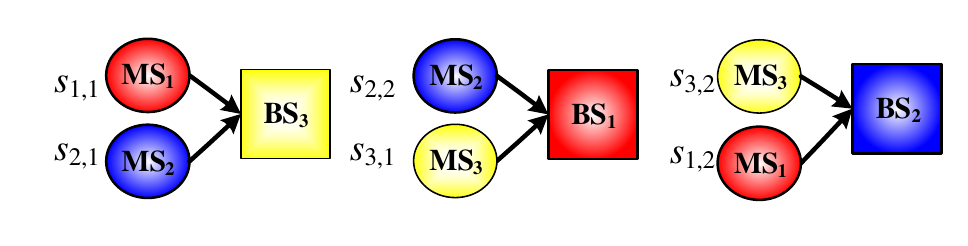}
\caption{Alignment graphs for \textbf{Ex 9}.} \label{fig:Aligned_matrix_Ex9}
\end{figure}

Since the height of genie branches is one, we can obtain the alignment graphs from the genie branches directly, which are shown in Fig. \ref{fig:Aligned_matrix_Ex9}.

According to the alignment graphs in Fig. \ref{fig:Aligned_matrix_Ex9}, we can obtain the alignment equations as follows,
\begin{align*}
 &\pmb{H}_{3,1}\pmb{u}_{1,1}+\pmb{H}_{3,2}\pmb{u}_{2,1}=\pmb{0} \\
 &\pmb{H}_{1,2}\pmb{u}_{2,2}+\pmb{H}_{1,3}\pmb{u}_{3,1}=\pmb{0} \\
 &\pmb{H}_{2,3}\pmb{u}_{3,2}+\pmb{H}_{2,1}\pmb{u}_{1,2}=\pmb{0}
\end{align*}
which can be expressed as
\begin{align*}
\pmb{A}_{\ell}\pmb{W}^{U}_{\ell}=\pmb{0},~\ell=1,2,3
\end{align*}
where
\begin{align*}
  \pmb{A}_{1} &= [\pmb{H}_{3,1},\pmb{H}_{3,2}],~\pmb{W}^{U}_{1} = [\pmb{u}_{1,1}^{T},\pmb{u}_{2,1}^{T}]^{T}\\
  \pmb{A}_{2} &= [\pmb{H}_{1,2},\pmb{H}_{1,3}],~\pmb{W}^{U}_{2} = [\pmb{u}_{2,2}^{T},\pmb{u}_{3,1}^{T}]^{T}\\
  \pmb{A}_{3} &= [\pmb{H}_{2,3},\pmb{H}_{2,1}],~\pmb{W}^{U}_{3} = [\pmb{u}_{3,2}^{T},\pmb{u}_{1,2}^{T}]^{T}
\end{align*}

For each BS, the number of received ICIs is $(G-1)Kd=4$ ICIs. There are three alignment equations for three BSs. Therefore, the users can help each BS cancel one ICI. Meanwhile, each BS can cancel the remaining $M-Kd = 5-2=3$ ICIs. As a result, all the ICIs can be jointly eliminated by the users and BSs, so that $d^{\mathrm{Quan}}=2$ is achievable by the designed IA transceiver.

\subsubsection{\textbf{Ex 10}}
We consider \textbf{Ex 10} where $M=7$, $N=5$, and $d=3$.
Since $M/N \in \mathcal{R}^{\mathrm{II-A}}_{3}=[4/3,3/2)$, we know that the genie tree is a full tree with height of three. According to the genie tree, we can obtain three genie branches with height of two.

\begin{figure}[htb!]
\centering
\includegraphics[width=1.00\linewidth]{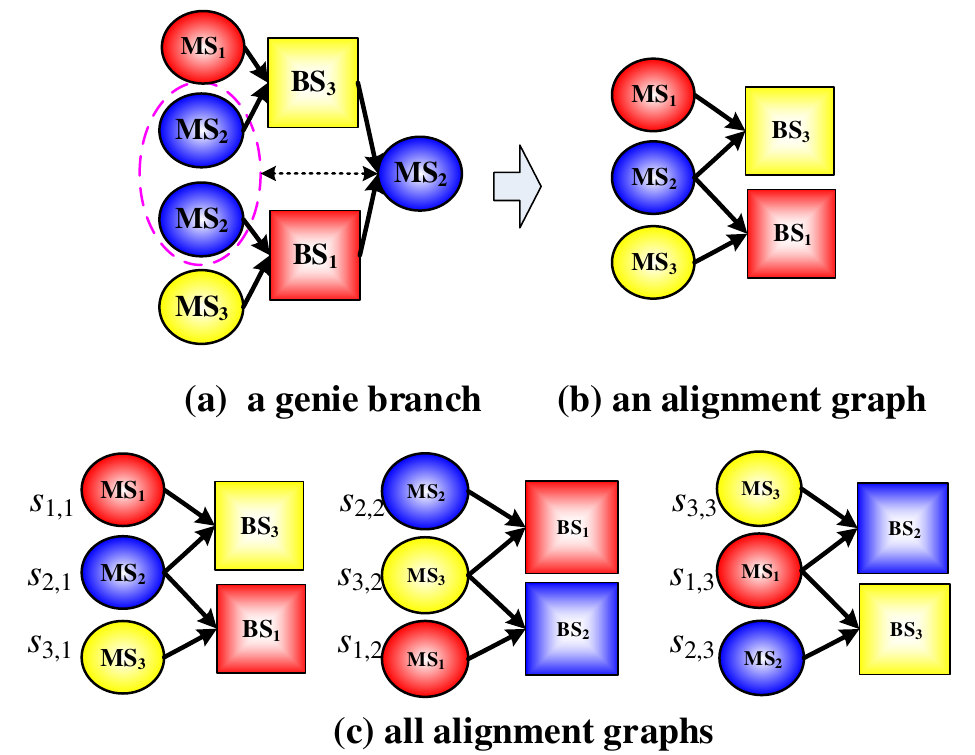}
\caption{Genie branch and alignment graphs for \textbf{Ex 10}.} \label{fig:Aligned_matrix_Ex10}
\end{figure}

As shown in Fig. \ref{fig:Aligned_matrix_Ex10}(a), the aligned ICIs at BS$_3$ and BS$_1$ can be further aligned at MS$_2$ in the 2nd round. It means that there is one redundant vertex between the two leaf vertexes labeled by MS$_{2}$. By merging one redundant vertex of the genie branch, we obtain the alignment graph in Fig. \ref{fig:Aligned_matrix_Ex10}(b).

Following the similar way, we can obtain three alignment graphs shown in Fig. \ref{fig:Aligned_matrix_Ex10}(c). According to these alignment graphs, we can obtain the alignment equations as follows,
\begin{align*}
&\left\{
\begin{array}{l}
\pmb{H}_{3,1}\pmb{u}_{1,1}+\pmb{H}_{3,2}\pmb{u}_{2,1}=\pmb{0}  \\
\pmb{H}_{1,2}\pmb{u}_{2,1}+\pmb{H}_{1,3}\pmb{u}_{3,1}=\pmb{0}
\end{array}
\right.\\
&\left\{
\begin{array}{l}
 \pmb{H}_{1,2}\pmb{u}_{2,2}+\pmb{H}_{1,3}\pmb{u}_{3,2}=\pmb{0} \\
 \pmb{H}_{2,3}\pmb{u}_{3,2}+\pmb{H}_{2,1}\pmb{u}_{1,2}=\pmb{0}
\end{array}
\right.\\
&\left\{
\begin{array}{l}
 \pmb{H}_{2,3}\pmb{u}_{3,3}+\pmb{H}_{2,1}\pmb{u}_{1,3}=\pmb{0} \\
 \pmb{H}_{3,1}\pmb{u}_{1,3}+\pmb{H}_{3,2}\pmb{u}_{2,3}=\pmb{0}
\end{array}
\right.
\end{align*}
which can be expressed as
\begin{align*}
\pmb{A}_{\ell}\pmb{W}^{U}_{\ell}=\pmb{0},~\ell=1,2,3
\end{align*}
where
\begin{align*}
  \pmb{A}_{1} &=
 \left [\begin{array}{lll}
     \pmb{H}_{3,1} & \pmb{H}_{3,2}  & \\
      & \pmb{H}_{1,2}  & \pmb{H}_{1,3}\\
   \end{array}\right],~\pmb{W}^{U}_{1} = [\pmb{u}_{1,1}^{T},\pmb{u}_{2,1}^{T},\pmb{u}_{3,1}^{T}]^{T}\\
  \pmb{A}_{2} &=
 \left [\begin{array}{lll}
     \pmb{H}_{1,2} & \pmb{H}_{1,3}  & \\
      & \pmb{H}_{2,3}  & \pmb{H}_{2,1}\\
   \end{array}\right],~\pmb{W}^{U}_{2} = [\pmb{u}_{2,2}^{T},\pmb{u}_{3,2}^{T},\pmb{u}_{1,2}^{T}]^{T}\\
  \pmb{A}_{3} &=
 \left [\begin{array}{lll}
     \pmb{H}_{2,3} & \pmb{H}_{2,1}  & \\
      & \pmb{H}_{3,1}  & \pmb{H}_{3,2}\\
   \end{array}\right],~\pmb{W}^{U}_{3} = [\pmb{u}_{3,3}^{T},\pmb{u}_{1,3}^{T},\pmb{u}_{2,3}^{T}]^{T}
\end{align*}

For each BS, the number of received ICIs is $(G-1)Kd=6$ ICIs. There are six alignment equations for three BSs. Therefore, the users can help each BS cancel two ICIs. Meanwhile, each BS can cancel the remaining $M-Kd = 7-3=4$ ICIs. As a result, all the ICIs can be jointly eliminated by the users and BSs, so that $d^{\mathrm{Quan}}=3$ is achievable by the designed IA transceiver.

\subsubsection{\textbf{Ex 11}}
We consider \textbf{Ex 11} where $M=9$, $N=7$, and $d=4$.
Since $M/N \in \mathcal{R}^{\mathrm{II-A}}_{4}=[5/4,4/3)$, we know that the genie tree is a full tree with height of four. According to the genie tree, we can obtain three genie branches with height of three.

\begin{figure}[htb!]
\centering
\includegraphics[width=1.00\linewidth]{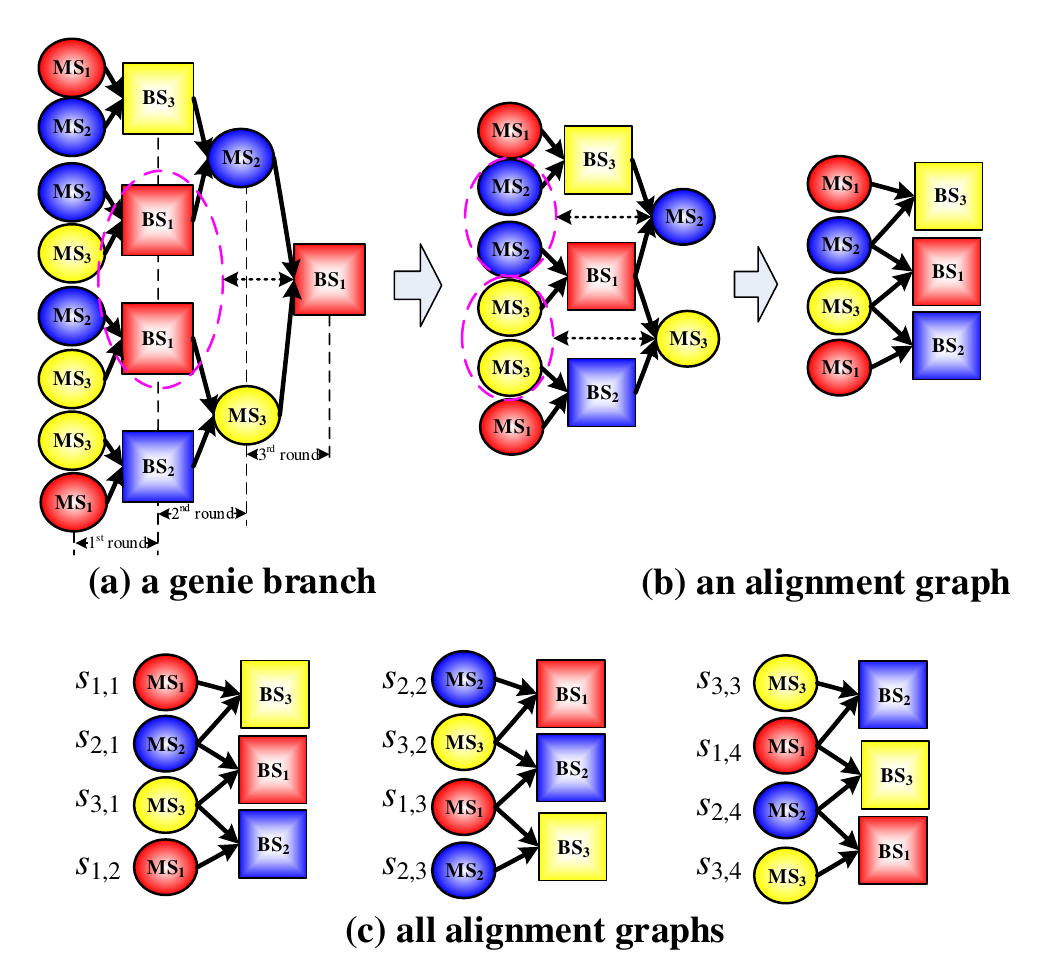}
\caption{Genie branch and alignment graphs for \textbf{Ex 11}.} \label{fig:Aligned_matrix_Ex11}
\end{figure}

As shown in Fig. \ref{fig:Aligned_matrix_Ex11}(a), the aligned ICIs at MS$_2$ and MS$_3$ can be further aligned at BS$_1$ in the 3rd round. By merging one redundant vertex labeled by BS$_{1}$, we obtain a genie branch with height of two. Moreover, the aligned ICIs at BS$_3$ and BS$_1$ can be further aligned at MS$_2$, and the aligned ICIs at BS$_1$ and BS$_2$ can be further aligned at MS$_3$. Therefore, the leaf vertexes labeled by MS$_{2}$ and MS$_{3}$ contain two redundant vertexes. By further merging the redundant vertexes, we obtain the alignment graph shown in Fig. \ref{fig:Aligned_matrix_Ex11}(b).

Following the similar way, we can obtain three alignment graphs shown in Fig. \ref{fig:Aligned_matrix_Ex11}(c). According to these alignment graphs, we can obtain the alignment equations as follows,
\begin{align*}
&\left\{
\begin{array}{l}
\pmb{H}_{3,1}\pmb{u}_{1,1}+\pmb{H}_{3,2}\pmb{u}_{2,1}=\pmb{0} \\
\pmb{H}_{1,2}\pmb{u}_{2,1}+\pmb{H}_{1,3}\pmb{u}_{3,1}=\pmb{0} \\
\pmb{H}_{2,3}\pmb{u}_{3,1}+\pmb{H}_{2,1}\pmb{u}_{1,2}=\pmb{0} \\
\end{array}
\right.\\
&\left\{
\begin{array}{l}
\pmb{H}_{1,2}\pmb{u}_{2,2}+\pmb{H}_{1,3}\pmb{u}_{3,2}=\pmb{0} \\
\pmb{H}_{2,3}\pmb{u}_{3,2}+\pmb{H}_{2,1}\pmb{u}_{1,3}=\pmb{0} \\
\pmb{H}_{3,1}\pmb{u}_{1,3}+\pmb{H}_{3,2}\pmb{u}_{2,3}=\pmb{0} \\
\end{array}
\right.\\
&\left\{
\begin{array}{l}
\pmb{H}_{2,3}\pmb{u}_{3,3}+\pmb{H}_{2,1}\pmb{u}_{1,4}=\pmb{0} \\
\pmb{H}_{3,1}\pmb{u}_{1,4}+\pmb{H}_{3,2}\pmb{u}_{2,4}=\pmb{0} \\
\pmb{H}_{1,2}\pmb{u}_{2,4}+\pmb{H}_{1,3}\pmb{u}_{3,4}=\pmb{0}
\end{array}
\right.
\end{align*}
which can be expressed as
\begin{align*}
\pmb{A}_{\ell}\pmb{W}^{U}_{\ell}=\pmb{0},~\ell=1,2,3
\end{align*}
where
\begin{align*}
  \pmb{A}_{1} &=
 \left [\begin{array}{llll}
     \pmb{H}_{3,1} & \pmb{H}_{3,2}  & &\\
      & \pmb{H}_{1,2}  & \pmb{H}_{1,3} &\\
      & & \pmb{H}_{2,3}  & \pmb{H}_{2,1}\\
   \end{array}\right],~\pmb{W}^{U}_{1} = [\pmb{u}_{1,1}^{T},\pmb{u}_{2,1}^{T},\pmb{u}_{3,1}^{T},\pmb{u}_{1,2}^{T}]^{T}\\
  \pmb{A}_{2} &=
 \left [\begin{array}{llll}
     \pmb{H}_{1,2} & \pmb{H}_{1,3}  & &\\
      & \pmb{H}_{2,3}  & \pmb{H}_{2,1} &\\
      & & \pmb{H}_{3,1}  & \pmb{H}_{3,2} \\
   \end{array}\right],~\pmb{W}^{U}_{2} = [\pmb{u}_{2,2}^{T},\pmb{u}_{3,2}^{T},\pmb{u}_{1,3}^{T},\pmb{u}_{2,3}^{T}]^{T}\\
  \pmb{A}_{3} &=
 \left [\begin{array}{llll}
     \pmb{H}_{2,3} & \pmb{H}_{2,1}  & & \\
      & \pmb{H}_{3,1}  & \pmb{H}_{3,2} & \\
      & & \pmb{H}_{1,2}  & \pmb{H}_{1,3} \\
   \end{array}\right],~\pmb{W}^{U}_{3} = [\pmb{u}_{3,3}^{T},\pmb{u}_{1,4}^{T},\pmb{u}_{2,4}^{T},\pmb{u}_{3,4}^{T}]^{T}
\end{align*}

For each BS, the number of received ICIs is $(G-1)Kd=8$ ICIs. There are nine alignment equations for three BSs. Therefore, the users can help each BS cancel three ICIs. Meanwhile, each BS can cancel the remaining $M-Kd = 9-4=5$ ICIs. As a result, all the ICIs can be jointly eliminated by the users and BSs, so that $d^{\mathrm{Quan}}=4$ is achievable by the designed IA transceiver.

Considering the subspace alignment chains in~\cite{Jafar_3cell} to obtain the achievable DoF for the above three examples, we can obtain the same alignment equations. It is because that the subspace alignment chains in \cite{Jafar_3cell} are the same as our alignment graphs. Moreover, we can obtain the same alignment equations from the block matrices in~\cite{Tse_3cell_2011}
\begin{align*}
 \bar{\pmb{H}}_{1}=\left[
 \begin{array}{cccccc}
 \pmb{H}_{3,1} & \pmb{H}_{3,2} & & & & \\
 & \pmb{H}_{1,2} & \pmb{H}_{1,3}& & & \\
 & & \pmb{H}_{2,3} & \pmb{H}_{2,1} & & \\
 & & & \ddots & \ddots \\
 \end{array} \right]\in \mathbb{C}^{nN \times (n+1)M}
\end{align*}
We can see that the block matrices in~\cite{Tse_3cell_2011} are the same as our alignment matrices.

Therefore, the achievable DoF and the feasible IA transceivers for the three-cell MIMO-IC network are consistent with that in~\cite{Jafar_3cell,Tse_3cell_2011}. Nevertheless, the subspace alignment chains in~\cite{Jafar_3cell} and the block matrices in~\cite{Tse_3cell_2011} cannot be extended into other general cases. It is because, for the three-cell MIMO-IC network, each node (either BS or user) only suffers the ICIs from two interfering nodes, so that the ICIs from two different nodes can be aligned. There is no other alternative. As a result, it is not difficult to obtain the alignment graph to achieve the information theoretical maximal DoF. When each node suffers the ICIs from more than two nodes, there are multiple ways to align ICIs. It is impossible to enumerate all possibilities to design the alignment graph. Therefore, the subspace alignment chain in~\cite{Jafar_3cell} cannot be extended into general cases. Moreover, by substituting $G=3$ and $K=1$ into \eqref{Eq:Fsequence_AB}, the generalized Fibonacci sequence reduces to the arithmetic sequence. However, the analysis about the block matrix in~\cite{Tse_3cell_2011} depends on the property of the arithmetic sequence, which cannot be extended into general cases.

\subsubsection{\textbf{Ex 4}}
In \cite{Jafar_IAchainKcell_all}, the authors conjectured that the maximal DoF in the region $N/M \in ((G-1)/(G(G-2)),(G-2)/(G^2-3G+1))$ is $d^{\mathrm{Decom}}=MN/(M+N)$ when $G\geq 4$ and $K=1$. Here, we illustrate an example in this region to show that there exist a achievable DoF higher than $d^{\mathrm{Decom}}=MN/(M+N)$.

\begin{figure}[htb!]
\centering
\includegraphics[width=1.00\linewidth]{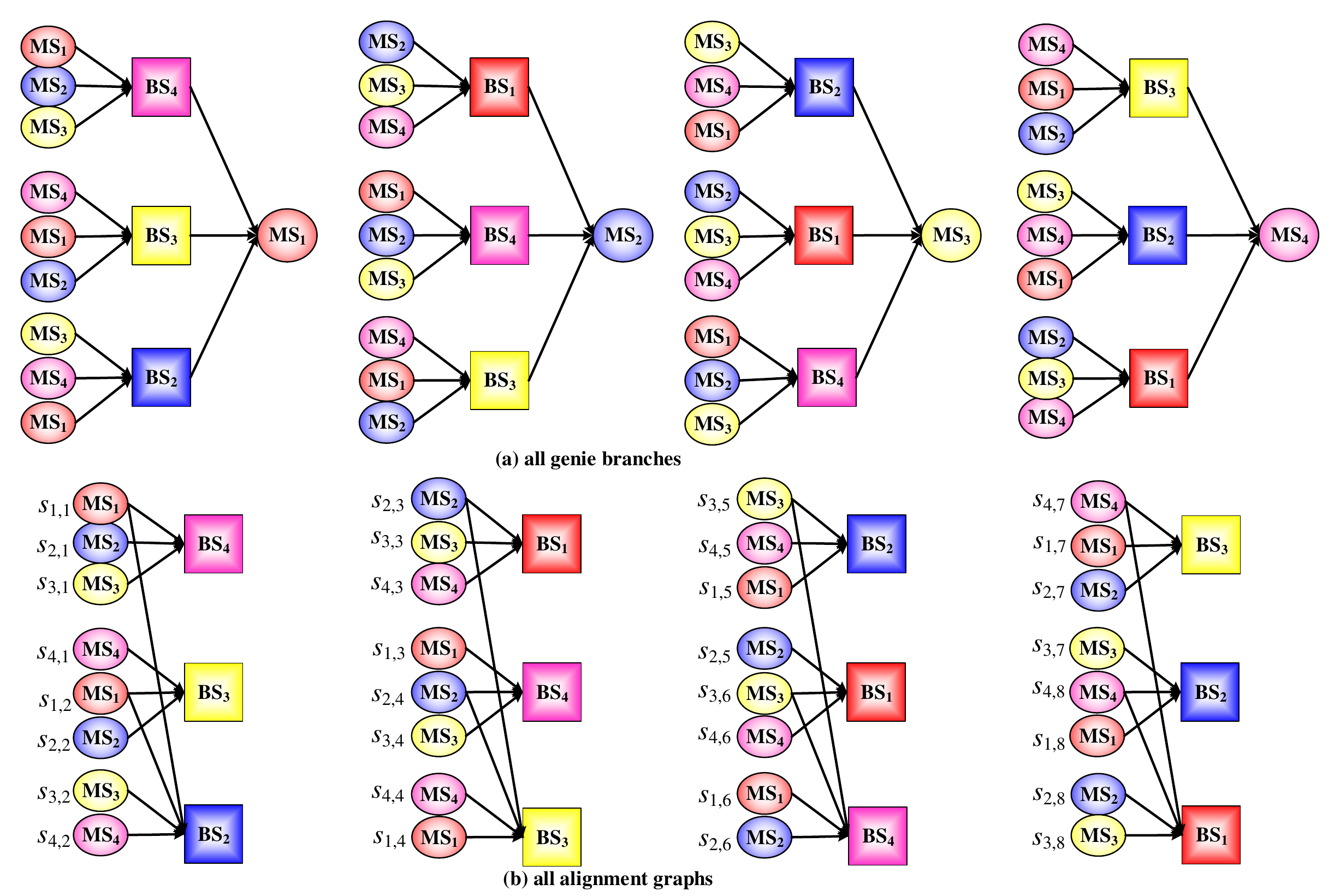}
\caption{Genie branches and alignment graphs for \textbf{Ex 4}.} \label{fig:Aligned_matrix_Ex4}
\end{figure}

We consider \textbf{Ex 4} where $G=4$, $K=1$, $M=29$, $N=11$.
Since $N/M \in ((G-1)/(G(G-2)),(G-2)/(G^2-3G+1))=(3/8,2/5)$, the conjectured  maximal DoF is $d^{\mathrm{Decom}}=319/40$. Considering $M/N \in \mathcal{R}^{\mathrm{II-A}}_{3}=[21/8,8/3)$, the achievable DoF is
$d^{\mathrm{Quan}}=8$ form Theorem \ref{Theorem:Achievable_DoF}. Since $M/N \in \mathcal{R}^{\mathrm{II-A}}_{3}$, we know that the genie tree is a full tree with height of three. According to the genie tree, we can obtain four genie branches with height of two as shown in Fig. \ref{fig:Aligned_matrix_Ex4}(a).

Following the similar analysis as \textbf{Ex 11}, we can obtain three alignment graphs shown in Fig. \ref{fig:Aligned_matrix_Ex11}(b). According to these alignment graphs, we can obtain the alignment equations as follows,
\begin{align*}
&\left\{
\begin{array}{l}
  \pmb{H}_{4,1}\pmb{u}_{1,1} + \pmb{H}_{4,2}\pmb{u}_{2,1} + \pmb{H}_{4,3}\pmb{u}_{3,1}=\pmb{0} \\
  \pmb{H}_{3,4}\pmb{u}_{4,1} + \pmb{H}_{3,1}\pmb{u}_{1,1} +
  \pmb{H}_{3,2}\pmb{u}_{2,1} =\pmb{0}\\
  \pmb{H}_{2,1}\pmb{u}_{1,1}+\pmb{H}_{2,1}\pmb{u}_{1,2} + \pmb{H}_{2,3}\pmb{u}_{3,2} + \pmb{H}_{2,4}\pmb{u}_{4,2}=\pmb{0}
\end{array}
\right.\\
&\left\{
\begin{array}{l}
  \pmb{H}_{1,2}\pmb{u}_{2,3} + \pmb{H}_{1,3}\pmb{u}_{3,3} + \pmb{H}_{1,4}\pmb{u}_{3,3}=\pmb{0} \\
  \pmb{H}_{4,1}\pmb{u}_{1,3} + \pmb{H}_{4,2}\pmb{u}_{2,4} +
  \pmb{H}_{4,3}\pmb{u}_{3,4} =\pmb{0}\\
  \pmb{H}_{3,2}\pmb{u}_{2,3} + \pmb{H}_{3,2}\pmb{u}_{3,3} + \pmb{H}_{3,4}\pmb{u}_{4,4} + \pmb{H}_{3,1}\pmb{u}_{1,4}=\pmb{0}
\end{array}
\right.\\
&\left\{
\begin{array}{l}
  \pmb{H}_{2,3}\pmb{u}_{3,5} + \pmb{H}_{2,4}\pmb{u}_{4,5} + \pmb{H}_{2,1}\pmb{u}_{1,5}=\pmb{0} \\
  \pmb{H}_{1,2}\pmb{u}_{2,5} + \pmb{H}_{1,3}\pmb{u}_{3,6} +
  \pmb{H}_{1,4}\pmb{u}_{4,6} =\pmb{0}\\
  \pmb{H}_{4,3}\pmb{u}_{3,5} + \pmb{H}_{4,3}\pmb{u}_{3,6} + \pmb{H}_{4,1}\pmb{u}_{1,6} + \pmb{H}_{4,2}\pmb{u}_{2,6}=\pmb{0}
\end{array}
\right.\\
&\left\{
\begin{array}{l}
  \pmb{H}_{3,4}\pmb{u}_{4,7} + \pmb{H}_{3,1}\pmb{u}_{1,7} + \pmb{H}_{3,2}\pmb{u}_{2,7}=\pmb{0} \\
  \pmb{H}_{2,3}\pmb{u}_{3,7} + \pmb{H}_{2,4}\pmb{u}_{4,8} +
  \pmb{H}_{2,1}\pmb{u}_{1,8} =\pmb{0}\\
  \pmb{H}_{1,4}\pmb{u}_{4,7} + \pmb{H}_{1,4}\pmb{u}_{4,8} + \pmb{H}_{1,2}\pmb{u}_{2,8} + \pmb{H}_{1,3}\pmb{u}_{3,8}=\pmb{0}
\end{array}
\right.
\end{align*}
which can be expressed as
\begin{align*}
\pmb{A}_{\ell}\pmb{W}^{U}_{\ell}=\pmb{0},~\ell=1,2,3,4
\end{align*}
where
\begin{align*}
  \pmb{A}_{1} &=
 \left [
\begin{array}{llllllll}
  \pmb{H}_{4,1} & \pmb{H}_{4,2} & \pmb{H}_{4,3} &  &  &  &  &  \\
   &  &  & \pmb{H}_{3,4} & \pmb{H}_{3,1} & \pmb{H}_{3,2} &  &  \\
  \pmb{H}_{2,1} &  &  &  &  \pmb{H}_{2,1} &  & \pmb{H}_{2,3} & \pmb{H}_{2,4}
\end{array}\right]\\
  \pmb{A}_{2} &=
 \left [
\begin{array}{llllllll}
  \pmb{H}_{1,2} & \pmb{H}_{1,2} & \pmb{H}_{1,4} &  &  &  &  &  \\
   &  &  & \pmb{H}_{4,1} & \pmb{H}_{4,2} & \pmb{H}_{4,3} &  &  \\
  \pmb{H}_{3,2} &  &  &  &  \pmb{H}_{3,2} &  & \pmb{H}_{3,4} & \pmb{H}_{3,1}
\end{array}\right]\\
  \pmb{A}_{3} &=
 \left [
\begin{array}{llllllll}
  \pmb{H}_{2,3} & \pmb{H}_{2,4} & \pmb{H}_{2,1} &  &  &  &  &  \\
   &  &  & \pmb{H}_{1,2} & \pmb{H}_{1,3} & \pmb{H}_{1,4} &  &  \\
  \pmb{H}_{4,3} &  &  &  &  \pmb{H}_{4,3} &  & \pmb{H}_{4,1} & \pmb{H}_{4,2}
\end{array}\right]\\
  \pmb{A}_{4} &=
 \left [
\begin{array}{llllllll}
  \pmb{H}_{3,4} & \pmb{H}_{3,1} & \pmb{H}_{3,2} &  &  &  &  &  \\
   &  &  & \pmb{H}_{2,3} & \pmb{H}_{2,4} & \pmb{H}_{2,1} &  &  \\
  \pmb{H}_{1,4} &  &  &  &  \pmb{H}_{1,4} &  & \pmb{H}_{1,2} & \pmb{H}_{1,3}
\end{array}\right]
\end{align*}

\begin{align*}
\pmb{W}^{U}_{1} = [\pmb{u}_{1,1}^{T},\pmb{u}_{2,1}^{T},\pmb{u}_{3,1}^{T},\pmb{u}_{4,1}^{T}, \pmb{u}_{1,2}^{T},\pmb{u}_{2,2}^{T},\pmb{u}_{3,2}^{T},\pmb{u}_{4,2}^{T}]^{T}\\
\pmb{W}^{U}_{2} = [\pmb{u}_{2,3}^{T},\pmb{u}_{3,3}^{T},\pmb{u}_{4,3}^{T}, \pmb{u}_{1,3}^{T}, \pmb{u}_{2,4}^{T},\pmb{u}_{3,4}^{T},\pmb{u}_{4,4}^{T}£¬\pmb{u}_{1,4}^{T}]^{T}\\
\pmb{W}^{U}_{3} = [\pmb{u}_{3,5}^{T},\pmb{u}_{4,5}^{T}, \pmb{u}_{1,5}^{T},\pmb{u}_{2,5}^{T}, \pmb{u}_{3,6}^{T},\pmb{u}_{4,6}^{T},\pmb{u}_{1,6}^{T},\pmb{u}_{2,6}^{T}]^{T}\\
\pmb{W}^{U}_{4} = [\pmb{u}_{4,7}^{T},\pmb{u}_{1,7}^{T},\pmb{u}_{2,7}^{T},\pmb{u}_{3,7}^{T}, \pmb{u}_{4,8}^{T},\pmb{u}_{1,8}^{T},\pmb{u}_{2,8}^{T},\pmb{u}_{3,8}^{T}]^{T}\\
\end{align*}

For each BS, the number of received ICIs is $(G-1)Kd=24$ ICIs. There are twelve alignment equations for four BSs. Therefore, the users can help each BS cancel three ICIs. Meanwhile, each BS can cancel the remaining $M-Kd = 29-8=21$ ICIs. As a result, all the ICIs can be jointly eliminated by the users and BSs, so that $d^{\mathrm{Quan}}=8$ is achievable by the designed IA transceiver.

From the examples, we can see that the genie branches can reflect how to align ICIs to make the transceivers to cancel the maximal number of ICIs. Therefore, the IA transceivers designed from the genie branches can achieve the DoF upper-bound for the networks with the antenna configurations in Region II.


\section{Conclusion}\label{Sec:Conclusion}
In this paper, we provided the information theoretic maximal DoF for the $G$-cell $K$-user $M\times N$ symmetric MIMO-IBC network. For most antenna configurations in Region I, the information theoretic maximal DoF per user are $MN/(M+KN)$, which can be achieved by the asymptotic IA. The sum DoF linearly increase with $G$ but at the cost of infinite symbol extensions. For all antenna configurations in Region II, the information theoretic maximal DoF are the piecewise linear function of $M$ and $N$, which can be achieved by the linear IA. The sum DoF are bounded by $M+N$. To derive the DoF upper-bound for the networks with arbitrary antenna configurations, we proposed a unified way to construct the genie to help each BS or user resolve the maximal number of ICIs. By converting an information theoretic DoF upper-bound problem into a linear algebra problem, we obtained the closed-form DoF upper-bound expression for general case. The constructed genies constituted a genie tree, which indicates how to introduce genies to derive
the tightest DoF upper-bound and how to align ICIs to derive the maximal achievable DoF. From the genie tree, we designed a non-iterative linear IA transceiver, which can achieve the information theoretic maximal DoF in Region II. The basic principles to derive the DoF upper-bound and design the feasible IA transceiver can be extended into more general asymmetric networks.

\bibliographystyle{IEEEtran}


\begin{biography}[{\includegraphics[width=1in,height=1.25in,clip,keepaspectratio]{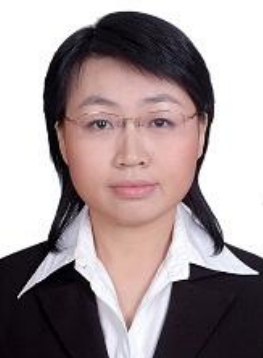}}]
{Tingting Liu} (S'09--M'11) received the B.S. and Ph.D. degrees in signal
and information processing from Beihang University, Beijing, China,
in 2004 and 2011, respectively. From December 2008 to January 2010, she was a visiting student with the School of Electronics and Computer Science, University of Southampton, Southampton, U.K.

She is currently an associate professor in the School of Electronics and Information Engineering, Beihang University (BUAA).  Her PhD thesis was received ¡°2013 National Excellent Doctoral Thesis Nomination¡±and ¡°2012 Excellent Beijing Doctoral Thesis Award¡±. Her research interests include wireless communications and signal processing, the interference management for ultra-dense networks.
\end{biography}

\begin{biography}[{\includegraphics[width=1in,height=1.25in,clip,keepaspectratio]{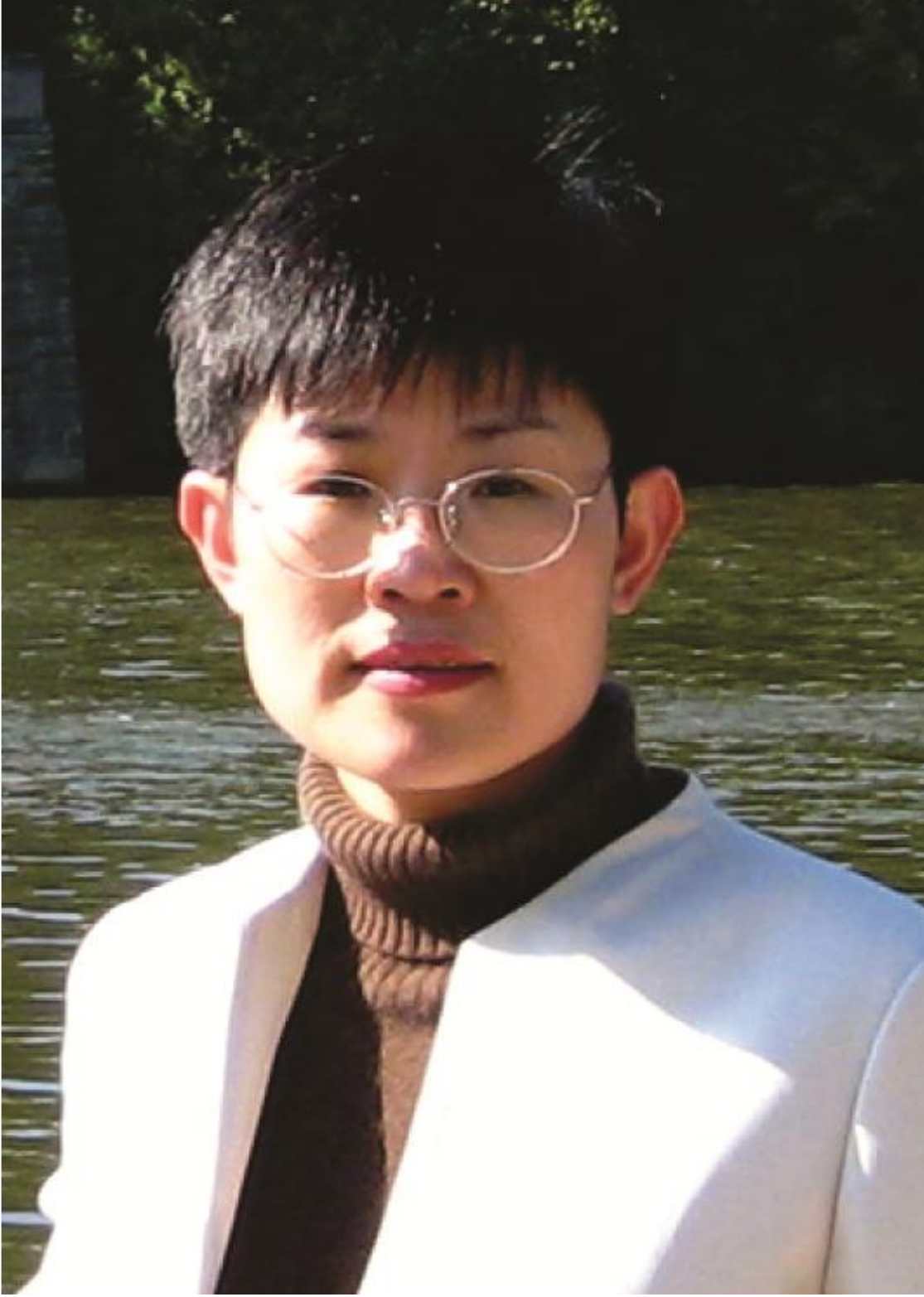}}]
{Chenyang Yang} (SM'08) received the M.S.E and Ph.D. degrees in
electrical engineering from Beihang University (formerly Beijing
University of Aeronautics and Astronautics), Beijing, China, in 1989
and 1997, respectively.

Since 1999, she has been a Full Professor with the School of Electronic and Information Engineering, Beihang University. She has published more than
200 international journal and conference papers and filed more than 60 patents in the fields of energy-efficient transmission, coordinated multi-point, interference management, cognitive radio, relay, etc. Her recent research interests include green radio, local caching, and other emerging
techniques for next generation wireless networks.

Prof. Yang was the Chair of the Beijing chapter of the IEEE Communications
Society during 2008-2012 and the Membership Development Committee
Chair of the Asia Pacific Board, IEEE Communications Society, during 2011-
2013. She has served as a Technical Program Committee Member for numerous
IEEE conferences and was the Publication Chair of the IEEE International
Conference on Communications in China 2012 and a Special Session Chair
of the IEEE China Summit and International Conference on Signal and
Information Processing (ChinaSIP) in 2013. She served as an Associate Editor
for the IEEE TRANSACTIONS ON WIRELESS COMMUNICATIONS during
2009-2014 and a Guest Editor for the IEEE JOURNAL ON SELECTED TOPICS
IN SIGNAL PROCESSING published in February 2015. She is currently an
Associate Editor-in-Chief of the \emph{Chinese Journal of Communications} and the \emph{Chinese Journal of Signal Processing}. She was nominated as an Outstanding Young Professor of Beijing in 1995 and was supported by the First Teaching and Research Award Program for Outstanding Young Teachers of Higher
Education Institutions by the Ministry of Education during 1999-2004.
\end{biography}

\end{document}